\documentclass[11pt,final]{article}
\usepackage{MRnotes}
\usepackage{tabularray}
\usepackage{placeins}
\usepackage{caption}
\usepackage{subcaption}
\usepackage{graphicx}

\title{The evaporation of near-extremal black holes
\\ through charged particle emission} 

\author{Ilija Rakic}

\affiliation{
Center for Quantum Mathematics and Physics (QMAP)\\
Department of Physics \& Astronomy, University of California, Davis, CA 95616 USA}

\emailAdd{irakic@ucdavis.edu}

\abstract{We compute the quantum rate for massless charged scalar emission by a near-extremal Reissner-Nordström black hole using Schwarzian theory as the effective description of the black hole. This is compared to the semi-classical Hawking rate which we also compute near extremality. We classify black holes into small and large, each with a unique spectrum. For small black holes, at energies below a particular quantum scale, the emission is captured by the quantum rate, which gives different predictions from the semi-classical. Furthermore, depending on how the energy compares to another scale associated with the phenomenon of superradiance, the radiation either comes out as mostly non-superradiant or mostly superradiant. For non-superradiant emission, it is found that the quantum rate is suppressed compared to the semi-classical in the same way as recently observed for neutral radiation. For superradiant emission, a novel behavior is observed, the quantum rate is enhanced compared to the semi-classical. For large black holes, we argue that the quantum rate always reduces to the semi-classical. In the limit of very large black holes, from our semi-classical rate, we recover the Gibbons result of Schwinger-like suppression. This gives a unified story of near-extremal charged emission rates, both quantum and semi-classical, which covers all sizes of black hole and all energy regimes. We use these rates to discuss the evaporation history of each type of black hole, from when it starts very near extremality, until it has left this regime. Finally, for the near-extremal Kerr black hole, we argue that the quantum rate always reduces to the semi-classical with the superradiant modes dominating. This rate is computed for spin $0,1,2$. Our analysis emphasizes the AdS$_2$ structure of the Reissner-Nordström and Kerr near-horizon regions, which enables a completely parallel treatment of the two.}

\begin{document}

\maketitle

\section{Introduction}\label{sec:intro}

Black holes radiate field quanta, as shown by Hawking \cite{Hawking:1974,Hawking:1975}. In this sense, they behave as ordinary matter at a temperature determined by their surface gravity, and have an emission spectrum such as that of a black-body up to a frequency-dependent term called the greybody factor. As a result, black holes lose energy, and possibly charge and angular momentum, which implies that they evaporate, i.e. they have a finite lifetime determined by the rate at which they radiate.

The greybody factor is found by solving the problem of scattering the field in question against the black hole background. This was done for the Reissner-Nordström charged black hole emitting charged scalar radiation by Gibbons in \cite{Gibbons:1975}. For the Kerr rotating black hole, the scattering problem involving a scalar field was solved by Starobinskii \cite{Starobinskii:1973a} in the context of classical wave scattering and was later extended to electromagnetic and gravitational radiation \cite{Starobinskii:1973b} (this work was based on the analysis of fields with spin in the Kerr background by Press \& Teukolsky in \cite{Teukolsky:1973,Press:1973,Teukolsky:1974}). The analytical results in these works were performed under various approximation conditions on the black hole and radiation parameters, but were subsequently extended numerically to the full range of parameters in a series of papers by Page \cite{Page:1976a,Page:1976b,Page:1977}, which resulted in a full evaporation history of these black holes in our Universe.

Recent work \cite{Brown:2024} has shown that there is more to this story for near-extremal black holes, those with mass near their extremal limit or, equivalently, black holes with a low temperature (for earlier work, see \cite{Bai:2023}, and for extensions to supergravity, see \cite{Lin:2025}). Before recounting their story, we introduce the effective quantum theory used to describe these black holes. Recent work on low-dimensional gravity has focused on a 2D dilaton-gravity theory called Jackiw-Teitelboim gravity, or JT gravity for short \cite{Jackiw:1984je,Teitelboim:1983ux,Almheiri:2014,Jensen:2016,Engelsoy:2016,Maldacena:2016}, which finds one of its main applications in the quantum description of near-extremal black holes \cite{Sachdev:2015,Almheiri:2016,Nayak:2018,Moitra:2018,Moitra:2019}. The spacetime of a near-extremal Reissner-Nordström black hole near its horizon (nearly) factorizes into AdS$_2$ and a sphere. Dimensional reduction on the sphere of the 4D gravity theory coupled to the scalar field gives rise to 2D JT gravity with additional matter fields. This theory can be quantized and exactly solved. At low temperatures, the dynamics is dominated by a strongly coupled mode, the Schwarzian mode. This mode gives important corrections to the low-temperature thermodynamics of the black hole \cite{Preskill:1991,Page:2000,Ghosh:2019,Iliesiu:2020,Iliesiu:2022,Kapec:2023,Rakic:2023,Kolanowski:2024}. Furthermore, its interactions with other matter fields can be used to study corrections to processes such as the decay of the black hole through emission of quanta, the scattering of fields by the black hole, and others.

The quantum-corrected emission spectra and resulting evaporation rates of near-extremal Reissner-Nordström based on this effective theory were worked out by \cite{Brown:2024} for the most dominant modes of the neutral scalar, electromagnetic and gravitational fields. It was found that these rates are different from the semi-classical ones predicted by Hawking's theory. In a low energy regime, the quantum evaporation rates are suppressed compared to the semi-classical ones, while at energies beyond this regime, it was shown that the quantum rates reduce to the semi-classical ones.

Computing the evaporation rate for these neutral radiation fields using the effective theory involves input from the same wave scattering problem used to compute the greybody factors for the semi-classical rate, and therefore we will refer to this as the wave method. For charged radiation being emitted by a large black hole (large compared to the charge and mass of the particle being emitted), one can employ a completely different technique for computing the rate, which we will refer to as the instanton method. The decay rate to a lower energy state through the emission of a charged particle-antiparticle pair involves finding a geometry called the instanton and comparing it to the original black hole geometry \cite{Coleman:1977a, Coleman:1977b, Coleman:1980}. This results in an exponentially suppressed decay rate, analogous to the Schwinger rate for pair production in a strong electric field \cite{Schwinger:1951}. It was shown in \cite{Brown:2024} that adding quantum corrections coming from the Schwarzian mode to the black hole and instanton geometries does not change the decay rate, implying that it is fully captured by the original theory.

These rates (corrected for neutral and unchanged for charged) were then used by \cite{Brown:2024} to give an evaporation history of the large black hole, different from the one based on the semi-classical Hawking rates. The story goes as follows. A black hole in our Universe emits quanta of all of these fields, with the neutral quanta being the photon and graviton, and the charged quanta being the electron or positron. All higher mass fields can safely be ignored for black holes at low temperatures, because their rates are suppressed due to their higher mass. Since the electron emission rate is highly suppressed, most of the radiation being emitted is neutral, and this causes the black hole to lose energy, and drives it closer to the extremal mass. Eventually, the black hole enters the low-energy quantum regime, and the neutral rates become small, smaller than predicted by the semi-classical theory. Every once in a while, on time scales of order the exponentially suppressed electron emission rate, the electron is emitted carrying both energy and charge. Since the extremal mass bound for the charged black hole is determined by its charge, when the electron is emitted, both the mass and the bound are lowered. It was shown that in this process more charge is carried away than mass, meaning that the black hole moves away from the extremal bound. In fact, this disparity is large enough that the black hole is launched out of the quantum regime. From this point onward, the story repeats itself until all of the charge is exhausted. Since the charged particle rate is unaffected by the corrections from the quantum effective theory, it takes the same amount of time for the black hole to emit all of its charge, but once it reaches that point, it will have emitted less energy (through neutral radiation) than the semi-classical theory would have predicted. Therefore, the lifetime of the black hole is longer, as there is more mass left over to be radiated. This also implies that the black hole spends a larger fraction of its lifetime in the near-extremal regime than the semi-classical theory would have predicted.

The quantum treatment of the neutral emission rates using the wave method, the rates found for charged radiation from the instanton method, and the application of these in deriving the evaporation history of a large near-extremal black hole motivate the following questions, which form the basis of this work.
\begin{itemize}[leftmargin=*]
    \item Can the quantum-corrected evaporation rates for a charged field be computed in an analogous manner to the neutral rates, i.e. by solving the charged field scattering problem and using this as input for the effective quantum theory? This would result in a unified treatment of neutral and charged radiation based on the wave method. Furthermore, can this be done for any combination of the background and field parameters, not just for the large black holes, which were needed to justify the use of the instanton method?
    
    \item For large black holes, does this quantum-corrected rate reduce to the semi-classical one, regardless of the energy regime, as the instanton calculation suggests, and does this semi-classical rate further reduce to the same exponentially suppressed Schwinger-like result the instanton method gives when the black hole is taken to be very large? The latter question was already partially answered by Gibbons in \cite{Gibbons:1975}, where it was shown that for a large black hole the exponentially suppressed rate can be obtained from the wave equation using the approximate technique of WKB theory.\footnote{
        In fact, this was the first work to derive this rate in the context of black holes, and the instanton calculations came later.}
    Our goal will be to fully solve the wave equation and to show that the WKB result is obtained in the appropriate limit.

    \item Using these quantum-corrected evaporation rates, we will be able to derive the evaporation history of a charged black hole of any size. Do these rates lead to a history different from the one that semi-classical theory would predict, at least for some combination of the parameters?
\end{itemize}

We will also consider the evaporation of a near-extremal Kerr black hole. As in the case of Reissner-Nordström, the near-horizon region of the Kerr spacetime is well approximated by the NHEK (Near-Horizon-Extremal-Kerr) geometry \cite{Bardeen:1999}, which also contains AdS$_2$ with the sphere fibered over it. As we will see, the (radial part of the) wave equation of a neutral field of any spin in the near-extremal Kerr geometry has the same form as that of a charged field in Reissner-Nordström, which in part has to do with the fact that both have an AdS$_2$ factor in their near-horizon region. To make this more obvious, the gravity plus matter theory on this spacetime can be dimensionally reduced on the sphere, causing the metric to break up into an AdS$_2$ metric and a U$(1)$ gauge potential corresponding to the $\text{SO}(2)$ symmetry of Kerr and NHEK. The modes of any field in the original 4D background act as massive AdS$_2$ fields that are charged under this gauge potential, with the charge being the azimuthal angular momentum of the original mode. This motivates the following question.
\begin{itemize}[leftmargin=*]
    \item Given the close analogy between near-extremal Reissner-Nordström and Kerr, is a parallel treatment of the two possible, and what does it imply for the evaporation rates and history of near-extremal Kerr?
\end{itemize}
We will answer all of these questions in the positive. Recent work on the first question includes \cite{Maulik:2025, Betzios:2025}, and on the fourth includes \cite{Maulik:2025}. We will indicate where our results recover or differ from the results of those works. We will also mention where our results reproduce the results of \cite{Brown:2024} and the classic results of Gibbons, Starobinskii, and Press \& Teukolsky in certain limits.

One distinctive feature of charged radiation is the phenomenon of superradiance. This was first discovered in the context of black hole scattering in \cite{Zel'dovich:1971,Zel'dovich:1972}, specifically for Kerr in \cite{Starobinskii:1973a,Starobinskii:1973b} and discussed for Reissner-Nordström in \cite{Gibbons:1975}. It can be defined as follows. For charged radiation, for frequencies below a certain threshold frequency, an incident wave can be amplified, meaning that it is reflected with an amplitude higher than the incident one. Later in the text, we will give a definition more suited to a black hole spontaneously emitting radiation. For now, we only mention the most important property of superradiance for our analysis, which is the fact that, unlike for neutral radiation, the evaporation rate of modes with superradiant frequencies does not go to zero as the temperature of the black hole is taken to zero, instead it goes to a constant. This implies that no matter how low the temperature is, there is always a constant flux of charged radiation in this range of frequencies. The same is true for field modes with angular momentum in the Kerr background. We will see that the fact that superradiance exists introduces an additional scale into the problem of charged radiation, which will give it a richer spectrum of solutions than was found for neutral radiation.

Before turning to our results, one comment is in order. A large Reissner-Nordström black hole, no matter at what energy it starts at, naturally approaches extremality, both in the original semi-classical calculation \cite{Gibbons:1975} and in the quantum-corrected one of \cite{Brown:2024}. This is a consequence of the fact that the charged particle emission is highly suppressed such that the neutral radiation has enough time to lower the temperature close enough to zero before the next charged particle is emitted. For small charged black holes, and for Kerr, this is not the case, both neutral and charged radiation is continuously being emitted. Whether the black hole moves towards or away from extremality depends on the ratio of energy emitted to charge/angular momentum emitted. In \cite{Page:1976a,Page:1976b,Page:1977} it was shown that for the entire evaporation history of both of these, charged emission drives the black hole away from extremality. Therefore, if there is any chance of seeing quantum corrections for these black holes, it would be for one starting as close to extremality as possible, as once it has left this region, it never comes back. This motivates the following two additional questions related to the third one listed above. First, do the quantum-corrected evaporation rates give a different ratio of emitted energy to charge than the semi-classical ones, and does it possibly drive the black hole towards extremality? We will find that the ratio is different, but it still drives the black hole away. Secondly, does the black hole spend more or less time in the near-extremal regime than the semi-classical theory predicts? This also touches upon a question raised in \cite{Rakic:2023} of how the time scale associated with the quantum corrections to the thermodynamics of a near-extremal black hole compares with the time it has to spend near extremality. Once the black hole has left the near-extremal regime, the subsequent evolution is just that of \cite{Page:1976a,Page:1976b,Page:1977}.

We now turn to an overview of our work and a presentation of the main results. For Reissner-Nordström emitting a nearly\footnote{
    For most of this work, we will take $m=0$ for simplicity, but to obtain the Gibbons exponential suppression result, we will need $m\neq 0$.}
massless charged scalar field, the scattering problem is defined by three parameters: the size of the black hole determined by its charge $r_\text{h} = \sqrt{G}Q$, a measure of near-extremality taken to be either its temperature $T$ or the energy above extremality $E$, and the charge of the quanta being emitted $e$. We only assume that the black hole is near-extremal, which means that $T$ and $E$ are small. We solve the wave equation for any allowed combination of these parameters and compute both the semi-classical Hawking and the quantum-corrected JT/Schwarzian emission rate.

We define small black holes as those with $e \leq Q \leq \frac{1}{2e}$ and large black holes as those with $Q \geq \frac{1}{2e}$. We will see that this determines whether the conformal dimension\footnote{
    We note that $\Delta$ plays two roles in this work. First, it is just a parameter in the 4D wave equation, which will take a set of possible values, for each of which the wave equation gives sensible physical solutions. Second, it can be though of as the usual conformal dimension of the AdS$_2$/CFT$_1$ correspondence, where we have to be careful with what value it takes.}
$\Delta$, a key parameter in the wave equation, which further determines the character of the spectra, is purely real or both real and complex on the interval of available frequencies. For small black holes, the semi-classical and quantum emission rates of a mode with frequency $\omega$ and angular momentum $l,m_l$ can be found in equations \eqref{eq:RNscrate} and \eqref{eq:RNqrate}, respectively. For large black holes, the semi-classical emission rate is given by \eqref{eq:RNscrate} only for the frequencies with real $\Delta$ and by \eqref{eq:RNscratei} for the rest.

The two parameters $Q$ and $e$ further define two energy scales, the breakdown scale $E_\text{b} = \frac{1}{\sqrt{G}Q^3}$, and the superradiant scale $\omega_\text{sr} = \frac{e}{\sqrt{G}}$. The breakdown scale defines where the quantum regime starts, i.e. where the quantum evaporation rate starts deviating from the semi-classical one. When at least one of the other two scales $E$ and $\omega_\text{sr}$ is above $E_\text{b}$, the quantum evaporation rate to leading order reduces to the semi-classical, so it is safe to compute only the semi-classical rate in this regime. When both of the other two scales are below $E_\text{b}$, the quantum rate gives different predictions for the spectra from the semi-classical. In this case we will compute both rates and compare them. We will find that all large black holes and some of the small ones always have $\omega_\text{sr} > E_\text{b}$, which implies that they are always in the semi-classical regime. The remainder of small black holes is where the novel quantum behavior will be found. A summary of the different sizes and corresponding regimes can be found in figure \ref{fig:logQloge}.

The superradiant scale determines the superradiant frequencies; all modes below are superradiant, and all above are not. 
Depending on how the scales $E$ and $\omega_\text{sr}$ compare to each other, the emission spectra will either be dominated by non-superradiant frequencies or dominated by superradiant ones. For the non-superradiant-dominated radiation, we will find the spectra to be similar to neutral radiation and will therefore also refer to it as neutral-like.

If we assume a separation between these three scales, $E_\text{b}$, $E$, and $\omega_\text{sr}$, the emission rates simplify significantly. For any choice of these scales, we will have a small or large black hole with a semi-classical or quantum, and a neutral-like or superradiant-dominated emission spectrum. We proceed to list our results for each regime (see table \ref{tab:Eregimes} for a summary). Starting with small black holes $e \leq Q \leq \frac{1}{2e}$ with $\omega_\text{sr} \ll E_\text{b} \ll E$ or $E_\text{b} \ll \omega_\text{sr} \ll E$, which defines the semi-classical neutral-like regime, the emission rate of the $l-$th angular momentum mode is found using the semi-classical formula to be
\begin{equation}\label{eq:resnlsc}
    \frac{dE}{dt}\bigg|_{\text{sc}}
    \ = \
    F_l(eQ) \,
    \bigg( \frac{\sqrt{G} E}{Q} \bigg)^{\!2\Delta_e}
    \frac{1}{G Q^2}
    \ \xrightarrow{eQ \ll 1} \
    C_l \,
    \bigg( \frac{\sqrt{G} E}{Q} \bigg)^{\!2l+2}
    \frac{1}{G Q^2} \, ,
\end{equation}
where we have also included the very small charge limit, which just reduces to ordinary neutral radiation. The unitless function $F_l(eQ)$ and the coefficient $C_l$ are given in equations \eqref{eq:Fl} and \eqref{eq:Cl}, and the power $\Delta_e$ in \eqref{eq:Deltae}. Plots of the emission spectrum at different values of $eQ$ are given in figure \ref{fig:QNLSCrate} together with a plot of the integrated rate as a function of $eQ$. Then, the evaporation rate of small black holes with $\omega_\text{sr} \ll E_\text{} \ll E_\text{b} $, which defines the quantum neutral-like regime, is found using the quantum-corrected formula to be
\begin{equation}\label{eq:resnlq}
    \frac{dE}{dt}\bigg|_{\text{q}}
    \ = \
    \hat{F}_l(eQ)\,
    \bigg( \frac{\sqrt{G} E}{Q} \bigg)^{\!2\Delta_e}
    \left( \frac{E_\text{}}{E_\text{b}} \right)^{\!\frac{3}{2}}
    \frac{1}{G Q^2}
    \ \xrightarrow{eQ \ll 1} \
    \hat{C}_l \,
    \bigg( \frac{\sqrt{G} E}{Q} \bigg)^{\!2l+2}
    \left( \frac{E_\text{}}{E_\text{b}} \right)^{\!\frac{3}{2}}
    \frac{1}{G Q^2} \, ,
\end{equation}
where we have again included the neutral limit, which reduces to the result of \cite{Brown:2024}. The coefficients $\hat{F}_l(eQ)$ and $\hat{C}_l$ are given in \eqref{eq:Fhl2} and \eqref{eq:Chl}. This is our first main result. Note that both the semi-classical and quantum rates have the same form, with the main difference being the extra suppression factor in the quantum rate, while The coefficients are roughly the same size. Both have the property that they go to zero as the black hole approaches extremality. Furthermore, both show that the dominant contribution comes from the lowest $l=0$ mode.

Turning to superradiant emission, for small black holes with $E_\text{} \ll E_\text{b} \ll \omega_\text{sr}$ or $E_\text{b} \ll E_\text{} \ll \omega_\text{sr}$, which defines the semi-classical superradiant-dominated regime, the semi-classical formula gives
\begin{equation}\label{eq:ressdsc}
    \frac{dE}{dt}\bigg|_{\text{sc}} =
    G_l(eQ) \,
    \frac{1}{G Q^2}
    \ \xrightarrow{eQ \ll 1} \
    D_l \,
    (eQ)^{4l+4} \,
    \frac{1}{G Q^2} \, ,
\end{equation}
where we have also included a small charge limit. The coefficients $G_l(eQ)$ and $D_l$ can be found in \eqref{eq:Gl} and \eqref{eq:Dl}. Plots are given in figure \ref{fig:QSRDSCrate}. For small black holes with $E_\text{} \ll \omega_\text{sr} \ll E_\text{b} $, which defines the quantum superradiant-dominated regime, the quantum formula gives
\begin{equation}\label{eq:ressdq}
    \frac{dE}{dt}\bigg|_{\text{q}} =
    \hat{G}_l
    \bigg(eQ,\frac{\sqrt{G} E_\text{b}}{e} \bigg) \,
    \frac{1}{G Q^2}
    \ \xrightarrow{eQ \ll 1} \
    \hat{D}_l \,
    (eQ)^{4l+4} \,
    \bigg( \frac{\sqrt{G} E_\text{b}}{e}
    \bigg)^{\!2l+\frac{1}{2}}
    \frac{1}{G Q^2} \, .
\end{equation}
Coefficients $\hat{G}_l\left(eQ,\frac{\sqrt{G} E_\text{b}}{e} \right)$ and $\hat{D}_l$ can be found in equations \eqref{eq:Ghl} and \eqref{eq:Dhl}. This is our second main result. Again, both have the same form, but now the quantum rate is enhanced as a result of the extra factor. Both are temperature-independent, giving a constant flux at extremality, which is a signature of superradiant emission.

For large black holes, as mentioned above, we expect the quantum rate to always reduce to the semi-classical one. For superradiant-dominated emission, the following rate is found using the semi-classical formula
\begin{equation}\label{eq:resl}
    \frac{dE}{dt}\bigg|_{\text{sc}} =
    \Big( G_l(eQ) + G'_l(eQ) \Big) \,
    \frac{1}{G Q^2}
    \ \xrightarrow{eQ \gg 1} \
    \frac{\left( eQ \right)^2}{2\pi} \,
    e^{-2\pi Q(e-\sqrt{e^2-m^2})} \,
    \frac{1}{G Q^2} \, .
\end{equation}
There are now two coefficient functions $G_l(eQ)$ and $G'_l(eQ)$, which can be found in \eqref{eq:lGl} and \eqref{eq:lGlpr}, the first coming from the subset of frequencies where the conformal dimension $\Delta$ is real and the second where it is complex. Each of these is computed by solving the wave equation in a different regime. We have also included the "very large" black hole limit which reproduces the Schwinger-like exponential suppression found in \cite{Gibbons:1975} using the wave method and WKB theory, and in \cite{Brown:2024} using the instanton method. We also plot the emission spectrum for different values of $eQ$ in figure \ref{fig:RNlrate1} where we can explicitly see how the rate transitions into the Gibbons result as $eQ$ is turned up. We do not compute the neutral-like rate for large black holes, as it should follow analogously and does not give any additional insight. With that, all sizes of black hole and all possible energy regimes are exhausted.

For near-extremal Kerr, where the scattering problem is defined by the black hole angular momentum $J$ and the temperature $T$ or energy above extremality $E$, we solve the wave equation and compute the semi-classical emission rate for the scalar, photon, and graviton in equation \eqref{eq:Kerrrate}. We argue that for any combination of these parameters, the quantum rate would reduce to the semi-classical, so we only compute the latter. Note that there is no analogous parameter to $e$, so there is no notion of small versus large black holes, but it turns out that the conformal dimension $\Delta$ can be complex for certain modes, and it turns out that it is so for the most dominant ones. Finally, we will argue that all modes, at all temperatures, as long as the black hole is near-extremal, must be in the superradiant-dominated regime, and therefore there is no neutral-like regime. Putting this all together, we find the emission rate of the $(l,m)$ angular momentum mode to be given by
\begin{equation}\label{eq:resK}
    \frac{dE}{dt} \bigg|^{(s)}_{\text{sc}} =
    \left( H_{slm} + H'_{slm} \right) \,
    \frac{1}{G J} \, ,
\end{equation}
where the coefficients $H_{slm}$ and $H'_{slm}$ are given in \eqref{eq:Hslm} and \eqref{eq:Hslmpr}. The first term comes from the set of frequencies for which $\Delta$ is real and the second from frequencies where it is complex, if such frequencies exist for that mode. We will see that the dominant contributions comes from the $l=m$ modes, and that they fall off with growing $l$, but slowly, implying that for the full rate we need to compute and add up the first few modes if we want any precision (for the real part, see tables \ref{tab:HRDelta0}, \ref{tab:HRDelta1} and \ref{tab:HRDelta2}, and for the complex, see \ref{tab:HCDelta0}, \ref{tab:HCDelta1} and \ref{tab:HCDelta2}).

The paper is organized as follows. We start by introducing the Reissner-Nordström background metric and gauge potential, and its near-extremal region in subsection \ref{ss:RNbckg}. We then solve the wave equation for real $\Delta$ in \ref{ss:RNeq} by breaking it up into an equation in a near region and one in a far region, then solve these two equations, and match their solutions in the overlapping region to obtain a full solution. In \ref{ss:RNsc} we use this solution to compute the greybody factor and the semi-classical evaporation rate. In \ref{ss:RNq} we introduce the effective string formalism, which is used to compute  the quantum-corrected evaporation rate based on Schwarzian theory. Putting this together with additional input from the wave equation, we compute the quantum evaporation rate. In subsection \ref{ss:class}, we define small versus large black holes and their relationship to $\Delta$. Then, we define the quantum and semi-classical regimes, and the neutral-like and superradiant-dominated regimes. In \ref{ss:nl} and \ref{ss:sd} we compute the evporation rates of small black holes in the neutral-like and superradiant-dominated regimes, respectively. In subsection \ref{ss:WKB}, we review the WKB computation of the exponentially suppressed rate of a very large black hole, which does not require us to solve the wave equation. Then, in \ref{ss:RNCDelta} we solve the wave equation for complex $\Delta$, compute the semi-classical evaporation rate, and show that it reduces to the Gibbons result in the very large limit. In section \ref{sec:evapK}, we repeat the analysis for the Kerr black hole, starting with the geometry in \ref{ss:Kbckg}, then turning to the wave equation for a spin-$s$ field and resulting evaporation rate in \ref{ss:Keq}, and ending with a calculation of the emission rates in \ref{ss:KRDelta} and \ref{ss:KCDelta}. In \ref{sec:disc}, we use the rates to discuss the evaporation histories for the different types of black holes. We end with a discussion of some potential future directions for our work.

\section{Evaporation of a Reissner-Nordström black hole}\label{sec:evapRN}

\subsection{Near-extremal Reissner-Nordström background}\label{ss:RNbckg}

The Reissner-Nordström solution of the Einstein-Maxwell system of equations describes a charged black hole with mass $M$ and charge $Q$. The gravitational field is described by the metric, which is given by
\begin{equation}\label{eq:RNmetric}
    ds^2_{\text{RN}} =
    -f(r) \, dt^2 +
    \frac{dr^2}{f(r)} +
    r^2 \, d\Omega^2_2 \, ,
\end{equation}
where
\begin{equation}\label{eq:fenbl}
    f(r) = 1 - \frac{2M}{r} + \frac{Q^2}{r^2} \, .
\end{equation}
We will set $G=1$ everywhere, until we reach our final results. An alternative parameterization of the metric is in terms of the roots of $f(r)$
\begin{equation}\label{eq:RNhor}
    r_\pm = M \pm \sqrt{M^2-Q^2} \, ,
\end{equation}
which also define the radii of the inner and outer horizons. So, we can also write
\begin{equation}\label{eq:fenbl2}
    f(r) = \frac{(r-r_+) \, (r-r_-)}{r^2} \, .
\end{equation}
The electromagnetic field is captured by the gauge potential, which is given by
\begin{equation}\label{eq:RNgpot}
    A_\mu \, dx^\mu = \left( \frac{Q}{r} - \frac{Q}{r_+} \right) \, dt \, ,
\end{equation}
from which the electric field strength can be calculated. The constant term is chosen such that a finite answer is obtained when we zoom into the near-horizon region later on.

The parameters $M$ and $Q$ must satisfy the following bound
\begin{equation}\label{eq:RNextbnd}
    M \geq Q \, .
\end{equation}
An extremal black hole is defined as one that saturates this bound, while for $M \approx Q$, we have a near-extremal one. As can be seen from \eqref{eq:RNhor}, at extremality, the inner and outer horizons meet at the extremal horizon
\begin{equation}\label{eq:RNexthor}
    \begin{split}
        r_\text{h} = Q \, .
    \end{split}
\end{equation}
For the rest of this work, we will parameterize the background in terms of the extremal horizon and a small parameter
\begin{equation}\label{eq:RNdr}
    \delta r = r_+-r_- \, ,
\end{equation}
which is a measure of how much the black hole deviates from near-extremality.

For a near-extremal black hole, we can zoom into the near-horizon region using the following shift plus rescaling of the radial and time coordinates
\begin{equation}\label{eq:RNNHRcoord}
    \begin{split}
        r & = r_+ + \delta r \, x \, , \\
        t & =  \frac{2 r_+^2}{\delta r} \, \tau \, .
    \end{split}
\end{equation}
Plugging into \eqref{eq:RNmetric} and ignoring terms of higher order in $\delta r \, x$ results in the near-horizon metric
\begin{equation}\label{eq:RNNHRmetric}
    ds^2_{\text{NHR}} =
    r_+^2 \left( - 4 x(x+1) \, d\tau^2 +
    \frac{dx^2}{x(x+1)} \right) +
    r_+^2 \, d\Omega^2_2 \, .
\end{equation}
We can see that the near-horizon spacetime factorizes into an AdS$_2$ factor (one can check that it has $\mathcal{R}_\text{Ricci} = -1/r_+^2$) and a sphere, both having radius $r_+$. This metric approximates the original background well up to a cut-off radius $x_c$, which has to satisfy the following two inequalities
\begin{equation}\label{eq:RNcutoff}
    1 \ll x_c \, , \qquad \delta r \, x_c \ll 1 \, .
\end{equation}
The first says that the cut-off radius can be taken to be large, while the second says that when it is multiplied by $\delta r$, this has to result in a small parameter, as these are the terms we ignored to obtain \eqref{eq:RNNHRmetric}. Outside of this radius, we would need to add terms higher-order in $\delta r \, x$ to obtain the original metric. Under this coordinate transformation, the gauge potential reduces to
\begin{equation}\label{eq:RNNHRgpot}
    A_\tau \, d\tau = - 2 r_+ x \, d\tau \, .
\end{equation}

To define the thermodynamics of the black hole, we introduce the temperature and entropy, both of which can be identified with geometric quantities as follows. The temperature can be written in terms of the surface gravity at the outer horizon
\begin{equation}\label{eq:RNtemp}
    T = \frac{\kappa}{2\pi} = \frac{f'(r_+)}{4\pi} = \frac{r_+-r_-}{4 \pi r_+^2} \, .
\end{equation}
The entropy is given by the famous Bekenstein-Hawking formula which gives it in terms of the area of the outer horizon
\begin{equation}\label{eq:BHentr}
    S = \frac{A}{4} \, .
\end{equation}
We also introduce the conjugate variable to the black hole charge, the electric potential at the outer horizon
\begin{equation}\label{eq:RNpot}
    \Phi = A_t (r=r_+) = \frac{Q}{r_+} \, .
\end{equation}
Finally, for convenience, instead of the full energy we will use the energy above extremality defined as follows
\begin{equation}\label{eq:Eabove}
    E = M - Q \, .
\end{equation}
We can also parameterize the black hole using the microcanonical variables $(Q,E)$ or the canonical $(Q,T)$. From \eqref{eq:Eabove} and \eqref{eq:RNtemp} we see that these are also small quantities for a near-extremal black hole, which we can now use to give a more precise definition of near-extremality
\begin{equation}\label{eq:nearextdef}
    Q T \ll 1 \, , \qquad
    \frac{E}{Q} \ll 1 \, .
\end{equation}
We can perform the following expansions that will be used
\begin{equation}\label{eq:drET}
    \begin{split}
        r_+ & = a + 2 \pi Q^2 T \, , \\
        \delta r & = 4 \pi Q^2 T \, , \\
        E & = 2\pi^2 Q^3 T^2 \, .
    \end{split}
\end{equation}
We also define a breakdown energy scale
\begin{equation}\label{eq:RNEbrk}
    E_{\text{b}} = \frac{1}{Q^3} \, .
\end{equation}
below which thermodynamics "breaks down" in the sense of \cite{Preskill:1991}. Below this scale, the earlier mentioned effective quantum theory gives important corrections to the thermodynamics. This is also where corrections to the emission spectra of neutral radiation show up \cite{Brown:2024}.

\subsection{Solving the wave equation}\label{ss:RNeq}

The Klein-Gordon wave equation for a mass $m$ scalar field with charge $e$ is given by the following
\begin{equation}\label{eq:RNscalwave}
    \left( (\nabla_\mu-ieA_\mu)\,(\nabla^\mu-ieA^\mu) + m^2 \, \right) \Phi = 0 \, .
\end{equation}
This equation is separable in the Reissner-Nordström metric \eqref{eq:RNmetric}, and given the time-translation and spherical symmetry, we can look for a solution in the following form
\begin{equation}\label{eq:RNansatz}
    \Phi(t,r,\theta,\phi) =
    e^{-i\omega t} \,
    R(r) \,
    Y_{lm}(\theta,\phi) \, ,
\end{equation}
where $Y_{lm}(\theta,\phi)$ are the standard spherical harmonics. Therefore, solving the wave equation amounts to solving the resulting radial equation
\begin{equation}\label{eq:RNradeq}
    \frac{d}{dr} \left( \Delta \frac{dR}{dr} \right) +
    \left(\frac{(\omega - e Q /r)^2 \, r^4}{\Delta} -
    l(l+1) -
    m^2 r^2 \right) R = 0 \, .
\end{equation}
where we introduced the function $\Delta(r) = r^2 f(r)$ for easier comparison with the Kerr wave equation to be solved in subsection \ref{ss:Keq}.

We start with a discussion of the boundary conditions used for this equation in the problem of black hole evaporation. To do this, we first transform the radial equation into a one-dimensional Schrodinger-like equation. Out at asymptotic infinity $r\to\infty$, we expect spherical waves, which have amplitude proportional to inverse radial distance. To obtain plane-wave-like solutions, we factor out this radial dependence by introducing the wavefunction
\begin{equation}\label{eq:RNSwavefcn}
    S(r) = \frac{R(r)}{r} \, .
\end{equation}
Furthermore, to remove the first derivative term from the resulting equation for $S(r)$, we require that $dr_* = \frac{dr}{f(r)}$, which defines the tortoise coordinate given by
\begin{equation}\label{eq:RNtort}
    r_* = r + \frac{r_+^2}{\delta r} \log (r-r_+) - \frac{r_-^2}{\delta r} \log (r-r_-) \, .
\end{equation}
The original radial coordinate has a range from the outer horizon at $r_+$ to $\infty$, which means that $r_*$ goes from $-\infty$ to $+\infty$. These transformations result in a Schrodinger-like equation
\begin{equation}\label{eq:RN1Dscatt}
    \frac{d^2 S}{dr_*^2} + V_{\text{eff}}(r_*) \, S = 0 \, ,
\end{equation}
with the following effective potential, which we write in terms of $r$, as $r_* = r_*(r)$ cannot be inverted in terms of elementary functions,
\begin{equation}\label{eq:RNVeff}
    V_\text{eff}(r) = \left( \omega - \frac{eQ}{r} \right)^2 -
    \frac{f(r) \left( \, l(l+1) + m^2 r^2 + r f'(r) \, \right)}{r^2} \, .
\end{equation}
We can now think of the problem as follows, an incident wave comes in from infinity and is scattered by the potential, which produces a reflected wave and a transmitted wave, which goes into the black hole. Far away from the potential, the solution to \eqref{eq:RN1Dscatt} takes the following form
\begin{equation}\label{eq:RNbndcond}
        S(r_*) =
        \begin{cases}
            C_\text{trans} \, e^{-i \omega_\text{eff} r_*} \, , &
            r_* \rightarrow r_*(r_+)=-\infty \\
            C_\text{inc} \, e^{-i k_\infty r_*} +
            C_\text{refl} \, e^{+ik_\infty r_*} \, , &
            r_* \rightarrow r_*(+\infty) = +\infty
        \end{cases}
\end{equation}
which defines the amplitudes of the three waves $C_\text{inc}$, $C_\text{refl}$, and $C_\text{trans}$, which in subsection \ref{ss:RNsc} we will use to compute the reflection and transmission coefficients. The wavenumber at asymptotic infinity is given by 
\begin{equation}\label{eq:RNks}
    k_\infty = \sqrt{V_{\text{eff}}(\infty)} = \sqrt{\omega^2 - m^2} \, ,
\end{equation}
whereas the wavenumber at the horizon, which we will also refer to as the effective near-horizon frequency is given by
\begin{equation}\label{eq:freqeff}
    \omega_\text{eff} =
    \sqrt{V_{\text{eff}}(r_+)} =
    \omega - \frac{eQ}{r_+} \, .
\end{equation}
This equation defines an energy scale which we will refer to as the superradiant scale. For a near-extremal black hole it is given by
\begin{equation}\label{eq:RNsrscale}
    \omega_\text{sr} = \frac{eQ}{r_+}  = e \, .
\end{equation}
Note that $\omega_\text{eff}$ can be negative, depending on how $\omega$ compares to $\omega_\text{sr}$. The significance of this scale and the sign of $\omega_\text{eff}$ will be discussed in subsection \ref{ss:RNsc} in the context of superradiance.

The wave equation \eqref{eq:RNscalwave} with these boundary conditions was recently solved in \cite{Maulik:2025,Betzios:2025} using techniques involving the Heun equation \cite{Bonelli:2021}. We will solve the equation using the more conventional technique \cite{Starobinskii:1973a} of breaking it up into an equation in a near region and an equation in a far region, with the regions overlapping. We will then solve each of these equations and require their solutions to match in the overlapping region. Starting with the near region, we perform the coordinate transformation
\begin{equation}\label{eq:RNrnear}
    r = r_+ + \delta r \, x \, .
\end{equation}
Using $\Delta(x) = \delta r^2 \, x \, (x+1)$, the radial equation in this coordinate is given by
\begin{equation}\label{eq:RNneareq}
    \begin{split}
        \frac{d}{dx} \! \left(
        x (x \!+\! 1) \frac{dR}{dx} \right) \!+\!
        \left(
        \frac{\left( \omega \!-\!
        eQ / (r_+ \!+\! \delta r x) \right)^2
        (r_+ \!+\! \delta r x)^4}
        {\delta r^2 \, x \,(x+1)} \!-\! 
        l(l \!+\! 1) \!-\!
        m^2 (r_+ \!+\! \delta r \, x )^2
        \right) \!
        R \!=\! 0 \, .
    \end{split}
\end{equation}
In the near region, we will treat $\delta r \, x$ as a small quantity and ignore terms that are higher-order in $\delta r \, x$, as we did when defining the near horizon region in \eqref{eq:RNNHRmetric}. Note that in the $m^2$ term, we can ignore the terms that are first and second order in $\delta r \, x$, but in the $\omega$ term, we can only ignore the terms that are third and fourth order in $\delta r \, x$, since there is an extra $\delta r$ in the denominator, and $x$ by itself is not necessarily small. This results in the near region equation given by
\begin{equation}\label{eq:RNneareqx}
    \begin{split}
        & \frac{d^2R}{dx^2} +
        \frac{(2x+1)}{x\,(x+1)} \,
        \frac{dR}{dx} +
        \frac{a_0 + a_1 x + a_2 x^2}{x^2 \, (x+1)^2} \, R = 0 \, , \\
        & a_0 =
        \frac{r_+^4 \left( \omega - \frac{eQ}{r_+} \right)^2}
        {\delta r^2} \, , \\
        & a_1 =
        \frac{2 r_+ (2r_+^2\omega^2-3eQr_+\omega+e^2Q^2)}{\delta r} -
        l(l+1) - m^2 r_+^2 \, , \\
        & a_2 =
        -l(l+1) + e^2 Q^2 - m^2 r_+^2 - 6 r_+ \omega \, (eQ-r_+\omega) \, .
    \end{split}
\end{equation}
This equation has three regular singular points, $0$, $-1$ and $\infty$. The solutions can therefore be written in terms of ordinary hypergeometric functions $F(a, \,b; \,c; \,x)$ (defined in appendix \ref{app:math}), where the $a$, $b$, and $c$ are determined by the parameters $a_0$, $a_1$ and $a_2$. These parameters assemble into the following three combinations
\begin{equation}\label{eq:RNnearparam2}
    \begin{split}
        p & =
        \sqrt{a_0} \, , \\
        q & =
        \sqrt{a_0 - a_1 + a_2} \, , \\
        \beta & =
        \sqrt{\tfrac{1}{4} - a_2} \, .
    \end{split}
\end{equation}
For a near-extremal black hole we can further expand these parameters to leading order in $\delta r$ which gives
\begin{equation}\label{eq:RNnearparam3}
    \begin{split}
        p & =
        \frac{Q^2 \left( \omega - e \right)}{\delta r} -
        \frac{1}{2} Q (e-2\omega) \, , \\
        q & = \frac{Q^2 \left( \omega - e \right)}{\delta r} +
        \frac{1}{2} Q (e-2\omega) \, , \\
        \beta & =
        \sqrt{1/4 + l(l+1) - e^2 Q^2 + m^2 Q^2 +
        6 Q^2  \omega (e - \omega)} \, .
    \end{split}
\end{equation}
Notice that $\beta$ can be real or imaginary depending on the combination of the black hole and radiation parameters. For now we will not impose any restrictions on it. We introduce two more parameters $\mu = q-p$ and $F = \tfrac{1}{2} (p+q)$, which can be used instead of $p$ and $q$. For small $\delta r$ these are given by
\begin{equation}\label{eq:RNnearparam4}
    \begin{split}
        \mu & =
        Q(e - 2 \omega) \, , \\
        F & =
        \frac{Q^2 (\omega - e)}{\delta r} \, .
    \end{split}
\end{equation}
We can now write the general solution to our equation in terms of two independent hypergeometric functions
\begin{equation}\label{eq:RNnearsol}
    \begin{split}
        R(x) =
        \sum_{\pm}
        A_\mp \,
        x^{\mp i p} \, &
        (1+x)^{\pm i q} \,
        F \, \left(
        \tfrac{1}{2} - \beta \pm i \mu, \, 
        \tfrac{1}{2} + \beta \pm i \mu; \,
        1 \mp 2 i F \pm i \mu; \,
        -x
        \right) \, .
    \end{split}
\end{equation}
To impose the ingoing boundary conditions at the horizon as defined in \eqref{eq:RNbndcond}, we need to change coordinates in our solution to $r_*$ and expand it around the horizon. The $r_*$ coordinates written in terms of $x$ near the horizon and to leading order in $\delta r$ is given by
\begin{equation}\label{eq:RNturtx}
    \frac{\delta r}{Q^2} r_* = \log x \, .
\end{equation}
Using the small argument limit of the hypergeometric function \eqref{eq:Fdef}, we can expand our solution around the horizon as follows
\begin{equation}\label{eq:RNApm}
    R(x\to0)=
    A_\mp \,
    x^{\mp ip} =
    A_\mp \,
    e^{\mp i\frac{\delta r}{Q^2} \, p \, r_*} =
    A_\mp \, e^{\mp i \, \omega_{\text{eff}} \, r_*} \, .
\end{equation}
Comparing to \eqref{eq:RNbndcond}, we see that the function with the $A_-$ coefficient is the ingoing wave at the horizon, and the $A_+$ function is the outgoing. Therefore, we require $A_+=0$. Relabeling $A_-\rightarrow A$, the solution to the near equation is given by
\begin{equation}\label{eq:RNnearsolfin}
    \begin{split}
        R(x) =
        A \,
        x^{- i p} \, &
        (1+x)^{+ i q} \,
        F \, \left(
        \tfrac{1}{2} - \beta + i \mu, \, 
        \tfrac{1}{2} + \beta + i \mu; \,
        1 - 2 i F + i \mu; \,
        -x
        \right) \, .
    \end{split}
\end{equation}
In equation $\eqref{eq:RNnearparam3}$, when expanding $p$ and $q$ in $\delta r$, we assumed that $\omega \geq \frac{eQ}{r_+}$. If we had assumed the opposite, we would have had an extra minus sign next to $(\omega-e)$, but then the boundary conditions would eliminate the opposite function in our solution, which would effectively cancel this minus sign. Therefore, the solution to the near equation \eqref{eq:RNnearsolfin} is valid for all $\omega$, with $p$ and $q$ as defined in \eqref{eq:RNnearparam3}.

Turning to the far region, we perform the following coordinate shift
\begin{equation}\label{eq:RNrfar}
    r = r_+ + y \, .
\end{equation}
Taking $y$ to be large, and expanding the parameters in the radial equation to second order results in the far equation given by
\begin{equation}\label{eq:RNfareq}
    \begin{split}
        & R_{yy} +
        \frac{2}{y} R_y +
        \frac{b_0 + b_1 \, y + b_2 \, y^2}{y^2} R = 0 \, , \\
        & b_0 = -l(l+1) + e^2 Q^2 - m^2 r_+^2 -
        6 r_+ \omega \, (eQ-r_+\omega) \\
        & \ \ \ \ \, \ \ \ \,
        + 2 \omega
        \left( eQ - 2 r_+ \omega \right) \, \delta r + 
        \omega^2 \, \delta r^2 \, , \\
        & b_1 = \,
        \left( 4 r_+ -\delta r \right) \, \omega^2 -
        2 Q e \omega -
        2 r_+ m^2  \, , \\
        & b_2 = \omega^2 - m^2 \, .
    \end{split}
\end{equation}
The solutions can be written in terms of the Kummer confluent hypergeometric function $M(a,b;x)$ (defined in appendix \ref{app:math}). The parameters $b_0$, $b_1$ and $b_2$ assemble into the following three combinations
\begin{equation}\label{eq:RNfarparam1}
    \begin{split}
        \beta_\text{far} & =
        \sqrt{\tfrac{1}{4} - b_0} \, , \\
        \alpha & =
        \frac{b_1}{2 \sqrt{b_2}} \, ,\\
        k & =
        \sqrt{b_2} \, .
    \end{split}
\end{equation}
For a near-extremal black hole, these simplify to
\begin{equation}\label{eq:RNfarparam2}
    \begin{split}
        \beta & =
        \sqrt{\tfrac{1}{4} + l(l+1) - e^2Q^2 + m^2Q^2 + 6Q^2\omega(e-\omega)} \, ,\\
        \alpha & =
        \frac{Q(2\omega^2-e\omega-m^2)}{\sqrt{\omega^2 -m^2}} \, ,\\
        k & =
        \sqrt{\omega^2 -m^2} \, .
    \end{split}
\end{equation}
The first parameter is the same $\beta$ we had in the near equation \eqref{eq:RNnearparam3} and the last parameter is just the wavenumber at asymptotic infinity \eqref{eq:RNks}. The general solution to the far equation can be written as follows
\begin{equation}\label{eq:RNfarsol}
    \begin{split}
        R(y) =
        e^{-iky} \,
        \sum_{\pm} \, 
        C_\pm \,
        y^{-1/2 \pm \beta} \,
        M \left( \tfrac{1}{2} \pm \beta + i \alpha,
        1 \pm 2\beta,
        2iky \right) \, .
    \end{split}
\end{equation}
We can check that this solution has the right spherical wave behavior at infinity by using the asymptotic expansion for the confluent hypergeometric function \eqref{eq:Mlarge}. This gives four terms, two of which are
\begin{equation}\label{eq:RNfarasymp}
    \begin{split}
        \left(
        C_+
        \frac{\Gamma(1 \!+\! 2\beta)}
        {\Gamma(\tfrac{1}{2} \!+\! \beta \!+\! i\alpha)}
        (2ik)^{-1/2-\beta+i\alpha}
        +
        C_-
        \frac{\Gamma(1 \!-\! 2\beta)}
        {\Gamma(\tfrac{1}{2} \!-\! \beta \!+\! i\alpha)}
        (2ik)^{-1/2 + \beta + i\alpha}
        \right)
        \frac{e^{+iky}}{y}
        y^{+i\alpha} \, ,
    \end{split}
\end{equation}
which is the outgoing or reflected wave as defined in \eqref{eq:RNbndcond} ($r_* = r = y$ to leading order). The remaining two terms are the ingoing or incident wave
\begin{equation}\label{eq:RNfarasymp2}
    \begin{split}
        \left(
        C_+
        \frac{\Gamma(1 \!+\! 2\beta)}
        {\Gamma(\tfrac{1}{2} \!+\! \beta \!-\! i\alpha)}
        (-2ik)^{-1/2-\beta-i\alpha} +
        C_-
        \frac{\Gamma(1 \!-\! 2\beta)}
        {\Gamma(\tfrac{1}{2} \!-\! \beta \!-\! i\alpha)}
        (-2ik)^{-1/2+\beta-i\alpha}
        \right)
        \frac{e^{-iky}}{y}
        y^{-i\alpha} \, .
    \end{split}
\end{equation}
The extra position-dependent phase shift $y^{\pm i \alpha} = e^{\pm i \alpha \ln y}$ is characteristic for potentials that at infinity are proportional to $1/r$ and will not contribute to the reflection and transmission coefficients.

We now match the two solutions in the overlapping region to find the relationship between the coefficients $A$, $C_+$, and $C_-$. We expand the near solution at large distances and the far solution at small. Starting with the former, we use the large argument expansion of the hypergeometric function, which is given in the appendix. Applying this to the near solution results in
\begin{equation}\label{eq:RNnearasymp}
    \begin{split}
        R(x\to\infty) =
        \sum_{\pm}
        A \,
        \frac{\Gamma(\pm2\beta) \,
        \Gamma(1-2iF+i\mu)}
        {\Gamma(1/2 \pm \beta + i\mu) \,
        \Gamma(1/2 \pm \beta-2iF)} \,
        x^{-1/2 \pm \beta} \, .
    \end{split}
\end{equation}
Then, for the far solution we use the small argument limit of the confluent hypergeometric functions \eqref{eq:Mdef}, which gives
\begin{equation}\label{eq:RNfarsmalllarg}
    \begin{split}
        R(y \to 0) =
        \sum_{\pm}
        C_\pm \,
        y^{-1/2 \pm \beta} \, .
    \end{split}
\end{equation}
Using the fact that $y = \delta r \, x$, we match the solutions in \eqref{eq:RNnearasymp} and \eqref{eq:RNfarsmalllarg} and obtain the following relationship between the coefficients
\begin{equation}\label{eq:RNC12}
    \begin{split}
        C_{\pm} =
        \frac{\Gamma(\pm 2\beta) \,
        \Gamma(1 - 2iF + i\mu)}
        {\Gamma(\tfrac{1}{2} \pm \beta + i\mu) \,
        \Gamma(\tfrac{1}{2} \pm \beta - 2iF)} \,
        \delta r^{+1/2 \mp \beta} \,
        A \, .
    \end{split}
\end{equation}
The overlapping region is at $x \gg 1$ and $y \ll 1$. Since $y = \delta r \, x$, this implies that $1 \ll x \ll 1/\delta r$, which is only possible if $\delta r \ll 1$, that is, if the black hole is near-extremal. To be more explicit, we can also compare the power of the two solutions. For the near solution, the power is given by $\beta_\text{near}$ from equation \eqref{eq:RNnearparam2}, while the far solution has $\beta_\text{far}$ given by \eqref{eq:RNfarparam1}. These two are only equal if $\delta r \ll 1$.

The solutions given by equations \eqref{eq:RNnearsolfin} and \eqref{eq:RNfarsol} together with the matching conditions \eqref{eq:RNC12} represent a full solution of the wave equation. Once $C_\pm$ are eliminated from this solution, we are left with only one undetermined constant $A$, as should be the case for a scattering problem. In the next subsection, these will be used to compute the reflection and transmission coefficients.

We also note that our wave equation includes the neutral scalar as a special case. In the limit of $e=0$, the equation in the near region has $\mu=0$, which gives real parameters $a$ and $b$ in the hypergeometric functions. If we further assume a massless field $m=0$, these parameters become integers, which means that the hypergeometric function reduces to the Jacobi polynomials. In the far region, the two confluent hypergeometric functions reduce to the spherical Bessel functions. For work on higher spin ($s=1,2$) neutral fields and their spectra, see \cite{Brown:2024}.

\subsection{Semiclassical emission rate}\label{ss:RNsc}

The number of quanta of a given field emitted by the black hole per unit time is found using the Hawking formula \cite{Hawking:1975}
\begin{equation}\label{eq:scNdot}
    \frac{dN}{dt}\Bigg|_{\text{semi-classical}} =
    \frac{1}{2\pi}
    \int_{0}^{\infty} d\omega \,
    \frac{\Gamma(\omega)}{e^{\omega_\text{eff}/T}-1} \, .
\end{equation}
We will refer to this quantity as the emission or evaporation rate, whereas the integrand we will refer to as the spectral (meaning per unit frequency) emission rate. This formula was obtained by treating the quantum fields in a fixed curved background provided by the black hole, so we will also refer to this as the semi-classical rate, as opposed to the quantum-corrected one that we will derive in the next subsection. The rate of emitting other quantities that the quanta carry, such as charge or angular momentum, is found by simply multiplying this integral by that quantity. The energy emitted, or energy flux, is found by multiplying the integrand in \eqref{eq:scNdot} by the energy $\omega$ of a quantum before integrating
\begin{equation}\label{eq:scEdot}
    \frac{dE}{dt}\Bigg|_{\text{semi-classical}} =
    \frac{1}{2\pi}
    \int_{0}^{\infty} d\omega \,
    \frac{\omega \, \Gamma(\omega)}
    {e^{\omega_\text{eff}/T}-1} \, .
\end{equation}
In these equations, $T$ is the black hole temperature given by \eqref{eq:RNtemp}. $\omega_{\text{eff}}$ is the effective frequency introduced in \eqref{eq:freqeff}. $\Gamma(\omega)$ is the greybody factor, a measure of how much the black hole emission spectrum deviates from that of a blackbody. It is equal to the transmission coefficient of the scattering problem set up in the previous subsection.

The transmission and reflection coefficients are defined as follows
\begin{equation}\label{eq:T&R}
    \begin{split}
        \mathcal{R} =
        \left| \frac{C_\text{refl}}{C_\text{inc}} \right|^2
         \, ,
        \qquad
        \mathcal{T} =
        \frac{\omega_\text{eff}}{k_{}}
        \Big| \frac{C_{\text{trans}}}{C_{\text{inc}}} \Big|^2 \, .
    \end{split}
\end{equation}
Before computing these two, we first discuss the sign of the transmission coefficient, which is related to the phenomenon of superradiance. As mentioned earlier, $\omega_\text{eff}$ can be negative, which implies that the transmission coefficient can be negative. The unitarity condition
\begin{equation}\label{eq:RNprobconserv}
    \mathcal{R} + \mathcal{T} = 1 \, ,
\end{equation}
implies that when this is the case, the reflection coefficient has to be greater than one i.e. the reflected wave comes out with a higher amplitude than the incident. This is the definition of superradiance. The transmitted wave in equation \eqref{eq:RNbndcond} looks like it has a negative phase velocity meaning that it is actually moving away from the horizon, but it turns out that the group velocity is positive, which implies that a physical disturbance moves towards the horizon (see \cite{Brito:2015} for a detailed review).

An inspection of the emission rate \eqref{eq:scEdot} gives us an idea of the role of superradiance in black hole evaporation. Consider a black hole at low temperature, which is when superradiance becomes significant. Equation \eqref{eq:scEdot} shows that for non-superradiant modes, i.e. those with $\omega_\text{eff}>0$, the spectral emission rate is exponentially suppressed. We will see that this implies that the emission rate is proportional to a power of temperature, which means that no non-superradiant quanta are emitted at zero temperature. For the superradiant modes, the opposite is true, the exponential in \eqref{eq:scEdot} goes to zero at low temperatures, and, as we shall see, the rest of the integrand goes to a constant. Therefore, at low temperatures, the spectrum of the black hole is fully superradiant. Note that when $\omega_\text{eff} < 0$, the statistical factor in \eqref{eq:scEdot} is negative, but this is also when the transmission coefficient \eqref{eq:T&R} is negative, so the emission rate is always positive. A more careful analysis of the integral in the emission rate in the low temperature limit is left for subsection \ref{ss:class}.

To compute the reflection and transmission coefficients, we will need the amplitudes of the incident, reflected, and transmitted waves defined in equation \eqref{eq:RNbndcond}. The amplitude of the transmitted wave can be found by expanding the solution of the near equation around the horizon, which was done in \eqref{eq:RNApm}. Comparing to \eqref{eq:RNbndcond} gives
\begin{equation}\label{eq:RNCtrans}
    C_\text{trans} =
    r_+ \,
    A \, ,
\end{equation}
where the $r_+$ comes from the relationship between $R(r)$ and $S(r)$ at the horizon. The amplitudes of the incident and reflected waves can be found by expanding the solution of the far equation at asymptotic infinity, which was done in equations \eqref{eq:RNfarasymp} and \eqref{eq:RNfarasymp2}. From the matching equation \eqref{eq:RNC12}, we can see that  that $C_- / C_+$ is of the order of $\delta r^{2\beta}$, which is very small for near-extremal black holes, and therefore $C_-$ can be safely ignored relative to $C_+$ in both the incident and reflected wave amplitudes. For the incident wave we have
\begin{equation}\label{eq:RNCinc}
    \begin{split}
        C_\text{inc} =
        C_+ \,
        \frac{\Gamma(1 + 2\beta)}
        {\Gamma(\tfrac{1}{2} + \beta - i\alpha)} \,
        (-2ik)^{-1/2 - \beta - i\alpha} \, ,
    \end{split}
\end{equation}
and for the reflected wave we find
\begin{equation}\label{eq:RNCrefl}
    C_\text{refl} =
    C_+ \,
    \frac{\Gamma(1 + 2\beta)}
    {\Gamma(\tfrac{1}{2} + \beta + i\alpha)} \,
    (2ik)^{-1/2-\beta+i\alpha} \, .
\end{equation}
In these two equations, $C_+$ can further be written in terms of $A$ using the matching equation \eqref{eq:RNC12}. The next step is to square these three expressions. As mentioned earlier, $\beta$ can be real or imaginary, and this will lead to very different expressions after the squaring. For now, we will focus on the case of real $\beta$ for all frequencies, and then in subsection \ref{ss:class}, we will discuss under what conditions this is true, and finally in \ref{ss:RNCDelta}, we will rederive the reflection and transmission coefficients for imaginary $\beta$. To square the amplitudes, we will use the following identity
\begin{equation}\label{eq:itoi}
    |\pm i^{i\alpha}| =
    |e^{(\pm i \frac{\pi}{2})(i\alpha)}| =
    e^{\mp \frac{\pi \alpha}{2}} \, .
\end{equation}
We find for the incident wave
\begin{equation}\label{eq:RNCinc2}
\begin{split}
    & |C_\text{inc}|^2 =
    \delta r^{+1-2\beta} \,
    (2k)^{-1-2\beta} \,
    \frac{\Gamma(1+2\beta)^2 \,
    \Gamma(2\beta)^2}
    {|\Gamma(\tfrac{1}{2}+\beta-i\alpha)|^2 \,
    |\Gamma(\tfrac{1}{2}+\beta+i\mu)|^2} \,
    e^{-\pi\alpha} \, \\
    & \qquad \qquad \qquad \qquad \qquad \qquad \qquad \ \ \ \,
    \frac{|\Gamma(1-2iF+i\mu)|^2}
    {|\Gamma(\tfrac{1}{2}+\beta-2iF)|^2} \,
    |A|^2 \, ,
\end{split}
\end{equation}
and for the reflected wave
\begin{equation}\label{eq:RNCrefl2}
\begin{split}
    & |C_\text{refl}|^2 =
    \delta r^{+1-2\beta} \,
    (2k)^{-1-2\beta} \,
    \frac{\Gamma(1+2\beta)^2 \,
    \Gamma(2\beta)^2}
    {|\Gamma(\tfrac{1}{2}+\beta+i\alpha)|^2 \,
    |\Gamma(\tfrac{1}{2}+\beta+i\mu)|^2} \,
    e^{-\pi\alpha} \, \\
    & \qquad \qquad \qquad \qquad \qquad \qquad \qquad \ \ \ \,
    \frac{|\Gamma(1-2iF+i\mu)|^2}
    {|\Gamma(\tfrac{1}{2}+\beta-2iF)|^2} \,
    |A|^2 \, .
\end{split}
\end{equation}
The final step is to plug into equation \eqref{eq:T&R} which for the reflected coefficient gives
\begin{equation}\label{eq:RNRfin}
    \mathcal{R} = 1\, .
\end{equation}
This is the answer to leading order in $\delta r$, and it tells us that most of the incident wave is reflected. To find the correction, we would had to have included the $C_-$ terms in the amplitudes. For the transmission coefficient we find
\begin{equation}\label{eq:RNTfin}
    \begin{split}
        \mathcal{T} = 
        2^{2\beta+1} \,
        & \delta r^{2\beta-1} \,
        r_+^2 \,
        k^{2\beta} \,
        \omega_\text{eff} \,
        \frac{|\Gamma(\tfrac{1}{2}+\beta-i\alpha)|^2 \,
        |\Gamma(\tfrac{1}{2}+\beta+i\mu)|^2}
        {\Gamma(2\beta+1)^2 \,
        \Gamma(2\beta)^2} \\
        & \ \ \ \ \  \frac{|\Gamma(\tfrac{1}{2}+\beta-2iF)|^2}
        {|\Gamma(1-2iF+i\mu)|^2} \,
        e^{\pi\alpha} \, .
    \end{split}
\end{equation}
The greybody factor is just equal to this coefficient. To make this equation more suitable for later calculations, we will rewrite $\delta r$ in terms of $T$ using equation \eqref{eq:drET}. Furthermore, for easier comparison to the quantum-corrected rate, we introduce the conformal dimension
\begin{equation}\label{eq:Deltadef}
    \Delta = \tfrac{1}{2} + \beta \, ,
\end{equation}
which we will use instead of $\beta$, going forward. In the next subsection, in the calculation of the quantum-corrected rate, this parameter will play the role of an operator conformal dimension in the sense of AdS$_2$/CFT$_1$, but it is important to note that in this subsection, in the context of the semi-classical emission rate, its role is just that of a parameter in the wave equation. With these changes, the greybody factor is found to be
\begin{equation}\label{eq:RNGamma}
    \begin{split}
        \Gamma\left( Q, T \, ; m, e \, ; \omega, l, m_l \right) = 
        2^{6\Delta-4} \,
        & \pi^{2\Delta-2} \,
        Q^{4\Delta-2} \,
        T^{2\Delta-2}
        k^{2\Delta-1} \,
        \omega_\text{eff} \,
        \frac{|\Gamma(\Delta-i\alpha)|^2 \,
        |\Gamma(\Delta+i\mu)|^2}
        {\Gamma(2\Delta)^2 \,
        \Gamma(2\Delta-1)^2} \\
        & \ \
        \frac{|\Gamma(\Delta-2iF)|^2}{|\Gamma(1-2iF+i\mu)|^2} \,
        e^{\pi\alpha} \, .
    \end{split}
\end{equation}
We have explicitly listed all the variables that determine the greybody factor; $(Q,T)$ label the black hole state, $(m,e)$ define the field being radiated, and $(\omega,l,m_l)$ the mode which it is being emitted (the equation is independent of $m_l$, but it has been included for completeness). We can now put this together with the remaining factors in the Hawking formula to obtain the final formula for the semi-classical spectral evaporation rate
\begin{equation}\label{eq:RNscrate}
    \begin{split}
        \frac{d^2E}{d\omega \, dt} \bigg|_{\text{sc}} =
        2^{6\Delta-5} \,
        \pi^{2\Delta-3} \,
        Q^{4\Delta-2} \, &
        T^{2\Delta-2} \,
        \frac{|\Gamma(\Delta-i\alpha)|^2 \,
        |\Gamma(\Delta+i\mu)|^2}
        {\Gamma(2\Delta)^2 \,
        \Gamma(2\Delta-1)^2} \,
        e^{\pi\alpha} \\
        \frac{\omega \, \omega_\text{eff} \, k^{2\Delta-1} \,}{e^{\omega_{\text{eff}}/T}-1} \,
        & \frac{|\Gamma(\Delta-2iF)|^2}{|\Gamma(1-2iF+i\mu)|^2} \, .
    \end{split}
\end{equation}
The parameters $\mu, F, \Delta, \alpha$ can be found in equations \eqref{eq:RNnearparam3}, \eqref{eq:RNnearparam4} and \eqref{eq:RNfarparam2}, which are all frequency-dependent in general. A systematic analysis of this equation is left for section \ref{sec:smBH}.

We note that this equation was recently derived in \cite{Betzios:2025}. If we set $\mu=0$ in our result \eqref{eq:RNscrate}, we find agreement with theirs. The non-zero value of $\mu=q-p$ comes from keeping subleading (in $\delta r$) terms in $p$ and $q$ in \eqref{eq:RNnearparam3}, which we believe should be kept as we can see that they don't go away in the $\delta r \ll 1$ limit in our final result.

This rate can also be compared to the one found in \cite{Gibbons:1975} for small black holes (the one for large will be recovered in \ref{ss:RNCDelta}), where non-extremal black holes were studied, but the radiation was taken to have low frequencies. Therefore, we need to take the low frequency limit of our rate and the near-extremal limit of the rate in \cite{Gibbons:1975} (which can be found in equation 4.33). For a small black hole, we can ignore $\mu$ and set $\Delta=l+1$, which gives
\begin{equation}\label{eq:RNscrateGibb}
    \begin{split}
        \frac{d^2E}{d\omega \, dt} \bigg|_{\text{sc}} =
        2^{6l+1} \,
        \pi^{2l-1} \,
        Q^{4l+2} \, &
        T^{2l} \,
        \frac{|\Gamma(l+1-i\alpha)|^2 \,
        \Gamma(l+1)^2}
        {\Gamma(2l+2)^2 \,
        \Gamma(2l+1)^2} \,
        e^{\pi\alpha} \\
        \frac{\omega \, \omega_\text{eff} \, k^{2l+1} \,}{e^{\omega_{\text{eff}}/T}-1} \,
        & \frac{|\Gamma(l+1-2iF)|^2}
        {|\Gamma(1-2iF)|^2} \, .
    \end{split}
\end{equation}
Using the Gamma function identities from appendix \ref{app:math}, we find that this can be further rewritten as follows
\begin{equation}\label{eq:RNscrateGibb2}
    \begin{split}
        \frac{d^2E}{d\omega \, dt} \bigg|_{\text{sc}}^{\text{}} =
        \frac{2^{6l+1} \,
        \pi^{2l-1} \,
        l!^4}
        {(2l+1)!^2 \,
        (2l)!^2} \,
        \frac{Q^{4l+2} \,
        T^{2l} \,
        \omega \,
        \omega_\text{eff} \,
        k^{2l+1} \,}
        {e^{\omega_{\text{eff}}/T}-1} \,
        e^{\pi\alpha} \, 
        \frac{\pi \alpha}{\sinh \pi \alpha} \,
        \prod_{j=1}^{l}
        \left( j^2 \!+\! \alpha^2 \right)
        \left( 1 \!+\! \frac{4F^2}{ j^2} \right)
        \, .
    \end{split}
\end{equation}
This agrees with equations 4.33 from \cite{Gibbons:1975} after writing the $\delta r$ in their result in terms of $T$. For the comparison, note that our $\alpha$ is negative of theirs.

\subsection{Quantum-corrected emission rate}\label{ss:RNq}

To compute the quantum-corrected evaporation rate, we will follow the effective string approach used to calculate black hole greybody factors \cite{Das:1996wn, Gubser:1996xe, Gubser:1996zp, Gubser:1997cm, Maldacena:1996ix, Maldacena:1997ih, Callan:1996tv}. It was first applied to 4D near-extremal black holes using JT gravity as the effective theory by \cite{Bai:2023,Brown:2024}.

We will treat both the black hole and the quantum field as two quantum systems that are coupled through an interaction Hamiltonian. This coupling enables spontaneous decay of the black hole from higher energy states to lower ones through the emission of field quanta. The setup is completely analogous to the atom-electromagnetic field system, where the coupling (e.g. dipole) comes in the form of the product of two operators, one acting on the Hilbert space of the atom, and having non-zero matrix elements between two energy eigenstates (e.g. the position operator), and the other, a creation/annihilation operator acting on the radiation space.\footnote{
    This also provides a framework for the study of quantum black hole scattering, where an incoming coherent wave, represented as a classical expectation value of the field operator, causes the processes of absorption and stimulated emission of quanta. For recent work, see \cite{Emparan:2025a,Emparan:2025b}.}

Once the interaction Hamiltonian is specified, we can use perturbation theory to compute the probability that a transition between two energy states will happen. Since the interaction has a time dependence coming from the radiation field, we need to use time-dependent perturbation theory, which for a harmonic perturbation (a single mode of the field) gives the rate of transition in the form of Fermi's golden rule as follows. For a transition between two black hole states with energies $E_\text{i}$ and $E_\text{f}$, through the emission of a single field quantum with energy $\omega$,
\begin{equation}\label{eq:transstates}
    | E_\text{i} \rangle^{\text{BH}} \otimes | \Omega \rangle^{\text{rad}}
    \rightarrow
    | E_\text{f} \rangle^{\text{BH}} \otimes | \omega \rangle^{\text{rad}} \, ,
\end{equation}
where $\Omega$ represents the vacuum states in the radiation space, the transition rate is given by
\begin{equation}\label{eq:Gammatrans}
    \Gamma_{i\rightarrow f} =
    2 \pi \,
    |\langle E_f, \omega | \hat{H}_\text{int} |
    E_i \rangle|^2 \,
    \delta(E_f-E_i-\omega) \, ,
\end{equation}
where $\hat{H}_\text{int}$ is the interaction Hamiltonian, which enables this transition in the quantum theory. The $\delta$-function ensures energy conservation in this process. For perturbation theory to be applicable, the matrix element of $\hat{H}_\text{int}$ has to be small. Once the transition rate is computed, we can find the total quantum-corrected evaporation rate by summing over all final states of both the black hole and radiation\footnote{
    We will be computing the emission rates in the microcanonical ensemble. For a discussion of both the microcanonical and canonical ensemble in this context, see \cite{Biggs:2025}.}
\begin{equation}\label{eq:qNdot}
    \frac{dN}{dt} \bigg|_{\text{quantum-corrected}} =
    \int \, d\omega
    \int \, dE_f \, \rho(E_f) \,
    \Gamma_{i\rightarrow f} \, ,
\end{equation}
which can further be used to find the emission rate of any quantity such as charge and angular momentum. Multiplying the rate with the energy of a quantum $\omega$ before integrating gives the energy flux
\begin{equation}\label{eq:qEdot}
    \frac{dE}{dt} \bigg|_{\text{quantum-corrected}} =
    \int \, d\omega \, \omega
    \int \, dE_f \, \rho(E_f) \,
    \Gamma_{i\rightarrow f} \, .
\end{equation}
In these equations, $\rho(E)$ is the density of states of the system, which is the product of the densities of both the black hole and the radiation.

Before we have a well-defined problem and the equations from above can be applied, we need to address two questions. First, how do we describe the energy states of the black hole, and find the matrix elements of operators between these states. And second, what operator do we use, in particular, what is the interaction Hamiltonian of the black hole-radiation system? To answer both of these questions, we will use holography in the form of AdS$_2$/CFT$_1$. A near-extremal Reissner-Nordström black hole can be dimensionally reduced in its near horizon region, which results in a 2D theory, JT gravity. The dynamics of the 2D metric is captured by a strongly-coupled boundary mode called the Schwarzian mode. This mode can be quantized, and the resulting 1D quantum mechanics can be exactly solved \cite{Mertens:2017,Yang:2018,Kitaev:2018,Suh:2019,Mertens:2022}. This mode describes the low-temperature quantum dynamics of the near-extremal black hole. Solving this quantum mechanics gives a continuous spectrum\footnote{
    It is commonly accepted that this theory actually describes an ensemble of theories of gravity. This is reflected in the fact that the density of states in this theory is continuous, instead of being discrete, which is what we would expect from a true bound quantum system like an atom for instance.}
with a density of states given by
\begin{equation}\label{eq:Schwrho}
    \rho(E) =
    \frac{e^{S_0}}{2 \pi^2 E_\text{b}} \,
    \sinh \left(
    2\pi \sqrt{2E/E_\text{b}}
    \right) \,
    \Theta(E) \, ,
\end{equation}
where $E_{\text{brk}}$ is the breakdown scale introduced in \eqref{eq:RNEbrk}. Furthermore, owing to the reparameterization invariance of this theory, operators can be classified by their conformal dimension $\Delta$, which also fully determines their matrix elements between the energy eigenstates. These are given by
\begin{equation}\label{eq:Schwmelem}
    \left|\langle E_f | \mathcal{O}_{\Delta} | E_i \rangle \right|^2 =
    \frac{E_\text{b}^{2\Delta} \, e^{-S_0}}{2^{2\Delta-1} \, \Gamma(2\Delta)} \,
    \Gamma\left(
    \Delta \pm
    i \sqrt{2E_f/E_\text{b}} \pm
    i \sqrt{2E_i/E_\text{b}}
    \right) \, ,
\end{equation}
where the two $\pm$ signs are shorts for multiplying four $\Gamma$-functions with all possible combinations of the signs. The four Gamma functions combine into two pairs of complex conjugates, so this matrix element is manifestly real. This covers the part of Schwarzian theory that we will have use for. Higher point matrix elements, which would appear for instance in higher-order perturbation theory, can be found in the references above.

To address the second question from above, note that in addition to the Schwarzian mode, the 2D near-horizon region also includes a modes coming from the dimensional reduction of the radiation field. Each of these modes represents a field in the 2D theory which interacts with the Schwarzian mode. The interaction term is given by a deformation of the 1D Schwarzian action by
\begin{equation}\label{eq:intact}
    I_{\text{int}} = \int dt \, \phi_\text{}(t) \, \mathcal{O}(t) \, ,
\end{equation}
where $\phi_\text{}(t)$ is the boundary value field, which can be thought of as a source and $\mathcal{O}(t)$ is its dual operator in the usual AdS/CFT sense. The conformal dimension of this operator, which we will need for the Schwarzian matrix element, is determined by the effective mass of the 2D field or, equivalently, it is the power of the non-normalizable mode in the asymptotic expansion of the solution to the 2D wave equation. From \eqref{eq:intact} we find the interaction Hamiltonian to be
\begin{equation}\label{eq:intHam}
    \hat{H}_\text{int} =
    \hat{\phi}_0 \,
    \hat{\mathcal{O}} \, .
\end{equation}
We only need the $t=0$ value of this operator, as the rest of the time dependence has been accounted for by Fermi's rule. Once this interaction Hamiltonian is plugged into the transition rate, the matrix elements breaks up into a matrix element between the black hole states given by \eqref{eq:Schwmelem} and a matrix element in radiation space, which we turn to next.

To quantize the 4D radiation field, we start by canonically normalizing the classical field modes of the scalar field as follows
\begin{equation}\label{eq:4Dwavecanon}
    \phi(t=0,r,\theta,\phi) \sim
    a_\text{far} \,
    \frac{r_+}{\sqrt{4 \pi k}} \,\
    \frac{e^{-ikr}}{r} \,
    r^{-i\alpha} \,
    Y_{lm_l}(\theta,\phi) \, .
\end{equation}
Once these operators are quantized, the creation $\hat{a}^\dagger$ and annihilation $\hat{a}$ operators define states with a definite number of field quanta at asymptotic infinity (past or future), and they satisfy the unit normalized commutation relations, which implies that the following matrix element gives one
\begin{equation}\label{eq:farmelem}
    \left|\langle \omega |
    \hat{a}^{\dagger}_\text{far} |
    \Omega \rangle \right|^2 =
    1 \, .
\end{equation}
Since the interaction Hamiltonian \eqref{eq:intHam} requires the matrix element of the 2D field, we need to dimensionally reduce the 4D field on the sphere of the Reissner-Nordström near-horizon region. We will canonically normalize the 2D fields by requiring that at AdS$_2$ infinity they behave as follows
\begin{equation}\label{eq:2Dwavecanon}
    \phi_{lm_l}(t=0,r) \sim
    {a}^{}_\text{near} \,
    \left(\frac{r_+^2}{r-r_+}\right)^{1-\Delta} +
    b_\text{near} \,
    \left( \frac{r_+^2}{r-r_+} \right)^\Delta \, ,
\end{equation}
where $\Delta$ is the conformal dimension of the dual operator $\mathcal{O}$, as mentioned above. Since we solved the 4D equation by breaking it up into a near and far region, we already have the solution to the 2D equation, which is just the near equation. Comparing the asymptotic behavior of our near solution \eqref{eq:RNnearasymp} to the equation from above, we see that they agree. As we argued in subsection \ref{ss:RNsc}, the coefficient $b_\text{near}$ can be ignored compared to $a_\text{near}$ for a near-extremal black hole. Furthermore, this comparison confirms that $\Delta = \tfrac{1}{2} + \beta$ as defined in subsection \ref{ss:RNsc}.

We can now evaluate the radiation space matrix element. Since the states are defined at asymptotic infinity and the field operator at AdS$_2$ infinity, there is no reason to expect this to be one. Since $\hat{a}^{\dagger}_\text{near}$ and $\hat{a}^{\dagger}_\text{far}$ are linear combinations of each other, in fact they're proportional to each other when $b_\text{near}=0$, we can calculate the matrix element as follows
\begin{equation}\label{eq:nearmelem}
    \left|\langle \omega |
    \hat{a}^{\dagger}_\text{near} |
    \Omega \rangle \right|^2 =
    |\mathcal{N}|^2 \,
    |\phi_0|^2 \,
    \left|\langle \omega |
    \hat{a}^{\dagger}_\text{far} |
    \Omega \rangle \right|^2 =
    |\mathcal{N}|^2 \,
    |\phi_0|^2 \, ,
\end{equation}
where $\phi_0$ is this proportionality constant. The normalization constant $\mathcal{N}$ we introduce when quantizing these fields such as to make the interaction term \eqref{eq:intact} unit normalized. To find $\phi_0$ we proceed as follows. The $\hat{a}^{\dagger}_\text{near}$ and $\hat{a}^{\dagger}_\text{far}$ operators will both be written in terms of $A$ from subsection \ref{ss:RNeq}, which is now also an operator, and then compared. First, by comparing equations \eqref{eq:4Dwavecanon} and \eqref{eq:RNfarasymp}, we find that $\hat{a}^{\dagger}_\text{far}$ is related to the coefficient of the reflected wave as follows
\begin{equation}\label{eq:farannih}
    \begin{split}
        \hat{a}^\dagger_\text{far} =
        \frac{\sqrt{4\pi k}}{r_+} \,
        \frac{\Gamma(1 + 2\beta)}
        {\Gamma(\tfrac{1}{2} + \beta + i\alpha)} \,
        (2ik)^{-1/2-\beta+i\alpha} \,
        \hat{C}_1 \, ,
    \end{split}
\end{equation}
which can further be written in terms of $\hat{A}$ using the matching equation \eqref{eq:RNC12}
\begin{equation}\label{eq:farannih2}
    \begin{split}
        \hat{a}^\dagger_\text{far} =
        \frac{\sqrt{4 \pi k}}{r_+}
        \frac{\Gamma(1 \!+\! 2\beta)
        \Gamma(2\beta)
        \Gamma(1 \!-\! 2ik_H)
        (2ik)^{-1/2-\beta+i\alpha}}
        {\Gamma(\tfrac{1}{2} \!+\! \beta \!+\! i\alpha)
        \Gamma(\tfrac{1}{2} \!+\! \beta \!+\! i\mu)
        \Gamma(\tfrac{1}{2} \!+\! \beta \!-\! i\mu \!-\! 2ik_H)}
        \delta r^{+1/2-\beta}
        \hat{A} \, .
    \end{split}
\end{equation}
Similarly, comparing equations \eqref{eq:2Dwavecanon} and \eqref{eq:RNnearasymp} gives us $\hat{a}_\text{near}$ in terms of $\hat{A}$ as follows
\begin{equation}\label{eq:nearannih}
    \hat{a}^\dagger_\text{near} =
    r_+^{-1+2\beta} \,
    \frac{\Gamma(2\beta) \,
    \Gamma(1-2ik_H)}
    {\Gamma(\tfrac{1}{2}+\beta+i\mu) \,
    \Gamma(\tfrac{1}{2}+\beta-i\mu-2ik_H)} \,
    \delta r^{+1/2-\beta}
    \hat{A} \, .
\end{equation}
Comparing $\hat{a}_\text{near}$ and $\hat{a}_\text{near}$, we can find the proportionality constant, which after some simplifying is given by
\begin{equation}\label{eq:phinaught}
    |\phi_0|^2 =
    \frac{1}{\pi} \,
    2^{2\Delta-2} \,
    r_+^{4\Delta-2} \,
    k^{2\Delta-1} \,
    \frac{|\Gamma(\Delta+i\alpha)|^2}{\Gamma(2\Delta)^2} \,
    e^{\pi \alpha} \, .
\end{equation}
We will find the normalization constant $\mathcal{N}$ once we have calculated the quantum-corrected evaporation rate by taking the limit of the quantum rate and comparing to the semi-classical rate.

We have obtained both the black hole and radiation matrix elements, which can now be plugged into the transition rate \eqref{eq:Gammatrans}. This results in
\begin{equation}\label{eq:Gammatrans2}
    \Gamma_{i\rightarrow f} =
    2 \pi \,
    |\mathcal{N}|^2 \,
    |\phi_0|^2 \,
    \left|\langle E_\text{f} |
    \mathcal{O} |
    E_\text{i} \rangle \right|^2 \,
    \delta(E_\text{f}-E_\text{i}-\omega_\text{eff}) \, .
\end{equation}
Note that we have written $\omega_\text{eff}$ instead of $\omega$, which accounts for one more effect that only appears for charged radiation. The $E$ in the Schwarzian equations stands for energy above extremality, but when a charged particle is emitted, the black hole charge also drops, which results in the extremal energy dropping as well. If the black hole started out with total energy $M_\text{i}$, and emits a quantum with energy $\omega$ (measured at asymptotic infinity), then it will end up with total energy $M_\text{f}$ which satisfies
\begin{equation}\label{eq:encons}
    M_\text{i} = M_\text{f} + \omega \, .
\end{equation}
This can be rewritten in terms of energy above extremality as follows
\begin{equation}\label{eq:encons2}
    E_\text{i} = E_\text{f} + (\omega - e) \, ,
\end{equation}
where the second term on the right side is precisely the effective frequency as introduced in equation \eqref{eq:freqeff}. This shows that the Delta function in the transition rate \eqref{eq:Gammatrans2} involves the effective frequency. It also implies that the range of frequencies for charged radiation goes from $0$ up to the biggest available energy which is $E_\text{i} + e$, not just $E_\text{i}$, with this extra energy coming from the drop of the extremal energy.

We can now find the spectral evaporation rate from the transition rate by integrating over the the final states. This gives
\begin{equation}\label{eq:RNqrate0}
    \frac{d^2E}{d\omega \, dt} \bigg|_{\text{q}} =
    2\pi \,
    \omega \,
    |\mathcal{N}|^2 \,
    |\Phi_0|^2 \,
    \rho(E_\text{}-\omega_{\text{eff}}) \, 
    \left|\langle E_\text{}-\omega_{\text{eff}} | \mathcal{O}_{\Delta} | E_\text{} \rangle \right|^2 \, ,
\end{equation}
where we are labeling the initial black hole energy with $E$.
The final step is to find the normalization constant $|\mathcal{N}|^2$ by comparing the quantum-corrected evaporation rate that we just found to the semi-classical rate \eqref{eq:RNscrate}. To do this, we need to take the high-energy/temperature limit. This means take $|E_\text{} - \omega_{\text{eff}}| \gg E_\text{b}$, and then zoom in on $\omega \ll E+\omega_{\text{sr}}$. A more careful discussion of different energy regimes will be given in \ref{ss:class}. The actual calculation of taking this limit is mostly technical and is done in appendix \ref{app:Nsq}. The following is found
\begin{equation}\label{eq:Nsq}
    |\mathcal{N}|^2 = \frac{1}{\pi} \, 2^{2\Delta-2} \,
    \frac{\Gamma(2\Delta)}{\Gamma(2\Delta-1)^2} \,
    \left|\Gamma(\Delta+i\mu) \right|^2 \,
    e^{-\text{sgn}(\omega_\text{eff}) \pi \mu} \, .
\end{equation}
Plugging this into equation \eqref{eq:RNqrate0}, we find the final formula for the quantum-corrected spectral evaporation rate
\begin{equation}\label{eq:RNqrate}
    \begin{split}
        & \frac{d^2 E}{d\omega \, dt} \bigg|_{\text{q}} =
        \frac{1}{\pi^3} \,
        2^{2\Delta-3} \,
        Q^{4\Delta-2} \,
        E_\text{b}^{2\Delta-1} \,
        \frac
        {\left| \Gamma(\Delta+i\alpha) \right|^2 \,
        \left| \Gamma(\Delta+i\mu) \right|^2}
        {\Gamma(2\Delta)^2 \,
        \Gamma(2\Delta-1)^2} \,
        e^{\pi(\alpha-\text{sgn}(\omega_\text{eff}) \mu)} \\
        & \qquad \qquad \ \ \,
        \omega \,
        k^{2\Delta-1} \,
        \sinh \! \left[ 2\pi
        \left( 2 \frac{E_\text{}-\omega_\text{eff}}{E_\text{b}} \right)^{\!\frac{1}{2}} \right] \\
        & \qquad \qquad \ \ \,
        \Gamma \! \left[
        \Delta \pm
        i
        \left( 2\frac{E_{}-\omega_\text{eff}}{E_\text{b}} \right)^{\!\frac{1}{2}} \! \pm i
        \left( \frac{2E_{}}{E_\text{b}} \right)^{\!\frac{1}{2}}  \right] \, .
    \end{split}
\end{equation}
As mentioned earlier, the range of frequencies in this equation goes from zero to $E_{} + \omega_{sr}$. The parameters $\mu, \Delta, \alpha$ are the same as before and can be found in equations \eqref{eq:RNnearparam3}, \eqref{eq:RNnearparam4} and \eqref{eq:RNfarparam2}. In the following section, we will compute this rate in different energy regimes, and compare it to the semi-classical rate.

This equation was recently derived in \cite{Betzios:2025}.
As mentioned before, if we set $\mu=0$ in our result \eqref{eq:RNqrate}, we find agreement with theirs.

\section{Radiation from small charged black holes}\label{sec:smBH}

The conformal dimension for a massive charged scalar field being radiated by a near-extremal Reissner-Nordström black hole is given by
\begin{equation}\label{eq:Deltaomega}
    \Delta(\omega) =
    \tfrac{1}{2} +
    \sqrt{\tfrac{1}{4} + l(l+1) - e^2Q^2 + m^2Q^2 + 6\omega(e-\omega)Q^2} \, .
\end{equation}
For simplicity, we will consider a massless field in this section. Depending on the relationship between $Q$ and $e$, this will either be real for all frequencies, or real for only a subset of frequencies and complex for the rest (it is never complex for all frequencies). The character of the greybody factor, and therefore the evaporation rate strongly depends on whether $\Delta$ is real or complex. In this section, we will study radiation with real $\Delta$. Inspecting the quadratic under the square root, we find that it is positive for all
frequencies when\footnote{
    Even for $eQ \leq \tfrac{1}{2} + l$ there will be a large frequency beyond which $\Delta$ goes complex, so here we really mean not all frequencies, but those between $0$ and $e+E_\text{}$, with $E$ being a small number. In the semi-classical rate, the integral does go from $0$ to $\infty$, but the contribution from the larger frequencies is negligible. We will explicitly see that this is true later in this section.
}
\begin{equation}\label{eq:smBHdef}
    eQ \leq \tfrac{1}{2} + l \,.
\end{equation}
Note that for all black holes, their charge has to be greater than the charge of the particles they are emitting, so we also have $Q\geq e$. These two conditions imply that $e \leq Q \leq \ (\tfrac{1}{2}+l) \frac{1}{e} $, so we will refer to these black holes as small\footnote{
    Of course $l$ can be arbitrarily high, but we will see that higher angular momentum modes contribute significantly less than the lowest one, and therefore it is safe to take $l=0$ in equation \eqref{eq:smBHdef}.
}
($Q$ can be equally thought of as the radius of the horizon, the charge, or the mass of the black hole). For large black holes, those with $ Q \geq (\tfrac{1}{2}+l) \tfrac{1}{e}$, the quadratic under the square root in \eqref{eq:Deltaomega} will still be positive on a subset of the available frequencies, so the equations that will be derived in this section will still hold on this subset. The rest of the frequency range will be covered in the next section.

\subsection{Classification of charged radiation}\label{ss:class}

We start with a discussion of the energy scales involved in the problem. For neutral radiation, if we look at the quantum-corrected evaporation rates, there are two scales, the initial energy of the black hole $E_\text{}$ (which in the semi-classical rate is more naturally thought of as the initial temperature) and the breakdown energy $E_\text{b}$. In the rate equation ($e=0$ limit of \eqref{eq:RNqrate}), these two scales appear in the following combination
\begin{equation}\label{eq:Eratio}
    \frac{E_\text{}-\omega}{E_\text{b}} \, ,
\end{equation}
with the frequency being at most $E_\text{}$. Requiring that this number be either very big or very small, \cite{Brown:2024} introduced two energy regimes, the semi-classical $E_\text{b} \ll E_\text{}$, where the quantum-corrected rate reduces to the semi-classical, and the quantum regime $E_\text{} \ll E_\text{b}$, where the quantum-corrected rate simplifies, but is not equal to the semi-classical. The latter regime is clearly the more interesting one, as this is where the corrections from Schwarzian theory give different predictions from the semi-classical theory.

Our goal will be to define a quantum regime for charged radiation, and then to compare the two rates in this regime. For charged radiation, there is an additional energy scale, the superradiant bound $\omega_\text{sr}=e$, which makes the analysis of the energy regimes more involved. From equation \eqref{eq:RNqrate}, we can see that these three scales appear in the combination
\begin{equation}\label{eq:Echratio}
    \frac{E_\text{}-\omega_\text{eff}}{E_\text{b}} =
    \frac{(E_\text{}+e)-\omega}{E_\text{b}}\, ,
\end{equation}
with the frequency now being at most equal to $e+E_\text{}$. Requiring this to be either big or small, we define the semi-classical regime to have $E_\text{b}$ smaller than \textit{either} (\textit{or both}) $E_\text{}$ and $\omega_\text{sr}$, and the quantum regime to have $E_\text{b}$ greater than \textit{both} $E_\text{}$ and $\omega_\text{sr}$.

Independently, in either regime, we can further require a separation between $E_\text{}$ and $\omega_\text{sr}$, which defines two more cases. When $\omega_\text{sr} \gg E_\text{}$, most of the emitted frequencies will be below the superradiant bound, so this will be referred to as the superradiant-dominated regime, while $\omega_\text{sr} \ll E_\text{}$ will be referred to as the neutral-like regime, since the character of the radiation, as we will see, will be similar (but not the same) as that found for neutral radiation in \cite{Brown:2024} for both the semi-classical and quantum rates. All of six possible permutations of these three energies, and the corresponding energy regimes are summarized in the following table.
\begin{table}[H]
    \centering
    \begin{tabular}{|m{2.5cm} | m{2.5cm}| m{2.5cm}|}
    \hline
    \ &
    Semi-Classical &
    \ \ \ Quantum \\ 
    \hline
    Neutral-Like &
    $\ e \ \textcolor{gray}{\ll E_\text{b}} \ll E_\text{}$
    $\textcolor{gray}{E_\text{b} \ll} \ e \ \ll E_\text{}$ &
    $e \ll E_\text{} \ll E_\text{b} $ \\ 
    \hline
    Superradiant-Dominated &
    $E_\text{} \ \textcolor{gray}{\ll E_\text{b}} \ll e$
    $\textcolor{gray}{E_\text{b} \ll} \ E_\text{} \ \ll e$ &
    $E_\text{} \ll e \ll E_\text{b} $ \\
    \hline
    \end{tabular}
    \captionsetup{width=.93\linewidth}
    \caption{A summary of the possible energy regimes for the emission of charged radiation from Reissner-Nordström.}
    \label{tab:Eregimes}
\end{table}
\noindent For the left column, it is safe to use the semi-classical rate \eqref{eq:RNscrate}, as the quantum rate would just reduce to it, while for the right column, the semi-classical \eqref{eq:RNscrate} and quantum \eqref{eq:RNqrate} rates give different answers. Note that $E_\text{b}$ is grayed out in the left column, since it is not "seen" by the semi-classical rate, so its position relative to the other two scales is irrelevant, as long as it is smaller than one of them.

Choosing the scales $E_\text{b} = \frac{1}{Q^3}$ and $\omega_\text{sr} = e$ amounts to choosing a charge for the black hole $Q$ and the particle it is radiating $e$. But, as we saw at the beginning of this section, there are further constraints on these two parameters, first $Q \geq e$, and second $eQ \lesssim 1$ (we are ignoring numbers of order one here) for the conformal dimension to be real. All possible values of these two parameters and the corresponding energy and $\Delta$ regimes they define are summarized in the following figure.
\begin{figure}[H]
    \centering
    \begin{tikzpicture}
    \draw[gray, thick, dashed] (0,-3) -- (0,+3);
    \filldraw[black] (+3.1,0) node[anchor=west]{$\log Q$};
    \draw[gray, thick, dashed] (-3,0) -- (+3,0);
    \filldraw[black] (-0.2,+3.3) node[anchor=west]{$\log e$};
    \draw[black, thick] (-2.5,-2.5) -- (+2.5,+2.5);
    \filldraw[black] (+2.5,+2.6) node[anchor=west]{$e=Q$};
    \draw[black, thick] (-2.5,+2.5) -- (+2.5,-2.5);
    \filldraw[black] (-2.57,+2.6) node[anchor=east]{$eQ \approx 1$};
    \draw[black, thick] (-1.1,+3.3) -- (+1.1,-3.3);
    \filldraw[black] (+0.9,-3.7) node[anchor=west]{$\omega_\text{sr} = E_\text{brk}$};
    \draw[<->, teal, semithick] (-135:2.5cm) 
    arc[radius=2.5, start angle=-135, end angle=-71.565];
    \filldraw[teal] (-0.3,-2.9) node[anchor=east]{\bf{Quantum}};
    \draw[<->, blue, semithick] (-71.565:2.5cm) 
    arc[radius=2.5, start angle=-71.565, end angle=-45];
    \filldraw[blue] (+1.4,-3.0) node[anchor=west]{\bf{Semi-Cl.,} $\Delta\in\mathbb{R}$};
    \draw[<->, blue, semithick] (-45:2.5cm) 
    arc[radius=2.5, start angle=-45, end angle=+45];
    \filldraw[blue] (+2.4,1.2) node[anchor=west]{\bf{Semi-Cl.,} $\Delta\in\mathbb{C}$};
    \draw[<->, red, semithick] (+45:2.5cm) 
    arc[radius=2.5, start angle=+45, end angle=+225];
    \filldraw[red] (-2.3,1.5) node[anchor=east]{\bf{Forbidden}};
    \end{tikzpicture}
    \captionsetup{width=.93\linewidth}
    \caption{A diagram of $\log Q$ vs. $\log e$ parameter space showing the available regions for radiation from a Reissner-Nordström black hole. The bottom two wedges (small black holes) always have real $\Delta$, the top wedge (large) has both real and complex $\Delta$. The top two wedges always have the quantum rate \eqref{eq:RNqrate} reduce to the semi-classical formula \eqref{eq:RNscrate}, while the bottom wedge can have (and this further depends on the size of $E$ which is not on this diagram) the quantum rate give different predictions from the semi-classical.}
    \label{fig:logQloge}
\end{figure}
\noindent Only points under $e=Q$ are possible. Everything under $eQ \approx 1$ are the small black holes\footnote{
    We will sometimes refer to the middle wedge in the figure as mid-sized black holes and the bottom as small, when we need to distinguish between them. The main difference is that the former always produce semi-classical radiation, while the latter can also radiate in the quantum regime.
}
with real $\Delta$ (the subject of this section), and everything above are large black holes with complex $\Delta$ (section \ref{sec:lBH}). Everything below the $\omega_\text{sr} = E_\text{b}$ line, which is $e = Q^{-3}$ in terms of $Q$ and $e$, is possibly in the quantum regime, and everything above is in the semi-classical regime. At each available $(Q,e)$ point on this plot, we still get to independently choose $E_\text{}$ (or $T$), which, depending on how it compares to $\omega_\text{sr}$, will further determine whether we are in the neutral-like or superradiant-dominated regime.

There are a few conclusions we can draw from this diagram. First, large black holes (the top wedge) necessarily radiate in the semi-classical regime. Therefore, we will not need to consider the quantum evaporation rates for these black holes in the next section. Second, if we consider the small black holes  in the bottom wedge, then that radiation will necessarily have small charge $eQ \ll 1$, which will allow us to take the small charge limits of the formulae to be derived in this section. Third, we can apply this analysis to our Universe where the smallest charged particle is the positron with $e=0.0854$. All near-extremal black holes then sit on a horizontal line slightly below the abscissa ($e=1$). This implies that the quantum-corrections would only appear for microscopic black holes which at most have a charge roughly double the positron.

The three parameters $(Q,T,e)$, or equivalently the three energy scales $(E_\text{b},E_\text{},\omega_\text{sr})$ define a unique emission spectrum in one of the four energy regimes given in table \ref{tab:Eregimes}. In the next two subsections, our goal will be to compute, analyze, and compare the spectra and rates in these regimes. Before turning to the details, we first point out some general features of the neutral-like and superradiant-dominated regimes. We start with the semi-classical evaporation rates. The integral in equation \eqref{eq:scEdot}, which goes from $0$ to $\infty$ can be split up into superradiant ($\omega \leq \omega_\text{sr}$) and non-superradiant frequencies ($\omega \geq \omega_\text{sr}$) frequencies. In each of the two regimes, superradiant (top left corner of table \ref{tab:Eregimes}, or just $T \ll e$) and neutral-like (bottom left corner of table \ref{tab:Eregimes}, or $e \ll T$), each of these two integrals can be estimated as follows. Starting with the lower frequencies in the superradiant regime, we only copy the temperature and frequency-dependent part of the semi-classical evaporation rate \eqref{eq:RNscrate} (ignoring the frequency-dependence of the parameters $\Delta, \mu, \alpha$ as these are irrelevant for the size of the integrals) which gives
\begin{equation}\label{eq:srint1}
    I^\text{SD}_{\omega\leq \omega_\text{sr}} \sim
    T^{2\Delta-2}
    \int_{0}^{e} d\omega \,
    \frac{\omega_\text{eff} \,
    \omega^{2\Delta}\,
    \gamma(\omega_\text{eff}/T)}
    {e^{\omega_\text{eff}/T} - 1} \,
    \sim
    T^{2\Delta-2}
    \int_{0}^{e} d\omega \,
    |\omega_\text{eff}| \,
    \omega^{2\Delta} \,
    \left( \frac{\omega_\text{eff}}{T} \right)^{2(\Delta-1)}
    \sim
    e^{4\Delta} \, .
\end{equation}
$\gamma(\omega_\text{eff}/T)$ consists of the $F$-dependent part of \eqref{eq:RNscrate} (the last fraction), and given that $F=\omega_\text{eff}/4\pi T$, it is a function of $\omega_\text{eff}/T$. In the second equality, we ignored the exponential as $|\omega_\text{eff}| \leq e$ and $e\gg T$. We also used the fact that $\gamma(\omega_\text{eff}/T)$ reduces to the fraction in the third integral in this limit, which we will explicitly show in subsection \ref{ss:sd}. As a result, the temperature dependence cancels out. In the third equality, we performed the integral to get the final estimate, which is true to leading order in $T/e$. For the higher frequencies, we have the following
\begin{equation}\label{eq:srint2}
    I^\text{SD}_{\omega\geq \omega_\text{sr}}
    \sim
    T^{2\Delta-2}
    \int_{e}^{\infty} d\omega
    \frac{\omega_\text{e}
    \omega^{2\Delta}
    \gamma(\omega_\text{e}/T)}
    {e^{\omega_\text{e}/T} - 1}
    \sim
    T^{2\Delta-2}
    \int_{0}^{\infty} d\omega'
    \frac{\omega'
    (\omega' \!+\! e)^{2\Delta}
    \gamma(\omega'/T)}
    {e^{\omega'/T} - 1}
    \sim
    e^{4\Delta}
    \left(\frac{T}{e} \right)^{2\Delta} \, .
\end{equation}
In the second equality, we shifted the frequency variable $\omega'=\omega-e$. In the resulting integral, all frequencies beyond $\omega'=T$ experience exponential suppression, and can be ignored, which means that $\omega'$ is at most of order $T$. As a result, in the third equality, we ignored $\omega'$ in $(\omega'+e)$, and did the integral to obtain the final estimate. What this analysis shows, is that, in this regime, the non-superradiant modes are strongly suppressed compared to the superradiant modes, and can therefore be ignored. In the neutral-like regime, we can perform an analogous analysis. Starting with the lower frequencies, we have
\begin{equation}\label{eq:nlint2}
    \begin{split}
        I^\text{NL}_{\omega\leq e}
        \sim
        T^{2\Delta-2}
        \int_{0}^{e} d\omega
        \frac{\omega_\text{e} \,
        \omega^{2\Delta} \,
        \gamma(\omega_e/T)}
        {e^{\omega_e/T}-1} \,
        \sim
        T^{2\Delta-2}
        \int_{0}^{e} d\omega \,
        \frac{|\omega_e| \,
        \omega^{2\Delta}}
        {\frac{|\omega_e|}{T}} \,
        \sim
        T^{4\Delta} \,
        \left( \frac{e}{T} \right)^{2\Delta+1} \, .
    \end{split}
\end{equation}
In the second equality, we approximated the denominator in the left integral to lowest order using the fact that $e \ll T$. We then did the integral. For the higher frequencies, we have
\begin{equation}\label{eq:nlint1}
    \begin{split}
        & I^\text{NL}_{\omega\geq e}
        \sim
        T^{2\Delta-2}
        \int_{0}^{\infty} d\omega' \,
        \frac{\omega' \,
        (\omega'+e)^{2\Delta} \,
        \gamma(\omega'/T)}
        {e^{\omega'/T}-1} \,
        \sim
        T^{4\Delta} \, ,
    \end{split}
\end{equation}
where we have performed the same shift as before. Here, we ignore $e$ in $(\omega'+e)$ and do the integral to find the final estimate. We find the opposite conclusion in this regime, the superradiant frequencies are suppressed compared to the non-superradiant ones.

For the quantum-corrected evaporation rate, the integral in equation \eqref{eq:qEdot} goes from $0$ to $e + E$, with the superradiant modes from $0$ to $e$, and the non-superradiant from $e$ to $e+E$. The emitted frequencies in the neutral-like ($e \ll E$) and superradiant-dominated ($E \ll e$) regimes parallel what we found for the semi-classical rate, which is much more straightforward to show. When $e\ll E$, the range of superradiant frequencies is very small and can be ignored, and when $E \ll e$, the opposite is true.

We also point out that in the quantum regime ($E + e \ll E_\text{b}$), we can make the following two approximations in the quantum-corrected evaporation rate. The $\sinh$ function in the Schwarzian density of states simplifies as follows
\begin{equation}\label{eq:sinhGamma}
    \begin{split}
        \sinh \left( 2 \pi \sqrt{2(E-\omega_\text{eff})/E_\text{b}} \right) & =
        2 \pi \sqrt{2(E-\omega_\text{eff})/E_\text{b}} \, ,
    \end{split}
\end{equation}
and the frequency-dependent part of the Schwarzian matrix element given by the four Gamma functions reduces to
\begin{equation}\label{eq:Gammaquart}
    \begin{split}
        \left| \Gamma\left(
        \Delta + i \sqrt{2(E-\omega_\text{eff})/E_\text{b}} \pm i \sqrt{2E_\text{}/E_\text{b}}
        \right) \,
        \right|^2 = \,
        \Gamma(\Delta)^4 \, .
    \end{split}
\end{equation}

Finally, we comment on the parameters $\Delta,\alpha,\mu$, which are all functions of frequency in general, but simplify in some of the energy regimes. First, for massless radiation, we have the following
\begin{equation}
    \mu = - \alpha = - Q(2\omega-1) \, .
\end{equation}
Then, in the neutral-like regime, as we saw, the superradiant frequencies can be ignored, and the frequency is at most $\omega \sim T$, which is small for a near-extremal black hole. This means that all terms of the order $Q\omega \sim QT \ll 1$ can be ignored, which gives $\Delta,\alpha,\mu$ the following frequency-independent values
\begin{equation}\label{eq:Deltae}
    \begin{split}
        \Delta_e & = \tfrac{1}{2} + \sqrt{\tfrac{1}{4} + l(l+1) - e^2\,Q^2} \, , \\
        \alpha_e & = - \mu_e = - eQ \, .
    \end{split}
\end{equation}
In the superradiant regime, the frequency is of order $\omega \sim e$, so terms such as $Q \omega \sim eQ$ cannot be necessarily ignored, and the full frequency-dependent form of $\Delta,\alpha,\mu$ must be taken
\begin{equation}\label{eq:Deltaalphaomega}
    \begin{split}
        \Delta(\omega) & =
        \tfrac{1}{2} +
        \sqrt{\tfrac{1}{4} + l(l+1) - e^2Q^2 + 6\omega(e-\omega)Q^2} \, , \\
        \alpha(\omega) & = - \mu(\omega) = Q (2\omega -e) \, .
    \end{split}
\end{equation}
One final limit that can be taken in either regime is the small charge limit $eQ\ll1$, where the parameters simplify as follows
\begin{equation}\label{eq:Deltanaught}
\begin{split}
    \Delta_0 & = l + 1 \, , \\
    \alpha_0 & = \mu_0 =0 \, .
\end{split}
\end{equation}

\subsection{Neutral-like regime}\label{ss:nl}

We start with the semi-classical regime, where the quantum-corrected evaporation rate just reduces to the semi-classical rate, and therefore we can just use equation \eqref{eq:RNscrate}. In the previous subsection, we showed that in the neutral-like regime we only need to consider the non-superradiant frequencies and that the parameters $\alpha=-\mu$ and $\Delta$ take the constant values given in equation \eqref{eq:Deltae}. Furthermore, we showed that for the non-superradiant frequencies $\omega \sim T \ll e$, and therefore we can ignore $e$ in all $\omega_\text{eff}$. Under these assumptions, the evaporation rate is given by
\begin{equation}\label{eq:nlscint}
    \begin{split}
        & \frac{dE}{dt} \bigg|_\text{sc} = 
        2^{6\Delta_e-5} \,
        \pi^{2\Delta_e-3} \,
        Q^{4\Delta_e-2} \,
        T^{2\Delta_e-2} \,
        \frac{|\Gamma(\Delta_e+ieQ)|^4}
        {\Gamma(2\Delta_e)^2 \,
        \Gamma(2\Delta_e-1)^2} \,
        e^{-\pi e Q} \\
        & \qquad \qquad
        \int_{0}^{\infty} d\omega \,
        \frac{\omega^{2\Delta_e+1}}
        {e^{\omega/T}-1} \,
        \left| \frac{ \Gamma(\Delta_e-i\omega/2\pi T) }
        {\Gamma(1-i\omega/2\pi T+ieQ)} \right|^2 \, ,
    \end{split}
\end{equation}
where we have pulled out all the frequency-independent terms in front of the integral. The only extra frequency dependence comes from $k=\omega$ and $\omega_\text{eff} = \omega$. Performing a change of variables $x=\omega/T$ under the integral, and writing $T$ in terms of $E$ using \eqref{eq:drET}, gives our final result for the semi-classical neutral-like evaporation rate\footnote{
    This is actually the evaporation rate of an $l, m_l$ angular momentum mode of which there are $2l+1$. To get the full evaporation rate, we would sum over these modes, which we will not explicitly do. The same will be true for all following evaporation rates that we write down.
}
\begin{equation}\label{eq:nlscfin}
    \frac{dE}{dt}\bigg|_{\text{sc}} =
    F_l(eQ) \,
    \left( \frac{E}{Q} \right)^{2\Delta_e}
    \frac{1}{Q^2} \, .
\end{equation}
The $E-$dependence factors out after changing variables and the rest of the rate can be collected into a function of a single variable $eQ$ given in terms of the following parametric integral
\begin{equation}\label{eq:Fl}
    \begin{split}
        & F_l(eQ) =
        \frac{2^{4\Delta_e-5} \,
        \pi^{-2\Delta_e-3} \,
        \left| \Gamma(\Delta_e+ieQ) \right|^4}
        {\Gamma(2\Delta_e)^2 \,
        \Gamma(2\Delta_e-1)^2} \,
        e^{-\pi eQ} \,
        \int_{0}^{\infty} dx \, \frac{x^{2\Delta_e+1}}
        {e^x-1} \,
        \left| \frac{\Gamma(\Delta_e+i\frac{x}{2\pi})}
        {\Gamma(1+i(\frac{x}{2\pi}-eQ))} \right|^2 \, .
    \end{split}
\end{equation}
In equation \eqref{eq:nlscfin}, we can reintroduce units by inserting factors of $G$ as follows. Replace $E/Q \to \sqrt{G} E/Q$ and $1/Q^2 \to 1/GQ^2$, where $r_\text{h} = \sqrt{G} Q$ is the horizon radius, which has units of distance.

\begin{figure}[H]
\centering
\includegraphics[width=1.0\textwidth]{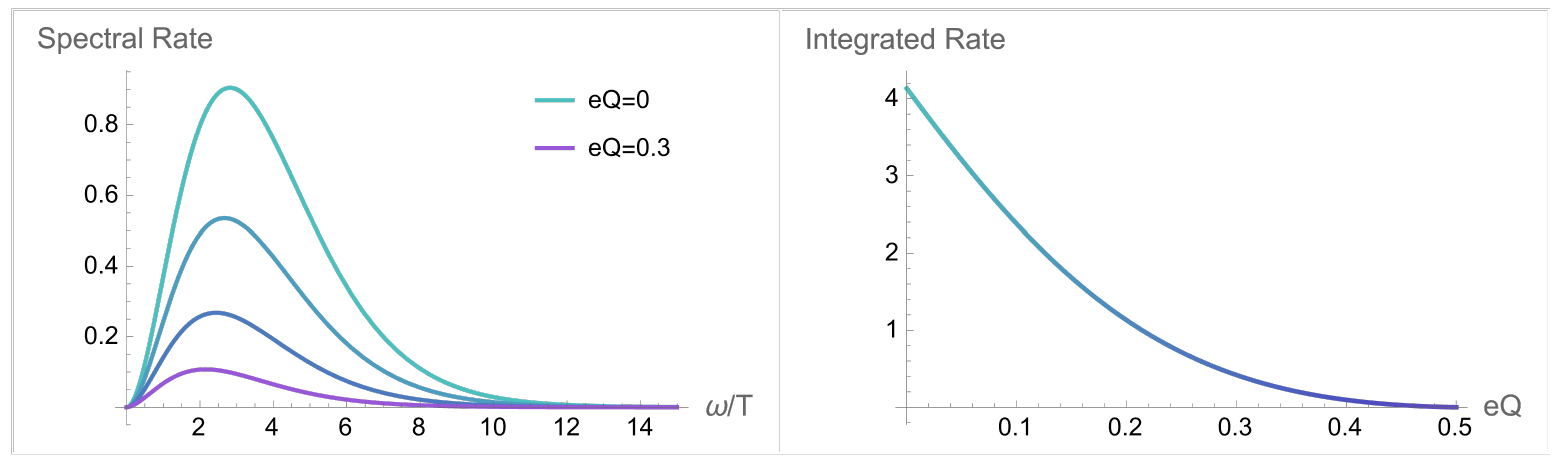}
\captionsetup{width=0.93\linewidth}
\caption{(Left) Spectral evaporation rate in quantum neutral-like regime at four different values of $eQ$ between zero and $0.3$. (Right) Integrated evaporation rate as a function of $eQ$.}
\label{fig:QNLSCrate}
\end{figure}

In the left panel of figure \ref{fig:QNLSCrate}, we plot the spectral evaporation rate, the integrand of \eqref{eq:nlscfin}, at different (non-integer) values of $eQ\leq1/2$ and $l=0$. We find that for $eQ\ll1/2$ it starts out large. This value corresponds to neutral radiation. It becomes smaller as $eQ$ is turned up, and gives zero at $eQ=1/2$. In the right panel, we also included a plot of the integrated evaporation rate as a function of $eQ$.
\noindent There are a number of competing factors in \eqref{eq:Fl} that lead to this dependence on $eQ$. A higher $eQ$ gives a lower conformal dimension $\Delta_e$, which gives a lower power of temperature in \eqref{eq:nlscfin}, and therefore a higher evaporation rate. The damping factor $\exp (-\pi eQ)$ and the following Gamma function property,
\begin{equation}\label{eq:Gammacomplbnd}
    |\Gamma(\Delta_e+i eQ)| \leq \Gamma(\Delta_e) \, ,
\end{equation}
contribute to lowering the rate. And finally, what further lowers the rate and causes it to go to zero at $eQ = 1/2$ is the fact that the Gamma function has a vertical asymptote at zero, which means that the denominator Gamma function
\begin{equation}\label{eq:Gammaasymp}
    \Gamma(2\Delta_e-1)^2 \sim \frac{1}{(2\Delta_e-1)^2} \, ,
\end{equation}
when $\Delta_e \rightarrow 1/2$ which happens when $eQ\rightarrow 1/2$ and causes the rate to go to zero, as all the other factors are non-zero in this limit. Apart from these extra suppression factors and the slightly smaller, but still constant conformal dimension, neutral-like radiation is in most ways similar to purely neutral radiation, hence its name.

We can also take the small charge or neutral limit $eQ\ll1$ of equation \eqref{eq:nlscfin}, which leads to further simplification. The conformal dimension simplifies to $\Delta_0=l+1$, an integer, and therefore the ratio of Gamma functions under the integral can be reduced to a polynomial in $x$ using the Gamma function identities from appendix \ref{app:math}. This leads to the following evaporation rate
\begin{equation}\label{eq:nlsclim}
    \begin{split}
        \frac{dE}{dt} \bigg|_{\text{sc}}^{} =
        C_l \,
        \left( \frac{E}{Q} \right)^{2l+2}
        \frac{1}{Q^2} \, ,
    \end{split}
\end{equation}
with all the numerical factors collected into the following coefficient
\begin{equation}\label{eq:Cl}
    \begin{split}
        C_l =
        \frac{2^{4l-1} \,
        \pi^{-2l-5} \,
        l!^4}
        {(2l+1)!^2 \, (2l)!^2} \,
        \int_{0}^{\infty} dx
        \frac{x^{2l+3}}{e^{x}-1} \,
        \prod_{j=1}^{l} \left( j^2 + \frac{x^2}{4\pi^2} \right) \, .
    \end{split}
\end{equation}
For a given $l$, the product can be expanded into a sum and each term can be solved using the integral formula for the Zeta function, but there is no simple closed form expression for writing this product in terms of a sum for arbitrary $l$.

From this formula it is clear that higher angular momentum modes are proportional to higher powers of temperature, and therefore contribute less to the evaporation rate. As an example, we compute the dominant lowest $l=0$ mode using \eqref{eq:nlsclim}. Setting $l=0$ and doing the integral gives
\begin{equation}\label{eq:nlscleq0}
    \frac{dE}{dt} \bigg|_{\text{sc}}^{(l=0)} =
    \frac{1}{30 \pi} \,
    \left( \frac{E}{Q} \right)^{2}
    \frac{1}
    {Q^2} \, .
\end{equation}
This agrees with the neutral rate found in \cite{Brown:2024}. For completeness and the purposes of later discussion, we also compute the rate of emission of charge in this limit for this dominant mode. By replacing one power of $\omega$ with a factor $e$ in the derivation from above, it is straightforward to show that the following is obtained for the emission rate of charge
\begin{equation}\label{eq:nlscleq0Q}
    \frac{dQ}{dt} \bigg|_{\text{sc}}^{(l=0)} =
    \frac{\sqrt{2} \, \zeta(3)}{\pi^4} \,
    \left( \frac{E}{Q} \right)^{\frac{3}{2}}
    \frac{e}{Q} \, .
\end{equation}

We now turn to the quantum neutral-like regime, where we have $e \ll E_\text{} \ll E_\text{b}$. Using the two approximations introduced in the previous subsection, equations \eqref{eq:sinhGamma} and \eqref{eq:Gammaquart}, the fact that $e \ll E_\text{}$, and the fact that $\alpha$ and $\Delta$ are the constants, we find that the quantum rate given by equation \eqref{eq:RNqrate} reduces to
\begin{equation}\label{eq:nlqint}
    \begin{split}
        \frac{dE}{dt}\bigg|_{\text{q}} =
        \frac{1}{\pi^2} \,
        2^{2\Delta_e-3/2} \,
        Q^{4\Delta_e-2} \,
        E_\text{b}^{2\Delta_e-3/2}
        \frac{\left| \Gamma(\Delta_e+ieQ) \right|^4 \,
        \Gamma(\Delta_e)^4}
        {\Gamma(2\Delta_e)^2 \,
        \Gamma(2\Delta_e-1)^2} \,
        e^{-2 \pi eQ} \,
        \int_{0}^{E_\text{}} d\omega \,
        \omega^{2\Delta_e} \,
        (E_\text{}-\omega)^{1/2} \, .
    \end{split}
\end{equation}
Performing a change of variables $x=\omega/E$, we obtain our final result for the quantum neutral-like evaporation rate
\begin{equation}\label{eq:nlqfin}
    \frac{dE}{dt}\bigg|_{\text{q}} =
    \hat{F}_l(eQ)\,
    \left( \frac{E}{Q} \right)^{2\Delta_e}
    \left( \frac{E_\text{}}{E_\text{b}} \right)^{\frac{3}{2}}
    \frac{1}{Q^2} \, .
\end{equation}
The $E-$dependence factors out with an extra factor of the energy scales $(E_\text{}/E_\text{b})^{3/2}$. The rest of the rate is collected into the following function
\begin{equation}\label{eq:Fhl1}
    \hat{F}_l(eQ) =
    \frac{1}{\pi^2} \,
    2^{2\Delta_e-3/2} \,
    \frac{\left| \Gamma(\Delta_e+ieQ) \right|^4 \,
    \Gamma(\Delta_e)^4}
    {\Gamma(2\Delta_e)^2 \,
    \Gamma(2\Delta_e-1)^2} \,
    e^{-2 \pi eQ}
    \int_{0}^{1} dx \,
    x^{2\Delta_e} \,
    (1-x)^{1/2} \, .
\end{equation}
This integral can be performed using the Beta function integral, which can then be written in terms of Gamma functions, and then further simplified using the Gamma function duplication formula. The following is obtained
\begin{equation}\label{eq:Fhl2}
    \hat{F}_l(eQ) =
    \frac{2^{6\Delta_e+1/2} \,
    \left| \Gamma(\Delta_e+ieQ) \right|^4 \,
    \Gamma(\Delta_e)^4 \,
    \Gamma(2\Delta_e+1) \,
    \Gamma(2\Delta_e+2)}
    {\pi^2 \,
    \Gamma(2\Delta_e)^2 \,
    \Gamma(2\Delta_e-1)^2 \,
    \Gamma(4\Delta_e+4)} \,
    e^{-2 \pi eQ} \, .
\end{equation}
As mentioned in the previous subsection, in the quantum regime, we cannot approach $eQ=1/2$ without leaving it. In fact $eQ\ll1$ for almost all of the bottom wedge in figure \ref{fig:logQloge}. This implies that the plots of the spectral evaporation rate would be no different from those for neutral radiation produce in \cite{Brown:2024}, so we will not repeat them here.

At this point, we can address the question of whether perturbation theory, which was used to derive the quantum evaporation rate, is valid for charged radiation, at least in the neutral-like regime. Since the evaporation rate is proportional to a power of $QT$ times a power of $E/E_\text{b}$, both of which are small quantities in this regime, it is safe to say that perturbation theory is valid.

Given that $eQ\ll1$ for most of the quantum regime, the small charge / neutral limit is even more significant in this case. Setting $eQ\ll1$ in equation \eqref{eq:nlqfin} we find the following rate
\begin{equation}\label{eq:nlqlim}
    \frac{dE}{dt}\bigg|_{\text{q}} =
    \hat{C}_l \,
    \left( \frac{E}{Q} \right)^{2l+2}
    \left( \frac{E_\text{}}{E_\text{b}} \right)^{\frac{3}{2}}
    \frac{1}{Q^2} \, ,
\end{equation}
with the numerical coefficient given by
\begin{equation}\label{eq:Chl}
    \begin{split}
        \hat{C}_l =
        \frac{2^{6l+6+1/2} \,
        l!^8 \, (2l+2)! \, (2l+3)! \, (2l+1)^2}
        {\pi^2 \,
        (2l+1)!^4 \, (4l+7)!} \, .
    \end{split}
\end{equation}
As an example, we evaluate this formula for the dominant $l=0$ mode. This gives the following
\begin{equation}\label{eq:nlqleq0}
    \frac{dE}{dt} \bigg|_{\text{q}}^{(l=0)} =
    \frac{16\sqrt{2}}{105 \, \pi^2} \,
    \left( \frac{E}{Q} \right)^{2} \,
    \left( \frac{E_\text{}}{E_\text{b}} \right)^{\frac{3}{2}}
    \frac{1}{Q^2} \, .
\end{equation}
This agrees with what was found in \cite{Brown:2024} for neutral scalar radiation. For completeness, we also give the charge emission rate
\begin{equation}\label{eq:nlqleq0Q}
    \frac{dQ}{dt} \bigg|_{\text{q}}^{(l=0)} =
    \frac{8}{15 \pi} \,
    \left( \frac{E}{Q} \right)^{\frac{3}{2}}
    \left( \frac{E_\text{}}{E_\text{b}} \right)^{} \,
    \frac{e}{Q}\, .
\end{equation}

When we compare the semi-classical and quantum emission rates, equations \eqref{eq:nlscfin} and \eqref{eq:nlqfin}, we find that the answers differ by the following ratio of scales
\begin{equation}\label{eq:nlratio}
    \frac{\dfrac{dE}{dt}\bigg|_\text{q}}
    {\,\dfrac{dE}{dt}\bigg|_\text{sc}}
    \sim
    \left( \frac{E_\text{}}{E_\text{b}} \right)^{3/2} \, ,
\end{equation}
which is true up to an order one function of $eQ$ in general, or an order on number in the small charge limit. Either way, we find that the quantum-corrected evaporation rate for neutral-like radiation is suppressed compared to the semi-classical at low energies, meaning that the black hole evaporates less than we would have found using the semi-classical Hawking formula. This is the same conclusion found for purely neutral radiation in \cite{Brown:2024}.

\subsection{Superradiant-dominated regime}\label{ss:sd}

\begin{figure}[H]
\centering
\includegraphics[width=1.0\textwidth]{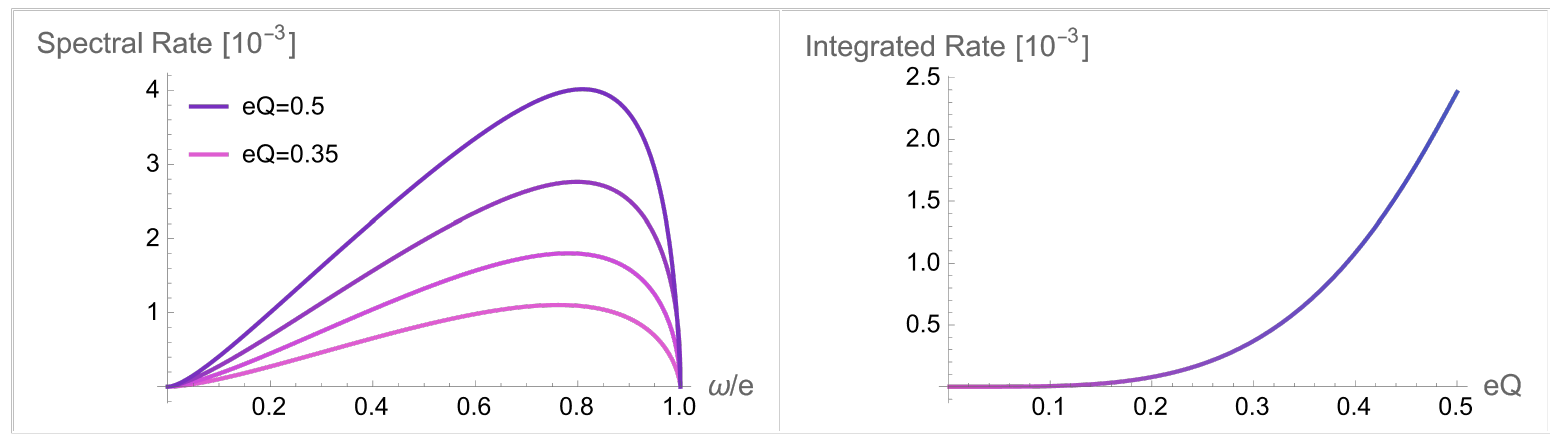}
\captionsetup{width=.93\linewidth}
\caption{(Left) Spectral evaporation rate in quantum superradiant-dominated regime at four different values of $eQ$ between $0.5$ and $0.35$, in equal increments. (Right) Integrated evaporation rate as a function of $eQ$.}
\label{fig:QSRDSCrate}
\end{figure}

We start with the semi-classical regime. In subsection \ref{ss:class}, we showed that the non-superradiant frequencies can be ignored, so the frequency range is between $0$ and $e$. In the semi-classical rate \eqref{eq:RNscrate}, the following approximation can be made. For the superradiant frequencies $\omega_\text{eff}<0$, so $F = - |F|$. Given that $e \gg T$ in the superradiant regime, we also have $|F| =|\omega_\text{eff}| / 4\pi T \sim e / T \gg 1$. As a result, the $F$-dependent Gamma functions in the semi-classical rate simplify as folllows
\begin{equation}\label{eq:Gammaratioapprox}
    \begin{split}
        & \left| \frac{\Gamma(\Delta+2i|F|)}
        {\Gamma(1+i(2|F|+\mu))} \right|^2
        \sim
        \left( \frac{\omega_\text{eff}}{2\pi T} \right)^{2(\Delta-1)} \,
        e^{\pi \mu} \, ,
    \end{split}
\end{equation}
where we used the asymptotic expansion for the complex argument Gamma function \eqref{eq:Gappimag}. Note that we have to be careful about the relative sign between $F$ and $\mu$, due to the absolute value in the asymptotic expansion formula. Plugging this into the evaporation rate, we find
\begin{equation}\label{eq:srscint}
    \begin{split}
        & \frac{dE}{dt} \bigg|_\text{sc} =
        \int_{0}^{e}
        \ 2^{4\Delta-3} \,
        \pi^{-1} \,
        Q^{4\Delta-2} \,
        \frac{|\Gamma(\Delta+i\alpha)|^4}
        {\Gamma(2\Delta)^2 \,
        \Gamma(2\Delta-1)^2} \,
        \omega^{2\Delta} \,
        \omega_\text{eff}^{2\Delta-1} \, .
    \end{split}
\end{equation}
Notice that the temperature dependence drops out, as we mentioned under \eqref{eq:srint2}. Furthermore, to simplify the exponential functions, we used the fact that $\mu = - \alpha$.\footnote{
    We are working with massless radiation $m=0$, which gives us $\alpha=-\mu$. When we take the limit of $\omega\to 0$, we get $\alpha\to -eQ$. If we took $m\neq0$, and we zoomed in onto the lower range of the frequencies $\omega\rightarrow m$, we would find that $\alpha\to-\infty$, which implies that there is an order of limits issue with $\alpha$. This doesn't cause any issues in the evaporation rate, since at $\omega\rightarrow m$ we have $|\Gamma(\Delta+i\alpha)|^2 \sim 2 \pi \alpha^{2\Delta-1} \, e^{-\pi |\alpha|} $, which when $m\rightarrow0$ is taken gives zero for the evaporation rate, which is exactly what we get if we take $m=0$ from the start, and then take $\omega\to 0$.}
Performing a change of variables $x=\omega/e$ under the integral, we find our final result for the semi-classical superradiant-dominated evaporation rate
\begin{equation}\label{eq:srscfin}
    \frac{dE}{dt}\bigg|_{\text{sc}} =
    G_l(eQ) \,
    \frac{1}{Q^2} \, ,
\end{equation}
where the coefficient function is given by the following integral
\begin{equation}\label{eq:Gl}
    \begin{split}
        & G_l(eQ) =
        \frac{1}{\pi} \int_{0}^{1} dx \,
        2^{4\Delta-3} \,
        (eQ)^{4\Delta} \,
        \frac{\left| \Gamma(\Delta+i\alpha) \right|^4}
        {\Gamma(2\Delta)^2 \, \Gamma(2\Delta-1)^2} \,
        x^{2\Delta} \, (1-x)^{2\Delta-1} \, ,
    \end{split}
\end{equation}
and the parameters $\Delta$, and $\alpha$ are now functions of $x$ and $eQ$
\begin{equation}\label{eq:Deltaalphax}
    \begin{split}
        \Delta(x,eQ) & =
        \tfrac{1}{2} + \sqrt{\tfrac{1}{4} + l(l+1) - (eQ)^2 + 6(eQ)^2\,x\,(1-x)} \, , \\
        \alpha(x,eQ) & = eQ \, (2x-1) \, .
    \end{split}
\end{equation}
Without further approximation, these integrals cannot be performed in terms of elementary functions. In the left panel of figure \ref{fig:QSRDSCrate}, we plot the spectral evaporation rate at different values of $eQ$ for the $l=0$ mode. In the right panel, we also plot the integrated rate as a function of $eQ$.
\noindent We can see that the rate gets larger with growing $eQ$. When $eQ=1/2$, the Gamma function $\Gamma(2\Delta-1)^2$ doesn't drives the total rate to zero, since the conformal dimension $\Delta=1/2$ only at the boundaries of the frequency interval, which is where the integrand is already zero due to the $\omega$ and $\omega_\text{eff}$ factors going to zero.

To further simplify these expressions, we can consider the small charge limit, $eQ\ll1$. Note that $e$ cannot be taken to be too small, as we would eventually leave the superradiant-dominated regime $T \ll e$. When $eQ\ll1$, from equations \eqref{eq:Deltaalphax} we can see that $\Delta$ and $\alpha$ reduce to the neutral values $\Delta_0=l+1$ and $\alpha=0$. As a result, the rate \eqref{eq:srscfin} simplifies to
\begin{equation}\label{eq:srsclim}
    \frac{dE}{dt}\bigg|_{\text{sc}} =
    D_l \,
    (eQ)^{4l+4} \,
    \frac{1}{Q^2} \, ,
\end{equation}
where the integral in the coefficient can be performed using the Beta function integral. This gives the following
\begin{equation}\label{eq:Dl}
    D_l =
    \frac{1}{\pi} \,
    \frac{2^{4l+1} \,
    l!^4 \, (2l+1)^2}
    {(2l+1)!^4} \,
    \frac{(2l+2)! \,
    (2l+1)!}
    {(4l+4)!} \, .
\end{equation}
Higher angular momentum modes are suppressed in the low charge limit. As an example, we use this equation to compute the contribution from the dominant $l=0$ mode. The following is found
\begin{equation}\label{eq:srscleq0}
    \frac{dE}{dt}\bigg|_{\text{sc}}^{(l=0)} =
    \frac{1}{6\pi} \,
    (eQ)^4 \,
    \frac{1}{Q^2} \, .
\end{equation}
We also compute the charge emission rate, by repeating the procedure from above and replacing one power of frequency with a factor of $e$. This gives
\begin{equation}\label{eq:srscleq0Q}
    \frac{dQ}{dt}\bigg|_{\text{sc}}^{(l=0)} =
    \frac{1}{3\pi} \,
    (eQ)^3 \,
    \frac{e}{Q} \, .
\end{equation}

In the quantum superradiant-dominated regime, we have $E_\text{} \ll e \ll E_\text{b}$. For the evaporation rate, the calculation proceeds as before, but now the frequency range goes from zero to $e$. We use the density of states and matrix element approximations given in equations \eqref{eq:sinhGamma} and \eqref{eq:Gammaquart} which results in
\begin{equation}\label{eq:srqint}
    \begin{split}
        & \frac{dE}{dt}\bigg|_{\text{q}} =
        \int_{0}^{e} d\omega \,
        2^{2\Delta-3/2} \,
        \pi^{-2} \,
        Q^{4\Delta-2} \,
        E_\text{b}^{2\Delta-3/2} \,
        \frac{\left| \Gamma(\Delta+i\alpha) \right|^4 \,
        \Gamma(\Delta)^4}
        {\Gamma(2\Delta)^2 \,
        \Gamma(2\Delta-1)^2} \,
        \omega^{2\Delta} \,
        (e-\omega)^{1/2} \, .
    \end{split}
\end{equation}
Note that there is no $E$ (and therefore $T$) to begin with. Performing a change of variables $x=\omega / e$ under the integral gives our final result for the quantum superradiant-dominated evaporation rate
\begin{equation}\label{eq:srqfin}
    \frac{dE}{dt}\bigg|_{\text{q}} =
    \hat{G}_l \!
    \left(  eQ,\frac{E_\text{b}}{e} \! \right) \,
    \frac{1}{Q^2} \, ,
\end{equation}
where the coefficient is a function of $eQ$, but also an extra ratio of energy scales $E_\text{b} / e$ like we had in the quantum neutral-like regime, only now this ratio cannot be pulled out of the integral, because of the frequency-dependence of $\Delta$. This coefficient is given by
\begin{equation}\label{eq:Ghl}
    \begin{split}
        \hat{G}_l \! \left(
        eQ,\frac{E_\text{b}}{e} \! \right) =
        \frac{1}{\pi^2} \!
        \int_{0}^{1} \! dx \,
        2^{2\Delta-3/2} \,
        (eQ)^{4\Delta}
        \left(
        \frac{E_\text{b}}{e}
        \right)^{2\Delta-3/2}
        \frac{\left| \Gamma(\Delta+i\alpha) \right|^4 \,
        \Gamma(\Delta)^4}
        {\Gamma(2\Delta)^2 \,
        \Gamma(2\Delta-1)^2} \,
        x^{2\Delta} \,
        (1-x)^{1/2} \, .
    \end{split}
\end{equation}
Notice that when $eQ\ll1$, the integrand of this equation is the same as the one in the quantum neutral-like evaporation rate \eqref{eq:Fhl1}, up to numerical factors and energy scales, which implies that both of their plots would be the same as the plot for neutral radiation given in \cite{Brown:2024}.

The size of the evaporation rate is determined by the two parameters under the integral, $eQ$ and the ratio $E_\text{b} / e$, the first of which is a small quantity in the quantum regime, for most of the bottom wedge in figure \ref{fig:logQloge}, but the second is large. The product of these two simplifies as follows
\begin{equation}\label{eq:scalealg}
    (eQ)^{4\Delta} \,
    \left( \frac{E_\text{b}}{e} \right)^{2\Delta} =
    \left( \frac{e}{Q} \right)^{2\Delta} \, ,
\end{equation}
which implies that as long as the black hole charge is reasonably greater than the particle it is radiating, which is true for most of parameter space, then the evaporation rate is a small quantity, and the use of perturbation theory is valid in this regime as well. This addresses another concern. Based on equation \eqref{eq:Ghl}, we could worry that higher angular momentum modes would contribute more, given that the conformal dimension $\Delta$ grows with $l$. Since the size of the evaporation rate is determined by $e/Q$, which is small, we find that higher $l$ gives smaller contributions, as we would expect.

We proceed to take the small charge limit of our evaporation rate, which results in
\begin{equation}\label{eq:srqlim}
    \frac{dE}{dt}\bigg|_{\text{q}} =
    \hat{D}_l \,
    (eQ)^{4l+4} \,
    \left( \frac{E_\text{b}}{e} \right)^{2l+1/2} \,
    \frac{1}{Q^2} \, ,
\end{equation}
where the integral in the coefficient can be done using another Beta function integral, which results in
\begin{equation}\label{eq:Dhl}
    \hat{D}_l =
    \frac{1}{\pi^2} \,
    2^{6l-7+1/2} \,
    \frac{l!^8 \, (2l+1)^2}{(2l+1)!^4} \,
    \frac{(2l+2)! \, (2l+3)!}{(4l+7)!} \, .
\end{equation}
As an example, we use this equation to calculation the evaporation rate of the dominant $l=0$ mode
\begin{equation}\label{eq:srqleq0}
    \frac{dE}{dt}\bigg|_{\text{q}}^{(l=0)} =
    \frac{16 \, \sqrt{2}}{105 \, \pi^2} \,
    (eQ)^{4} \,
    \left( \frac{E_\text{b}}{e} \right)^{1/2} \,
    \frac{1}{Q^2} \, .
\end{equation}
We also compute the charge emission rate, which we find to be
\begin{equation}\label{eq:srqleq0Q}
    \frac{dQ}{dt}\bigg|_{\text{q}}^{(l=0)} =
    \frac{4 \sqrt{2}}{15 \pi^2} \,
    (eQ)^{3} \,
    \left( \frac{E_\text{b}}{e} \right)^{1/2} \,
    \frac{e}{Q} \, .
\end{equation}

Comparing the small charge limit of the semi-classical \eqref{eq:srsclim} and quantum rate evaporation rates \eqref{eq:srqlim}, we find
\begin{equation}\label{eq:srratio}
    \frac{\dfrac{dE}{dt}\bigg|_\text{q}}
    {\ \dfrac{dE}{dt}\bigg|_\text{sc}} \sim
    \left( \frac{E_\text{b}}{e} \right)^{2l+1/2} \, .
\end{equation}
The same would roughly be true for finite $eQ$ and the integral in equations \eqref{eq:srscfin} and \eqref{eq:srqfin}. This shows that the quantum rate is enhanced compared to the semi-classical by a power of $\frac{E_\text{b}}{e}$. This agrees with what was found in \cite{Maulik:2025,Betzios:2025} by analyzing plots of evaporation spectra. This also shows that higher modes are more strongly enhanced, which does not mean that higher modes radiate at a large rate, just that when the two rates are compared, a higher enhancement is found at higher $l$.

\section{Radiation from large charged black holes}\label{sec:lBH}

Large black holes were defined in the previous section as those having\footnote{
    Our definition of small versus large black holes is related to the definition used in \cite{Gibbons:1975}, but not the same. There, large black holes were defined as those with $M \gg m$, which made the effective potential large, allowing for a Schwinger-like calculation. Here, the Schwinger result will appear for $eQ\gg1$, which is only a limiting case of all large black holes $eQ\geq \tfrac{1}{2}$.}
\begin{equation}\label{eq:lBHdef}
    eQ \geq \tfrac{1}{2} + l \, ,
\end{equation}
which is the condition for the conformal dimension $\Delta(\omega)$ to have complex values on some subset of the available frequencies $[0,e + E]$. Focusing on the superradiant regime where $E \ll e$, and inspecting the quadratic under the square root in the conformal dimension \eqref{eq:Deltaomega}, we find the following behavior, as we vary $eQ$ from small to large values. At very small $eQ$, it starts out flat and positive. As $eQ$ approaches $1/2$, it starts bending downward. At $eQ=1/2$, the roots are exactly at $0$ and $e$. Beyond $eQ=1/2$, the roots, given by (for $l=0$)
\begin{equation}\label{eq:omegapm}
    \omega_\pm =
    \tfrac{1}{2} \,
    \left( 1 \pm \sqrt{\tfrac{1}{3}+\tfrac{1}{6e^2Q^2}} \right) \, e \, ,
\end{equation}
start moving inward and approaching the limiting values of $\tfrac{1}{2}(1\pm1/\sqrt{3}) e$. Therefore, for large black holes, $\Delta(\omega)$ is real for frequencies inside of an interval bounded by $[\omega_-,\omega_+]$, and complex outside of this interval $[0,\omega_-]$ and $[\omega_+,\omega_\text{sr}]$. In the neutral-like regime, $e\ll E$, and $\Delta$ is a frequency-independent number that is complex beyond $eQ\geq 1/2$.

In subsection \ref{ss:class}, using figure \ref{fig:logQloge}, we showed that we always have $E_\text{b} \gg e$ for large black holes, which implies that they are never in the quantum regime. Therefore, even if we calculate the quantum-corrected evaporation rate, we would expect it to reduce to the semi-classical formula, so we will only consider this rate in this section. Furthermore, we will only cover the superradiant regime, which is the more intricate for large black holes. The neutral-like regime would follow analogously, and would provide no additional insight.

The greybody factor that we computed in subsection \ref{ss:RNsc} assumed that $\beta$ and $\Delta$ are real. This result can still be used in the middle frequency interval for large black holes, but on the left and right intervals, where $\Delta$ is complex, and $\beta$ is imaginary, we will have to recompute the greybody factor, which we will find to be very different.

We will find significant simplification in our results, if we take the limit of very large black holes, those with $eQ \gg 1$. The greybody factor in this limit was derived in \cite{Gibbons:1975} where a Schwinger-like exponential suppression was found. This result was obtained using WKB theory, which does not require solving the wave equation. We will start by repeating this calculation, and then we will show that our greybody factor, which is valid on the full range of $eQ$, reduces to the Gibbons result in the large $eQ$ limit.

\subsection{Greybody factor from WKB theory}\label{ss:WKB}

We will repeat the calculation done in \cite{Gibbons:1975}, which in our terminology is valid for very large black hole, $eQ \gg 1$. For the rest of this section, we will consider a nearly massless scalar, as a non-zero $m$, no matter how small, will be required to give us the right results in the $eQ \gg 1$ limit.

In 1D quantum mechanics, if the potential in a "tunneling through the barrier" scattering problem is large, the transmission coefficient can be found using the quasi-classical approximation or WKB theory as follows
\begin{equation}\label{eq:greyWKB}
    \mathcal{T} = \exp\left( - 2 \int_{x_1}^{x_2} dx \sqrt{V(x)} \right) \, ,
\end{equation}
where the integral goes over the classically-forbidden region, which is defined by $x_{1,2}$, the roots of $V(x)$. As we can see, this does not require solving the Schrodigner equation.

In the 1D scattering problem set up in subsection \ref{ss:RNeq}, when $eQ\gg$ the effective potential \eqref{eq:RNVeff} is large and given by
\begin{equation}\label{eq:Veffapprox}
    V_\text{eff}(r) =
    \left( \omega - \frac{eQ}{r} \right)^2 -
    m^2 \, \frac{(r-Q)^2}{r^2} \, ,
    \qquad
    r_{\pm} =
    \frac{e \mp m}
    {\omega\mp m} \, ,
\end{equation}
where we have ignored the term involving $f'(r)$ as it is subleading in $eQ$. We have also included the roots $r_\pm$ of the potential, which we can see would coincide, if we had not taken $m \neq 0$, so there would be no classically-forbidden region, and the WKB approximation would not be possible. Plugging into \eqref{eq:greyWKB} and integrating gives
\begin{equation}\label{eq:WKBint}
    \begin{split}
        \Gamma =
        \exp \left( -2
        \int_{r_{*,-}}^{r_{*,+}} dr_*
        \sqrt{V_\text{eff}(r_*)} \right)
        =
        \exp \left(
        -2\pi Q
        \left( 2\omega-1 -
        \tfrac{2\omega^2-\omega-m^2}
        {\sqrt{\omega^2 -m^2}} \right)
        \right) \, ,
    \end{split}
\end{equation}
where we have taken $\delta r =0$ in $f(r)$. Also note that the 1D scattering problem is in terms of the $r_*$ coordinate, but to solve the integral, it is much more convenient to change to $r$. Following \cite{Gibbons:1975}, the final step is to evaluate this result at $\omega=e$. This can be justified on the basis of this formula being exponentially peaked at $\omega=e$, but we will also see, in the next subsection, that this is true for the full greybody factor. Therefore, we find the Gibbons greybody factor
\begin{equation}\label{eq:greyGibb}
    \Gamma = e^{-2\pi Q(e-\sqrt{e^2-m^2})} \, .
\end{equation}
Before proceeding, we note that if we had taken $\omega=e$ before the integral is done, the roots of the potential $r_{\pm}$ would be equal, and WKB theory could not be applied. We can take a further limit of \eqref{eq:greyGibb} for $m\ll e$, which results in the classical Schwinger result
\begin{equation}\label{eq:greySchw}
    \Gamma = e^{-\frac{\pi Q m^2}{e}} \, .
\end{equation}
Both of these equations, \eqref{eq:greyGibb} and \eqref{eq:greySchw}, tell us that there is a large suppression of charged radiation for very large black holes. Without assuming the size of $m$, equation \eqref{eq:greyGibb} tells us that the suppression rate is exponentially suppressed at $eQ \gg 1$. When $m\ll e$, \eqref{eq:greySchw} tells us that there is suppression when $eQ \, \frac{m^2}{e^2} \gg 1$, which defines even larger black holes given by
\begin{equation}\label{eq:vlBHdef}
    eQ \geq eQ_* = \frac{e^2}{\pi m^2} \, .
\end{equation}

If one could solve the wave equation and compute the exact transmission coefficient in terms of a parameter $\lambda$ which controls the size of the potential, then at large $\lambda$ the transmission coefficient can be quasi-classically expanded into terms of higher and higher order of $\exp(-\lambda)$. On general grounds, it can be argued that the coefficient of the first term is given by unity, which recovers the WKB result \eqref{eq:greyWKB}. See, for example, chapter 52 of \cite{Landau:1991}. One of the goals of the next subsection will be to confirm that this is true, by solving the wave equation, deriving the greybody factor, and showing that at $eQ\gg1$ it reduces to exponential in \eqref{eq:greyGibb}, with the coefficient exactly equal to one.

\subsection{Wave equation with complex Delta}\label{ss:RNCDelta}

The solution to the wave equation, given by equations \eqref{eq:RNnearsolfin}, \eqref{eq:RNfarsol} and the matching conditions \eqref{eq:RNC12} is valid for both real and imaginary $\beta$, which can be found in \eqref{eq:RNnearparam3}. To make the imaginary value of $\beta$ manifest, we will make the following replacement $\beta \rightarrow i \beta$, which amounts to pulling out a minus sign from the square root, which means that all the terms in the square root of the original $\beta$ will have their sign flipped. A first attempt at recomputing the transmission coefficient and the greybody factor would be to just make this same replacement in equation \eqref{eq:RNTfin}, but this would not work for the following reason. In subsection \ref{ss:RNsc}, we ignored $C_-$ compared to $C_+$, because they were proportional to different powers, determined by $\beta$, of $\delta r$, a small quantity, as can be seen in the matching conditions \eqref{eq:RNC12}. Since $\beta$ is now imaginary, the coefficients $C_+$ and $C_-$ take the following form
\begin{equation}\label{eq:RNCipm}
    \begin{split}
        C_{\pm} =
        \frac{\Gamma(\pm2i\beta) \,
        \Gamma(1-2iF+i\mu)}
        {\Gamma(\tfrac{1}{2}\pm i\beta+i\mu) \,
        \Gamma(\tfrac{1}{2}\pm i\beta-2iF)} \,
        \delta r^{+1/2\mp i\beta} \,
        A \, ,
    \end{split}
\end{equation}
where we can see that after squaring, this will give the same power of $\delta r$, and we cannot assume that one is small compared to the other in general. Therefore, we will have to use both in computing the greybody factor. The coefficient of the incident wave will now have a contribution from both given by
\begin{equation}\label{eq:RNCinci}
    C_{\text{inc, }\pm} =
    \frac{\Gamma(1\pm2i\beta)}
    {\Gamma(\tfrac{1}{2}\pm i\beta-i\alpha)} \,
    (-2ik)^{-1/2\mp i\beta-i\alpha} \,
    C_\pm \, ,
\end{equation}
Squaring each term gives
\begin{equation}\label{eq:RNCincisq}
    \begin{split}
        \left|C_{\text{inc, }\pm} \right|^2 =
        \frac{\delta r}{2k} \,
        \frac{|\Gamma(1+2i\beta)|^2 \,
        |\Gamma(2i\beta)|^2 \,
        |\Gamma(1-2iF+i\mu)|^2}
        {|\Gamma(\tfrac{1}{2}\pm i\beta-i\alpha)|^2 \,
        |\Gamma(\tfrac{1}{2}\pm i\beta+i\mu)|^2 \,
        |\Gamma(\tfrac{1}{2}\pm i\beta-2iF)|^2} \,
        e^{\pi(\mp\beta-\alpha)} \, 
        |A|^2 \, .
    \end{split}
\end{equation}
Comparing to the squared incident coefficient \eqref{eq:RNCinc2} from before, we see that in addition to replacing $\beta \rightarrow i \beta$, the powers of $\delta r$ and $k$ are different, and there is an extra $\beta$ term in the exponential function (coming from $i^{i\beta}$).\footnote{
    Note that we need to be careful when moving $\beta$ from the real part of the argument inside the Gamma functions to the imaginary part. In the transmission coefficient we have $|\Gamma(x+\beta+iy)|^2$ and we can freely change the sign of $y$, while in $C_\text{inc}$ we just have $\Gamma(x+\beta+iy)$, so, when moving $\beta$ to $i\beta$, we have to use the exact sign of $\beta$ that was found from solving the wave equation, which would give $\Gamma(x+i\beta+iy)$. Once this is squared, then we can change the overall sign of $i(\beta+y)$. In other words, we have to be careful with the order of these operations, otherwise we would end up with the wrong relative sign between $\beta$ and $y$.}

Of course, the squaring would also produce a cross term in the full $\left|C_{\text{inc}} \right|^2$ of the following form
\begin{equation}
    |C_{\text{inc, }+}| \,
    |C_{\text{inc, }+}| \,
    \cos(\psi) \, ,
\end{equation}
where $\psi$ is the relative phase between the two terms. This term has the following odd property, the relative phase has a term proportional to $\ln \delta r$ coming from the $\delta r^{\mp i \beta}$, which, as we take the black hole closer to extremality, causes the cross-term to oscillate indefinitely. This odd behavior was first observed in \cite{Starobinskii:1973a,Starobinskii:1973b} in the context of the Kerr black hole. In \cite{Teukolsky:1973,Press:1973,Teukolsky:1974}, also in the context of Kerr, an argument was given for why this cross term can be ignored, which eliminates this odd behavior. We will use that argument here, and also in the next section for Kerr, to argue-away the crossterm. As we shall see below, when $eQ\gg1$, which is the limit we are interested in, each of the two $\left|C_{\text{inc, }\pm} \right|$ will be exponentially small/large in $eQ$, with a different exponent, which means that one will always dominate. Depending on where we are in the frequency range, it can be a different term, so we cannot ignore both of them. But, we can definitely ignore the cross term, because regardless of which term dominates, the cross term will always be subdominant to that. In fact, even for finite $eQ\geq 1/2$, this is a very good approximation. Each Gamma function in \eqref{eq:RNCincisq} is either a $\sinh$ or $\cosh$, which is a sum of exponential functions with arguments of the form $\pi eQ f(\omega/e)$, where $f$ is of order one. For most $\omega$, this is a small/big number to a good degree, which again implies that each of $\left|C_{\text{inc, }\pm} \right|$ can be approximated by exponential functions with different exponents, so only one will always dominate, and the cross term can be ignored. Therefore, for all $eQ\geq 1/2$, to a high accuracy we have the following result
\begin{equation}
    \left|C_{\text{inc}} \right|^2 =
    \left|C_{\text{inc, }+} \right|^2 +
    \left|C_{\text{inc, }-} \right|^2 \, .
\end{equation}

Since we will only consider the superradiant regime, which enables further simplification of \eqref{eq:RNCincisq}, we will first do this, before writing down the final equation for the greybody factor. Using the fact that $e \gg T$, which implies that $|\omega_{\text{eff}}|/T \gg 1 $, the $F$-dependent Gamma functions in \eqref{eq:RNCincisq} can be approximated as follows
\begin{equation}\label{eq:Gammaratio2}
    \frac{|\Gamma(1-2iF+i\mu)|^2}
    {|\Gamma(\tfrac{1}{2} \pm i\beta-2iF)|^2} =
    \frac{|\Gamma(1+2i|F|+i\mu)|^2}
    {|\Gamma(\tfrac{1}{2} \pm i\beta+2i|F|)|^2} =
    2 |F| \,
    e^{\pi(-\mu\pm\beta)} \, .
\end{equation}
Note that this is the inverse of equation \eqref{eq:Gammaratioapprox} with $\Delta = 1/2$. Plugging this into the incident wave amplitude makes it simplify to the following
\begin{equation}\label{eq:RNCincisq2}
    \begin{split}
        \left|C_{\text{inc, }\pm} \right|^2 =
        \frac{\delta r \, |F|}{k} \,
        \frac{\cosh \pi (\alpha \mp \beta) \,
        \cosh \pi (\mu \pm \beta)}
        {\sinh^2 2 \pi \beta} \,
        e^{-\pi(\mu+\alpha)} \,
        |A|^2 \, .
    \end{split}
\end{equation}

The coefficient of the transmitted wave $C_\text{trans}$ remains unchanged when $\beta$ is imaginary, and is given by equation \eqref{eq:RNCtrans}. To find the greybody factor, we combine the incident and transmitted waves using \eqref{eq:T&R}, which gives\footnote{
    The transmission coefficient for charged radiation when $\Delta$ was real went to zero at $\omega=e$, but here we're finding that it goes to a constant. This happened because of the superradiant regime approximation we made - the $\omega_\text{eff}$ from the equation for $\mathcal{T}$ \eqref{eq:T&R} canceled with the $F$ in the approximation \eqref{eq:Gammaratio2}. If we approached $\omega=e$ from the right, where $\omega_\text{eff}>0$, we would find that the transmission coefficient approaches the same constant, but positive. This implies, that there is a discontinuous jump across $\omega=e$. If we had not made the approximation, the transmission coefficient would continuously transition from negative to positive, going through zero at $\omega=e$.}
\begin{equation}\label{eq:RNgreyi}
    \Gamma = -
    \frac{\sinh^2 2\pi\beta}
    {\cosh \pi (\alpha - \beta) \,
    \cosh \pi (\mu + \beta) +
    \cosh \pi (\alpha + \beta) \,
    \cosh \pi (\mu - \beta)} \,
    e^{\pi(\mu+\alpha)} \, .
\end{equation}
The minus sign comes from the $|F|$ in \eqref{eq:RNCincisq2} canceling with the $\omega_\text{eff}<0$ in \eqref{eq:T&R}. Plugging into the Hawking formula \eqref{eq:scEdot}, which amounts to multiplying by $-\omega/2\pi$ gives the final formula for the semi-classical spectral evaporation rate for complex $\Delta$ in the superradiant regime
\begin{equation}\label{eq:RNscratei}
    \frac{d^2 E}{d\omega \, dt} \bigg|^{}_\text{sc} = \frac{1}{2\pi}
    \frac{\omega \,
    \sinh^2 2\pi\beta}
    {\cosh \pi (\alpha - \beta) \,
    \cosh \pi (\mu + \beta) +
    \cosh \pi (\alpha + \beta) \,
    \cosh \pi (\mu - \beta)} \,
    e^{\pi(\mu+\alpha)} \, .
\end{equation}
The parameters $\mu,\beta,\alpha$ can be found in equations \eqref{eq:RNnearparam3}, \eqref{eq:RNnearparam4} and \eqref{eq:RNfarparam2} (with $\beta$ having all signs flipped under the square root). The extra minus sign comes from the statistical factor in the Hawking formula in the superradiant regime. These expression are only true for frequencies $\omega<e_-$ and $\omega>e_+$, while \eqref{eq:RNscrate} applies in between.

We will now use these formulae to analyze the spectrum at finite $eQ$ and to take the $eQ\gg1$ limit. Starting with the latter, we will take the $eQ\gg1$ limit directly for the greybody factor, which will result in a different limit in each of the three frequency intervals $[0,\omega_-]$, $[\omega_-,\omega_+]$, and $[\omega_+,e]$. Each $\sinh$ and $\cosh$ function will be replaced by an exponential function, and to compare these, we will need to know how the parameters $\alpha$, $\beta$, and $\mu$ compare along this range. We graph these in the figure below. The vertical asymptote in $\alpha$ come from a non-zero $m$. Also note that $\beta$ is always smaller than $\alpha$ and $\mu$, in terms of absolute values.
\begin{figure}[H]
\centering
\includegraphics[width=0.6\textwidth]{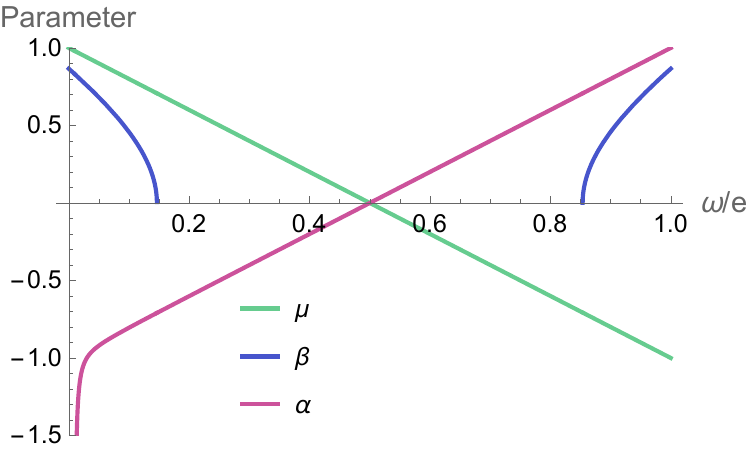}
\captionsetup{width=0.8\linewidth}
\caption{Parameters $\mu$, $\beta$, and $\alpha$ as a function of $\omega$ in units of $e$.}
\label{fig:mubeal}
\end{figure}

In the interval $\omega>\omega_+$, which is where the Gibbons result lies, we have $\alpha \geq 0$ and $-\mu = |\mu| \geq 0$, and also $\alpha \geq \beta$ and $ |\mu| \geq \beta$. Going back to the incident wave, as this way it will be easier to see which term is dominant in which interval, we replace $\mu = -|\mu|$ in \eqref{eq:RNCincisq2} which gives
\begin{equation}\label{eq:RNCincipl}
    \begin{split}
        \left|C_{\text{inc, }\pm} \right|^2 =
        \frac{\delta r \, |F|}{k} \,
        \frac{\cosh \pi (\alpha\mp\beta) \,
        \cosh \pi (|\mu|\mp\beta)}
        {\sinh^2 2 \pi \beta} \,
        e^{\pi(|\mu|-\alpha)} \,
        |A|^2 \, .
    \end{split}
\end{equation}
We can now take the $eQ \gg 1$ limit of this equation. Using \eqref{eq:hypapp}, we can replace each hyperbolic function with its asymptotic exponential function. In the following equation, we summarize what we get for each hyperbolic functions
\begin{equation}\label{eq:expalg}
\begin{split}
    e^{\pi(\alpha \mp \beta)} \,
    e^{\pi(|\mu| \mp \beta)} \,
    e^{\pi(|\mu| - \alpha)} \,
    e^{-4\pi\beta} \, .
\end{split}
\end{equation}
Due to the absolute value in \eqref{eq:hypapp}, the signs of the exponents crucially depend on the relative size of the parameters $\beta,\mu,\alpha$. The terms in the incident wave simplify to the following
\begin{equation}\label{eq:RNCincifin}
    \begin{split}
        \left|C_{\text{inc, }+} \right|^2 =
        \frac{\delta r \, |F|}{k} \,
        e^{\pi (2 |\mu| - 6 \beta )} \,
        |A|^2 \, , \\
        \left| C_{\text{inc, }-} \right|^2 =
        \frac{\delta r \, |F|}{k} \,
        e^{\pi(2|\mu|-2\beta)} \,
        |A|^2 \, .
    \end{split}
\end{equation}
We can clearly see that the second term is suppressed compared to the first. The extra suppression is given by $e^{-4\pi\beta}$, which is very small when $eQ\gg1$. Therefore, the greybody factor is given by
\begin{equation}\label{eq:greymubeta}
    \Gamma = - e^{-2\pi(|\mu|-\beta)} \, .
\end{equation}
This is strongly peaked at $\omega=e$. When evaluated at this value, it exactly reproduces the Gibbons result found earlier
\begin{equation}\label{eq:greyGibbwave}
    \Gamma = - e^{-2\pi Q(e-\sqrt{e^2-m^2})} \, .
\end{equation}
We can see that the coefficient in front of the exponential is (negative) one, as required by WKB theory. Even though the full Gibbons result has been recovered, this was obtained from only a part of the frequency range. We still need to check that the rest gives corrections which can be ignored compared to this. We also point out that this type of result has been obtained in \cite{Chen:2012,Chen:2016} by working directly in the Reissner-Nordström near-horizon region, which doesn't account for the full frequency range, but only the superradiant frequency.

In $\omega \leq \omega_-$, we can perform the same type of calculation, but now we have $-\alpha = |\alpha| \geq 0$ and $\mu \geq 0$. Replacing $\alpha=-|\alpha|$ in \eqref{eq:RNCincisq2}, we find
\begin{equation}\label{eq:RNCincimin}
    \begin{split}
        \left|C_{\text{inc, }\pm} \right|^2 =
        \frac{\delta r \, |F|}{k} \,
        \frac{\cosh \pi (|\alpha|\pm\beta) \,
        \cosh \pi (\mu\pm\beta)}
        {\sinh^2 2 \pi \beta}
        e^{\pi(-\mu+|\alpha|)} \,
        |A|^2 \, .
    \end{split}
\end{equation}
At $eQ\gg 1$ these two reduce to
\begin{equation}\label{eq:RNCincifinmin}
    \begin{split}
        \left|C_{\text{inc, }+} \right|^2 =
        \frac{\delta r \, |F|}{k} \,
        e^{\pi(2|\alpha| - 2 \beta)} \,
        |A|^2 \, , \\
        \left| C_{\text{inc, }-} \right|^2 =
        \frac{\delta r \, |F|}{k} \,
        e^{\pi(2|\alpha| - 6\beta)} \,
        |A|^2 \, .
    \end{split}
\end{equation}
The first coefficient is less suppressed, so we ignore the second. Therefore,
\begin{equation}\label{eq:greyalphabeta}
    \Gamma = - e^{-2\pi(|\alpha|-\beta)} \, ,
\end{equation}
which looks like the Gibbons result again, but there is an important difference. When $\omega \rightarrow m$, we have $|\alpha|  \rightarrow \infty$, which means that the greybody factor quickly goes to zero, instead of a constant as in the Gibbons result. Therefore, the dominant contribution from this interval is not at $\omega=0$, but at some small, but non-zero $\omega$, and this is smaller than the dominant contribution from the right interval, which was at $\omega=e$. What allows for this behavior is the asymmetry between $\alpha$ and $\mu$ (see figure \ref{fig:mubeal}), which is only possible when $m\neq 0$. If $m=0$, then $\alpha = -\mu$, and we could not obtain the Gibbons result, and the left and right frequency intervals would give the same contribution.

Finally, in $e_- < \omega < e_+$, we have to use the transmission coefficient for real $\beta$. Instead of using the full transmission coefficient, we want its superradiant limit, which can be found from the semi-classical evaporation rate \eqref{eq:srscfin}, by stripping off the $-\omega/2\pi$ factor from the Hawking formula. This gives the following
\begin{equation}\label{eq:greymid}
    \begin{split}
        \Gamma = 
        -
        2^{4\Delta-2} \,
        (eQ)^{4\Delta-2} \,
        \frac{\left| \Gamma(\Delta+i\alpha) \right|^4}
        {\Gamma(2\Delta)^2 \, \Gamma(2\Delta-1)^2} \,
        x^{2\Delta-1} \, (1-x)^{2\Delta-1} \, ,
    \end{split}
\end{equation}
where $x=\omega/e$. The parameters $\Delta$ and $\alpha$ are found by taking the large $eQ$ limit of equation \eqref{eq:Deltaalphax}, which gives
\begin{equation}\label{eq:DeltaxeQinf}
    \begin{split}
        \Delta(x) & =
        eQ \, \sqrt{- 1 + 6\,x\,(1-x)} \, , \\
        \alpha(x) & = eQ \, (2x-1) \, .
    \end{split}
\end{equation}
Also note that we can use $\mu=-\alpha$ in this interval, since their difference only appears near $\omega=0$ (see figure \ref{fig:mubeal}). We can take the large $eQ$ limit of the Gamma functions in the \eqref{eq:greymid} using Stirling's approximation \eqref{eq:GappStir} for the ones in the denominator, and \eqref{eq:Gappimag} for the one in the numerator. This gives
\begin{equation}\label{eq:greymid2}
    \Gamma = -
    \left( 
    \frac{e^2 \, (2x-1)}
    {12\sqrt{x(1-x)}}
    \right)^{4\Delta-2} \,
    e^{-2\pi|\alpha|}
    \, .
\end{equation}
The first factor is a function of $x\in[x_-,x_+]$ raised to a very large power. One can check that this function never crosses $1$ on this interval, which implies that the greybody factor is at least exponentially suppressed in $e^{-2\pi|\alpha|}$, which is already a much higher degree of suppression than the contributions from the left \eqref{eq:greymubeta} and right \eqref{eq:greyalphabeta} intervals. This completes our derivation of the Gibbons result.

We now turn to the finite values of $eQ\geq1/2$. Using spectral rate equation \eqref{eq:srscfin} for the inner interval, and equation \eqref{eq:RNscratei} for the outer intervals, after changing variables to $x=\omega/e$, we find our final result for the semi-classical superradiant-dominated evaporation rate for large black holes
\begin{equation}\label{eq:lsrscfin}
    \frac{dE}{dt}\Big|_{\text{sc}} =
    \left( G^{}_l(eQ) + G'_l(eQ) \right) \,
    \frac{1}{Q^2} \, ,
\end{equation}
where the coefficient functions are given by the following  two integrals. For the first we have
\begin{equation}\label{eq:lGl}
\begin{split}
        & G^{}_l(eQ) =
        \frac{1}{\pi} \int_{x_-}^{x_+} dx \,
        2^{4\Delta-3} \,
        (eQ)^{4\Delta} \,
        \frac{\left| \Gamma(\Delta+i\alpha) \right|^4}
        {\Gamma(2\Delta)^2 \, \Gamma(2\Delta-1)^2} \,
        x^{2\Delta} \, (1-x)^{2\Delta-1} \, ,
\end{split}
\end{equation}
where the bounds of integrations now go between $x_\pm = \omega_\pm/e$. And the second term is given by
\begin{equation}\label{eq:lGlpr}
\begin{split}
    G'_l(eQ) \!=\!
    \frac{(eQ)^2}{2\pi} \!
    \int_{ x \in I_\pm}^{} \! dx
    \frac{x
    \sinh^2 2 \pi \beta}
    {\cosh \pi (\alpha \!-\! \beta)
    \cosh \pi (\mu \!+\! \beta) \!+\!
    \cosh \pi (\alpha \!+\! \beta)
    \cosh \pi (\mu \!-\! \beta)}
    e^{-\pi(\mu+\alpha)} \, , 
\end{split}
\end{equation}
where the integration interval is given by $I_\pm = [0,\omega_-] \cup [\omega_+,e]$. The parameters $\Delta, \alpha, \beta$ are given by
\begin{equation}\label{eq:Deltaalphax2}
    \begin{split}
        \Delta(x,eQ) & =
        \tfrac{1}{2} + \sqrt{\tfrac{1}{4} + l(l+1) - (eQ)^2 + 6(eQ)^2\,x\,(1-x)} \, , \\
        \alpha(x,eQ) & = eQ \, (2x-1) \, , \\
        \beta(x,eQ) & = \sqrt{-\tfrac{1}{4} - l(l+1) + (eQ)^2 - 6(eQ)^2\,x\,(1-x)} \, .
    \end{split}
\end{equation}
Note the sign flip in the square root in $\Delta$ and $\beta$!

\begin{figure}[H]
    \centering
    \begin{subfigure}[t]{0.45\textwidth}
        \centering
        \includegraphics[width=1\linewidth]{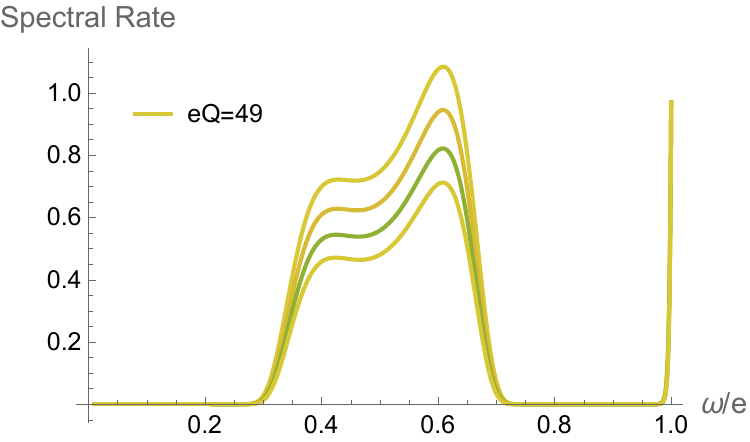} 
    \end{subfigure}
    \captionsetup{width=0.93\linewidth}
    \caption{Spectral evaporation rate of a large black hole in the semi-classical superradiant-dominated regime at different values of $eQ$. Top left panel, $eQ$ is goes from $0.6$ to $1.8$. Top right, $eQ$ is between $3$ and $9$. Bottom, $eQ$ is between $49$ and $55$. The mass of the particle is $m = 0.01 e$.}
    \label{fig:RNlrate1}
\end{figure}

We plot these rates in figure \ref{fig:RNlrate1} for a broad range of $eQ\geq1/2$. Starting with the top left panel at $eQ$ close to $1/2$, we can see that the $eQ=0.6$ spectrum, with maximum value of $\sim0.01$ nicely matches onto to $eQ = 0.5$ found in figure \ref{fig:QSRDSCrate}, which has a maximum value of $\sim 0.005$. We can also clearly see that the interval is broken up into three regions, and that the rate is zero at $x_\pm$ as this is where $\Delta = 1/2$. As we start increasing $eQ$, the spectral rate starts rising on all three intervals, and the $x_\pm$ start moving inward and eventually settle at the limiting value, which agrees with \eqref{eq:omegapm}. In the top right panel, we see that the rate on the middle interval keeps growing, while the left and right intervals start approaching zero, attaining their maxima at $\omega\approx0$ for the left interval \eqref{eq:greyalphabeta}, and at $\omega=e$ for the right interval \eqref{eq:greyGibbwave}, which is the Gibbons result.
\begin{figure}[H]
    \centering
    \begin{subfigure}[t]{0.45\textwidth}
        \centering
        \includegraphics[width=1\linewidth]{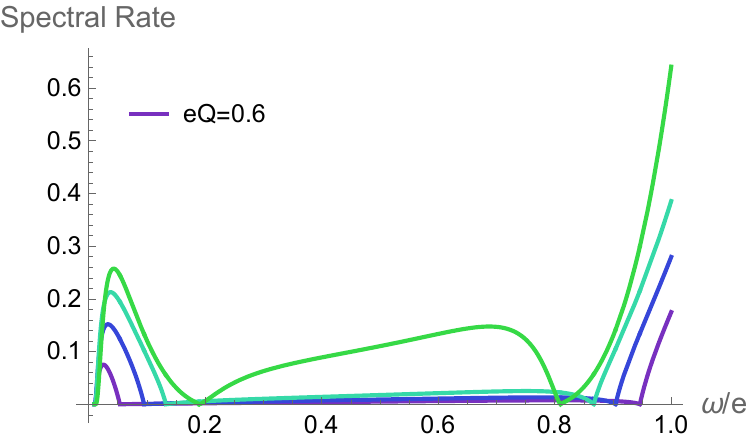} 
    \end{subfigure}
    \hfill
    \begin{subfigure}[t]{0.45\textwidth}
        \centering
        \includegraphics[width=1\linewidth]{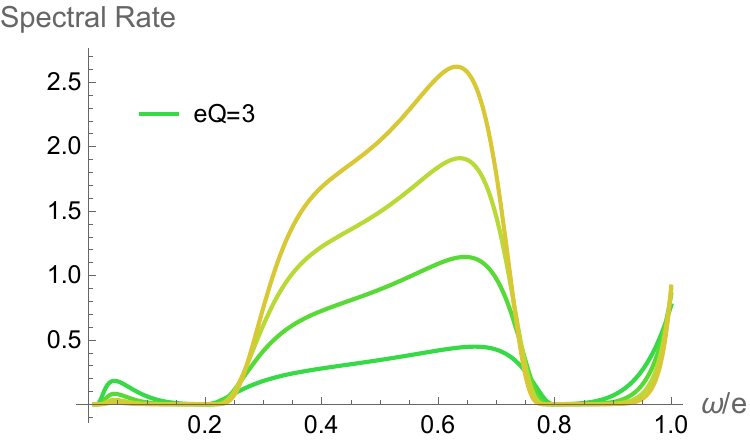} 
    \end{subfigure}
\end{figure}
\vspace{-2mm}
\noindent On the bottom panel, we can see that eventually the rate on the middle interval also starts going towards zero. Notice that the rate on the right end appears to have reached a limiting value, which is slightly below one. This number is given by $\exp(-\pi eQ \frac{m^2}{e^2})$. It is one on our plots, as the black hole $eQ$ is still not large enough to overcome the small mass ratio $\frac{m^2}{e^2}$. If we kept increasing $eQ$, we would reach $Q = Q_* = \frac{e}{\pi m^2}$, equation \eqref{eq:vlBHdef}, and this part of the rate would also start approaching zero.

\section{Evaporation of a Kerr black hole}\label{sec:evapK}

We repeat the same analysis for the near-extremal Kerr black hole, starting with the wave equation, which will have the same form. We will find that there are two important differences. First, the parameters $\mu$ and $\alpha$ in the solution of the wave equation will also have an imaginary spin-dependent part for both real and complex $\Delta$. Secondly, the equation for the transmission coefficient for the fields with spin will be more involved.

\subsection{Near-extremal Kerr Background}\label{ss:Kbckg}

The Kerr solution describes a rotating black hole with mass $M$ and angular momentum $J$. It is conventional to use the rotational parameter $a = J / M$ instead of $J$. The metric is found to be
\begin{equation}\label{eq:Kerrmetric}
    ds^2_{\text{Kerr}} =
    -\frac{\rho^2 \, \Delta}{\Xi} dt^2 +
    \frac{\rho^2}{\Delta} \, dr^2 +
    \rho^2 \, d\theta^2 +
    \frac{\Xi}{\rho^2} \, \sin^2\theta \,
    \Big( d\phi -
    \frac{2 a M r}{\Xi} \, dt \Big)^2 \, ,
\end{equation}
with the functions given by
\begin{equation}\label{eq:metricfcns}
    \begin{split}
        & \Delta = r^2 - 2Mr + a^2 \, , \qquad
        \rho^2 = r^2 + a^2 \, \cos^2\theta \, , \\
        & \qquad \ \ \
        \Xi = (r^2 + a^2)^2 - a^2 \, \Delta \, \sin^2\theta \, .
    \end{split}
\end{equation}
The radii of the inner and outer horizons are given by
\begin{equation}\label{eq:Kerrhor}
    r_\pm = M \pm \sqrt{M^2-a^2} \, .
\end{equation}

To define an extremal and near-extremal Kerr black hole, we start with the extremality bound involving the mass and rotational parameter given by the following
\begin{equation}\label{eq:Kerrextbnd}
    M \geq a \, ,
\end{equation}
from which the story proceeds as before. The extremal horizon and the horizon measure of near-extremality are given by
\begin{equation}\label{eq:Kerrexthor}
    \begin{split}
        r_h & = a \, , \\
        \delta r & = r_+ - r_- \, .
    \end{split}
\end{equation}
We can zoom into the near-horizon region using
\begin{equation}\label{eq:KerrNHRcoord}
\begin{split}
    r & = r_+ + \delta r \, x \, , \\
    t & = \frac{r_+^2}{\delta r} \, \tau \, , \\
    \phi & = \phi_\text{NHR} + \frac{a}{r_+^2+a^2} \frac{r_+^2}{\delta r} \, \tau \, ,
    \end{split}
\end{equation}
which when plugged into \eqref{eq:Kerrmetric} and expanded in $\delta r$ leads to the NHEK metric (Near-Horizon Extremal Kerr) 
\begin{equation}\label{eq:KerrNHRmetric}
    ds^2_{\text{NHR}} =
    2 a^2 \, \Omega^2(\theta) \,
    \left(
    - 4x(x+1) \, d\tau^2 +
    \frac{dx^2}{x(x+1)} +
    d\theta^2 +
    \Lambda(\theta) \left( d\phi_\text{NHR} + 2x \, dt \right)^2
    \right) \, ,
\end{equation}
with functions given by
\begin{equation}\label{eq:KerrNHRfcns}
    \Omega^2(\theta) = \frac{1+\cos^2\theta}{2} \, , \qquad
    \Lambda(\theta) = \frac{2\sin\theta}{1+\cos^2\theta} \, .
\end{equation}
We can see that when $\theta=\frac{\pi}{2}$, the metric reduces to AdS$_2$ with size $l_\text{AdS} = 2 a^2$. At arbitrary $\theta$, the sphere is fibered over the factor of AdS$_2$.

We can define the thermodynamics of the Kerr black hole starting with the temperature and entropy which are defined as follows
\begin{equation}\label{eq:KerrT}
\begin{split}
    T & = \frac{r_+-r_-}{4 \pi \, (a^2 + r_+^2)} \, , \\
    S & = \pi (r_+^2 + a^2) \, .
\end{split}
\end{equation}
The parameter conjugate to the angular momentum is the angular velocity of the horizon and is given by
\begin{equation}\label{eq:Kerrangvel}
    \Omega = \frac{a}{r_+^2+a^2} \, .
\end{equation}
The energy above extremality is given by
\begin{equation}\label{eq:KEaboveext}
    E = M - a \, .
\end{equation}
The parameters $\delta r$, $E$ and $T$ are all small, if the black hole is near-extremal. This allows us to perform the following expansions
\begin{equation}\label{eq:KerrTexp}
    \begin{split}
        r_+ & = a +4 \pi a^2 T \, \\
        \delta r & = 8 \pi a^2 T \, , \\
        E & = 8 \pi^2 a^3 T^2 \, .
    \end{split}
\end{equation}
For the Kerr black hole, the thermodynamic breakdown scale is given by
\begin{equation}\label{eq:KerrEbrk}
    E_\text{b} = \frac{1}{2 a^3} \, .
\end{equation}

\subsection{Wave equation and emission rate}\label{ss:Keq}

Our goal is to solve the wave equation for a spin$-s$ field, and to compute the transmission coefficient. This was done by Press \& Teukolsky \cite{Teukolsky:1973,Press:1973,Teukolsky:1974} using the Newman-Penrose formalism, which works as follows. Out of our spin$-s$ field, a curvature invariant is built, which we will label with $\psi$ for any spin. $\psi$ satisfies a wave equation, which is separable. If one can solve this equation, the original spin$-s$ field perturbation can be extracted from the invariant, at least in principle. We will not repeat how the scalars $\psi$ are related to the original spin$-s$ field perturbation and how the transmission coefficient of the spin$-s$ field is calculated from the perturbation. Both of these can be found in \cite{Teukolsky:1973,Press:1973,Teukolsky:1974}. Instead, we will only solve the equation for $\psi$, and then directly quote how the transmission coefficient for the spin$-s$ field is written in terms of $\psi$.

We will also not repeat the full partial differential equation satisfied by the scalar $\psi$, which can also be found in these references. Instead, we will write down the ordinary differential equations that it separates into. Given the time translation and axial symmetries of the background, the solution to this equation has the following form
\begin{equation}\label{eq:Kerransatz}
    \psi(t,r,\theta,\phi) = e^{i\omega t} \, R(r) \, S_{slm}(\theta) \, e^{im\phi} \, .
\end{equation}
This results in the following equation for the radial function
\begin{equation}\label{eq:Kerrrads}
\begin{split}
    \Delta & \frac{d^2R}{dr^2} \!+\! 
    (s+1)\frac{d\Delta}{dr} \frac{dR}{dr} \!+\!
    \left( \frac{K^2 - 2 i s (r-M) K}{\Delta} \!+\!
    2 a m \omega \!-\!
    K_{slm}(a\omega) \!+\!
    4 i s \omega r \right) R = 0 \, , \\
    &
    \, K = (r^2 + a^2) \, \omega - am \, .
\end{split}
\end{equation}
The angular functions, known as the spin-weighted spheroidal harmonics, satisfy the following equation
\begin{equation}\label{eq:Kerrang}
    \begin{split}
        \frac{1}{\sin \theta} \! \frac{d}{d\theta} \!
        \left( \! \sin \theta  \frac{d}{d\theta} \!
        S_{slm} \! \right)
        \!+\!
        \left( \!
        K_{slm}(a\omega)
        \!-\! \frac{(m+s\cos\theta)^2}{\sin^2\theta}
        \!-\! (a\omega)^2 \sin^2\theta
        \!-\! 2 a\omega s \cos\theta
        \!+\! s \!
        \right) S_{slm} \!=\! 0 \, ,
    \end{split}
\end{equation}
where $K_{slm}(a\omega) = 2 m a\omega + \lambda_{slm}(a\omega)$, with $\lambda_{slm}(a\omega)$ being the eigenvalues associated with these functions\footnote{
    This follows the conventions defined in Press $\&$ Teukolsky. We will use the Mathematica package Black Hole Perturbation Toolkit \cite{BHPToolkit} to calculate $\lambda_{slm}(a\omega)$ with the command
    "SpinWeightedSpheroidal-
    \newline
    Eigenvalue[$s, l, m,a\omega$]".}.
The $(s+1)$ factor in the first derivative term of \eqref{eq:Kerrrads} can be removed with the following change of variable
\begin{equation}\label{eq:Rtorho}
    \rho(r) = \Delta(r)^{s/2} \, R(r) \, .
\end{equation}
which results in the following radial equation
\begin{equation}\label{eq:Kerrradnos}
     \Delta \frac{d^2\rho}{dr^2} \!+\!
    \frac{d\Delta}{dr} \frac{d\rho}{dr} \!+\!
    \left(
    \frac{K^2 \!-\! 2 i s (r \!-\! M) K}{\Delta} \!+\!
    2 a m \omega \!-\!
    K_{slm}(a\omega) \!+\!
    4 i s \omega r \!-\!
    \frac{s^2 \, \Delta'^{\,2}}{4 \, \Delta} \!-\!
    \frac{s}{2} \, \Delta''
    \right) \, \rho \!=\! 0 \, .
\end{equation}

We will solve this equation for near-extremal Kerr, for which we first have to set up the boundary conditions. A tortoise-like variable can be introduced using $ d r_* = (r^2 + a^2) \, dr / \Delta(r) $. Integrating this relationship gives
\begin{equation}\label{eq:Kerrrtort}
	r_* =
	r +
	\frac{r_+^2 + a^2}{\delta r} \ln (r-r_+) -
	\frac{r_-^2 + a^2}{\delta r} \ln (r - r_-) \, ,
\end{equation}
with $r_*$ going from $-\infty$ to $+\infty$. Following Press \& Teukolsky, we will define the boundary conditions in the scattering problem by requiring the following asymptotic behavior for the radial function\footnote{
    If we were just dealing with a scalar field, we would further transform this equation into a 1D Schrodinger-like equation by defining a wavefunction $S(r)$ as follows $R(r) = S(r) / \sqrt{r^2 + a^2}$. Then, we could just use the 1D transmission coefficient equation \eqref{eq:T&R}. We will not go down this route, as our goal is to use the spin$-s$ transmission coefficient equations from Press \& Teukolsky.}
\begin{equation}\label{eq:Kerrbndcond}
        R(r_*) =
        \begin{cases}
            \frac{Y_\text{trans}}{\Delta^{s}} \,
            e^{-i \omega_\text{eff} r_*} \, , &
            r_* \rightarrow r_*(r_+)=-\infty \\
            \frac{Y_\text{inc}}{r} \,
            e^{-i \omega r_*} +
            \frac{Y_\text{refl}}{r} \,
            \frac{1}{(\omega r)^{2s}} \,
            e^{+i \omega r_*} \, , &
            r_* \rightarrow r_*(+\infty) = +\infty
        \end{cases}
\end{equation}
This also defines the amplitudes of the incident, reflected and transmitted waves $Y_\text{inc}, Y_\text{refl}, Y_\text{trans}$. The wave number at the horizon is given by
\begin{equation}\label{eq:Kerrks}
    \omega_\text{eff} =
    \omega - \frac{m a}{r_+^2 + a^2} \, .
\end{equation}
Given that $\omega_\text{eff}$ can be negative, we find that the Kerr black hole can also emit superradiant modes. Evaluating the threshold frequency at near-extremality defines the Kerr superradiant scale
\begin{equation}\label{eq:Komegasr}
    \omega_\text{sr} = \frac{m}{2a} \, .
\end{equation}

In both the near and far regions, equation \eqref{eq:Kerrradnos} reduces to exactly the same equations we had in subsection \ref{ss:RNeq}, equation \eqref{eq:RNneareq} in the near region and \eqref{eq:RNfareq} in the far. We will not repeat these here, but instead we will just give the solution parameters $p$, $q$ (or equivalently $F$ and $\mu$), $\beta$ (or equivalently $\Delta$), and $\alpha$. These are summarized in the following
\begin{equation}\label{eq:Kerrparams}
    \begin{split}
        p & =
        \frac{2 a^2 \omega_{\text{eff}}}{\delta r} +
        a\omega-\tfrac{1}{2}is \, , \\
        q & =
        \frac{2 a^2 \omega_{\text{eff}}}{\delta r} -
        a\omega +\tfrac{1}{2} i s \, ,\\
        F & = \frac{2 a^2 \omega_{\text{eff}}}{\delta r} \, , \\
        \mu & =
        \mu_r + i \mu_i =
        -2 a \omega  + i s  \, , \\
        \beta & =
        \sqrt{\tfrac{1}{4} + K_{slm}(a\omega) + s(s+1) - 8 a^2 \omega^2} \, , \\
        \Delta & = \tfrac{1}{2} + \beta \, , \\
        \alpha & = \alpha_r + i \alpha_i = 2 a \omega + is \, .
    \end{split}
\end{equation}
Note the extra imaginary terms. It turns out that for certain combinations of the $(s,l,m)$ mode numbers, $\beta$ can be imaginary for on a subinterval of the frequencies. In the next subsection, we will discuss which modes have only real $\beta$ and which have both. The near equation has the following solution after implementing the boundary conditions
\begin{equation}\label{eq:Kerrnearsol}
    \begin{split}
        \rho(x) =
        A \,
        x^{- i p_r - s/2} \,
        (1+x)^{+ i q_r - s/2} \,
        F \left(
        \tfrac{1}{2} \!-\! \beta \!-\! s \!+\! i \mu_r, \, 
        \tfrac{1}{2} \!+\! \beta \!-\! s \!+\! i \mu_r; \,
        1 \!-\! s \!-\! 2 i F \!+\! i \mu_r; \,
        -x
        \right) \, .
    \end{split}
\end{equation}
We can check that this solution has the correct behavior near the horizon as follows. First, we can expand $\rho$ near the horizon taking $x$ to be small, then we can relate this to $R$ using \eqref{eq:Rtorho} and the fact that $\Delta(x) \sim \delta r^2 \, x$ at small $x$, and finally we can write $r$ in terms of $r_*$ using \eqref{eq:Kerrrtort} and evaluating this near the horizon. This results in
\begin{equation}\label{eq:Rxnearhor}
    \begin{split}
        R(x\to 0) =
        \frac{A}{\delta r^s \, x^{s}} \,
        x^{- i p_r} =
        \frac{A \, \delta r^{s}}{\Delta^s} \,
        e^{-i\omega_\text{eff} \, r_*} \, ,
    \end{split}
\end{equation}
which we can see agrees with the expected behavior defined in \eqref{eq:Kerrbndcond}. The equation in the far region has the following solution
\begin{equation}\label{eq:Kerrfarsol}
    \begin{split}
        \rho(y) =
        e^{-i\omega y} \,
        \sum_{\pm}
        C_\pm \,
        y^{-1/2\pm\beta} \,
        M\Big( \tfrac{1}{2}\pm\beta-s+i\alpha_r,
        1\pm2\beta,
        2i\omega y \Big) \, .
    \end{split}
\end{equation}
We can use the asymptotics of the confluent hypergeoemtric function \eqref{eq:Mlarge}, and the fact that far away $\Delta(y) \sim \, y^2$, to find the following behavior for $R(r)$ at infinity
\begin{equation}\label{eq:Kerrfarasymp}
    \begin{split}
        R(y\to \infty) =
        C^\infty_- \,
        \frac{e^{-i\omega y}}{y} \,
        y^{+s-s} \,
        y^{-i\alpha_r}
        +
        C^{\infty}_+ \,
        \frac{e^{+i\omega y}}{y} \,
        y^{i\alpha_r} \,
        y^{-s-s} \, .
    \end{split}
\end{equation}
which agrees with the boundary conditions defined in \eqref{eq:Kerrbndcond} with the first term being the incident and the second the reflected. In equation \eqref{eq:Kerrfarasymp}, we have introduced the following coefficients
\begin{equation}\label{eq:Cpminf}
        C_\pm^\infty \!=\! 
        C_+
        \frac{\Gamma(1 \!+\! 2\beta)}
        {\Gamma(\tfrac{1}{2} \!+\! \beta \!\mp\! s \!\pm\! i\alpha_r)}
        (\pm2i\omega)^{-1/2-\beta \mp s \pm i\alpha_r}
        \!+\!
        C_-
        \frac{\Gamma(1 \!-\! 2\beta)}
        {\Gamma(\tfrac{1}{2} \!-\! \beta \!\mp\! s \!\pm\! i\alpha_r)}
        (\pm2i\omega)^{-1/2 + \beta \mp s \pm i\alpha_r} \, .
\end{equation}
The two solutions for $\rho(r)$ in the near \eqref{eq:Kerrnearsol} and far \eqref{eq:Kerrfarsol} regions can be matched in the overlapping region. Adding the extra $s-$dependent terms to what was found in equation \eqref{eq:RNC12} results in
\begin{equation}\label{eq:KerrCpm}
    \begin{split}
        C_{\pm} =
        \frac{\Gamma(\pm2\beta) \,
        \Gamma(1-s-2iF+i\mu_r)}
        {\Gamma(\tfrac{1}{2}\pm\beta-s+i\mu_r) \,
        \Gamma(\tfrac{1}{2}\pm\beta-2iF)} \,
        \delta r^{+1/2\mp\beta} \,
        A \, .
    \end{split}
\end{equation}

Having solved the wave equation, we can proceed to compute the transmission coefficients or the greybody factors. From Press $\&$ Teukolsky, we quote the greybody factor for a spin$-s$ field (with $s=0,1,2$), which can be found in equations (4.26), (4.32), and (4.44) of \cite{Teukolsky:1974}. These are given in terms of the Kerr parameters $M$ and $a$, and the amplitudes of the incoming and transmitted waves, which are defined in \eqref{eq:Kerrbndcond}. Using the fact that we have a near-extremal black hole, so $M\approx a$, their equations can be written as follows
\begin{equation}\label{eq:Kerrgrey}
    \begin{split}
        & \Gamma_{s=0} = 2 a^2 \,
        \frac{\omega_\text{eff}}{\omega} \,
        \left|
        \frac{Y_\text{trans}}{Y_\text{inc}}
        \right|_{s=0}^2 \, , \qquad
        \Gamma_{s=1} = 2 a^2 \,
        \frac{\omega_\text{eff}}{\omega} \,
        \frac{\omega^2}{4a^4\omega_\text{eff}^2} \,
        \left|
        \frac{Y_\text{trans}}{Y_\text{inc}}
        \right|_{s=1}^2 \, , \\
        & \Gamma_{s=2} = 2 a^2 \,
        \frac{\omega_\text{eff}}{\omega} \,
        \frac{\omega^4}
        {(4a^4)^2 \omega_\text{eff}^2 (\omega_\text{eff}^2+(4\pi T)^2)} \,
        \left|
        \frac{Y_\text{trans}}{Y_\text{inc}}
        \right|_{s=2}^2 \, .
    \end{split}
\end{equation}
These can further be rewritten in terms of the geometric factors $\delta r$ and $F$ and brought under the same product as follows
\begin{equation}\label{eq:Kerrgrey2}
    \Gamma_s =
    2a^2 \,
    \frac{\omega_\text{eff}}{\omega} \,
	\frac{2^{2s} \, \omega^{2s}}
	{(\delta r)^{2s} \,
	\prod_{j=0}^{s-1}
    \left( j^2 + \left( 2F \right)^2 \right)} \, 
    \left| \frac{Y_\text{trans}}{Y_\text{inc}} \right|_{s}^2,
\end{equation}
with the product absent when $s=0$.

To compute the greybody factors, we will need to square the wave amplitudes which, as we saw in \ref{sec:lBH}, depends on whether $\beta$ is real or imaginary. For the rest of this section, we will assume real $\beta$, and we will work out the factor for imaginary $\beta$ in \ref{ss:KCDelta}. Comparing equations \eqref{eq:Kerrbndcond} with \eqref{eq:Rxnearhor}, we find the amplitude of the transmitted wave to be\footnote{
    Note that, for the scalar, this is different from the transmitted wave amplitude for Reissner-Nordström \eqref{eq:RNCtrans} due to the different way of defining the boundary conditions in \eqref{eq:RNbndcond} and \eqref{eq:Kerrbndcond}.}
\begin{equation}\label{eq:Ytrans}
    Y_\text{trans} =
    \delta r^{s} \,
    A \, ,
\end{equation}
which can then straightforwardly be squared. For the incident wave, we can ignore $C_-$ relative to $C_+$ as it is subleading in $\delta r$. Comparing equation \eqref{eq:Kerrbndcond} with \eqref{eq:Kerrfarasymp}, we find that the amplitude of the incident wave is given by
\begin{equation}\label{eq:Yinc}
    Y_\text{inc} =
    \frac{\Gamma(1+2\beta)}
    {\Gamma(1/2+\beta+s-i\alpha_r)} \,
    (-2i\omega)^{-1/2-\beta+s-i\alpha_r} \,
    C_{+} \, .
\end{equation}
This can be further written in terms of $A$ using the matching equation \eqref{eq:KerrCpm}. Squaring the incident then gives
\begin{equation}\label{eq:Yincsq}
    \begin{split}
        &
        \left| Y_\text{inc} \right|^2 =
        2^{-2\Delta+2s} \,
        \delta r^{-2\Delta+2} \,
        \omega^{-2\Delta+2s} \,
        \frac{\Gamma(2\Delta)^2 \,
        \Gamma(2\Delta-1)^2}
        {|\Gamma(\Delta+s-i\alpha_r)|^2 \,
        |\Gamma(\Delta-s+i\mu_r)|^2} \\
        & \qquad \ \ \ \ \ \ \ \,
        \frac{|\Gamma(1-s-2iF+i\mu_r)|^2}
        {|\Gamma(\Delta-2iF|^2} \,
        e^{-\pi\alpha_r} \, .
        \end{split}
\end{equation}
In the last fraction, we can factor out the spin dependence as follows
\begin{equation}\label{eq:spinmanip}
    \begin{split}
        \frac{|\Gamma(1-s-2iF+i\mu_r)|^2}
        {|\Gamma(\Delta-2iF)|^2} =
        \frac{|\Gamma(1-2iF+i\mu_r)|^2}
        {|\Gamma(\Delta-2iF)|^2} \,
        \prod_{j=0}^{s-1} \frac{1}{j^2 + (2F-\mu_r)^2} \, ,
    \end{split}
\end{equation}
where we have used some of the Gamma function identities \eqref{eq:Gimag1} and \eqref{eq:Gimag2}. Plugging the incident amplitude \eqref{eq:Yincsq} and the transmitted amplitude \eqref{eq:Ytrans} into the greybody factor \eqref{eq:Kerrgrey2} and using this identity, we find the greybody factor for a spin-$s$ field to be
\begin{equation}\label{eq:Kerrgreyfin}
    \begin{split}
        & \Gamma^{(s)}(a,T;\omega,l,m_l) \!=\!
        2^{8\Delta-5}
        \pi^{2\Delta-2}
        a^{4\Delta-2}
        T^{2\Delta-2}
        \omega^{2\Delta-1}
        \omega_\text{eff}
        \frac{|\Gamma(\Delta \!+\! s \!-\! i\alpha_r)|^2 \,
        |\Gamma(\Delta \!-\! s \!+\! i\mu_r)|^2}
        {\Gamma(2\Delta)^2 \,
        \Gamma(2\Delta \!-\! 1)^2}
        e^{\pi\alpha_r} \\
        & \qquad \qquad \qquad \qquad \qquad \,
        \frac{|\Gamma(\Delta \!-\! 2iF)|^2}
        {|\Gamma(1 \!- \! 2iF \!+\! i\mu_r)|^2} \,
        \prod_{j=0}^{s-1} \,
        \frac{j^2 + (2F-\mu_r)^2}
        {j^2 + 4 F^2} \, ,
        \end{split}
\end{equation}
where we have also written $\beta$ in terms of $\Delta$, and $\delta r$ in terms of $T$. This can now be plugged into the Hawking formula \eqref{eq:scEdot}, which gives the final evaporation rate for a massless spin-$s$ field being radiated by a near-extremal Kerr black hole
\begin{equation}\label{eq:Kerrrate}
    \begin{split}
        & \frac{d^2E}{d\omega \, dt} \Big|^{(s)}_{\text{sc}} =
        2^{8\Delta-6} \,
        \pi^{2\Delta-3} \,
        a^{4\Delta-2} \,
        T^{2\Delta-2} \,
        \frac{|\Gamma(\Delta+s-i\alpha_r)|^2 \,
        |\Gamma(\Delta-s+i\mu_r)|^2}
        {\Gamma(2\Delta)^2 \,
        \Gamma(2\Delta-1)^2} \,
        e^{\pi\alpha_r} \\
        & \qquad \qquad \ \ \ \
        \frac{\omega^{2\Delta} \,
        \omega_\text{eff}}
        {e^{\omega_{\text{eff}}/T}-1} \,
        \frac{|\Gamma(\Delta-2iF)|^2}
        {|\Gamma(1-2iF+i\mu_r)|^2} \,
        \prod_{j=0}^{s-1} \,
        \frac{j^2 + (2F-\mu_r)^2}
        {j^2 + \left( 2 F \right)^2} \, .
    \end{split}
\end{equation}
The parameters $\mu_r$, $\Delta$ and $\alpha_r$ can be found in \eqref{eq:Kerrparams}. When we set $s=0$ and compare with the Reissner-Nordström rate \eqref{eq:RNscrate}, there is an extra $2^{2\Delta-2}$ from the extra factor of $2$ in $\delta r$, and there is an extra factor of $2$ in the greybody factor flux \eqref{eq:Kerrgrey2}. We will find it convenient to write the spin-$s$ evaporation rate in terms of the spin-$0$ rate as follows
\begin{equation}\label{eq:Kerrrate2}
    \begin{split}
        \frac{d^2E}{d\omega \, dt} \Big|^{(s)}_{\text{sc}} =
        \frac{|\Gamma(\Delta \!+ \!s \!-\! i\alpha_r)|^2 \,
        |\Gamma(\Delta \!-\! s \!+\! i\mu_r)|^2}
        {|\Gamma(\Delta \!-\! i\alpha_r)|^2 \,
        |\Gamma(\Delta \!+\! i\mu_r)|^2} \,
        \prod_{j=0}^{s-1} \,
        \frac{j^2 + (2F-\mu_r)^2}
        {j^2 + \left( 2 F \right)^2} \,
        \frac{d^2E}{d\omega \, dt} \Big|^{(0)}_{\text{sc}} \, .
    \end{split}
\end{equation}

We can take the small frequency limit of this equation and check that it reproduces known results from the literature. In this limit, the spin-weighted spheroidal harmonic eigenvalues reduce to the following
\begin{equation}\label{eq:spheig0}
    K_{slm} (a\omega=0) = \lambda_{slm} (a\omega=0) = l(l+1) - s(s+1) \, ,
\end{equation}
so $\Delta=l+1$. Furthermore, $\mu_r,\,\alpha_r=0$. We first consider the $s=0$ rate and plug these parameters into \eqref{eq:Kerrrate}, which results in
\begin{equation}\label{eq:Kerrratelim}
    \begin{split}
        & \frac{d^2E}{d\omega \, dt} \Big|^{(0)}_{\text{sc}} =
        2^{8l+2} \,
        \pi^{2l-1} \,
        a^{4l+2} \,
        T^{2l} \,
        \frac{l!^4\cdot
        l!^2}
        {(2l)!^2 \,
        (2l+1)!^2} \,
        \frac{\omega^{2l+2} \, \omega_\text{eff}}
        {e^{\omega_{\text{eff}}/T}-1} \,
        \prod_{j=1}^{l}
        \left( 1 +
        \left( \frac{\omega_\text{eff}}{2 j \pi T} \right)^2 \right) \, .
    \end{split}
\end{equation}
After using the double factorial identity, this result exactly matches what was found in \cite{Starobinskii:1973a}, after taking the $M\approx a$ in their paper. For the case of $s=1,2$, we only need to evaluate the extra spin-dependent factor in \eqref{eq:Kerrrate2} in this limit. This results in
\begin{equation}\label{eq:Kerrratelim2}
    \frac{d^2E}{d\omega \, dt} \Big|^{(s)}_{\text{sc}} =
    \frac{(l+s)!^2 \,
    (l-s)!^2}
    {l!^4} \,
    \frac{d^2E}{d\omega \, dt} \Big|^{(0)}_{\text{sc}} \, ,
\end{equation}
which matches what was found in \cite{Starobinskii:1973b}.

One could ask why the wave equation has the same form for the Reissner-Nordström and Kerr black holes. As mentioned in the introduction, the near wave equation is determined by the symmetries of the near-horizon region, which for both black holes is given by the enhanced AdS$_2$ symmetry. It was explicitly shown\footnote{
    These references fully worked in NHEK, which amounts to setting the frequency to the superradiant bound. We also point out that in the context of Kerr/CFT \cite{Guica:2008}, the wave equation was solved in the full near-extremal Kerr geometry (not just NHEK) by breaking it up into a near and far region, but still at frequencies approximately equal to the superradiant bound \cite{Bredberg:2009,Hartman:2009}.}
in \cite{Bardeen:1999} for the scalar and \cite{Dias:2009, Amsel:2009} for higher spin fields that the radial equation in NHEK \eqref{eq:KerrNHRmetric} has the same form as that in the near-horizon region of Reissner-Nordström \eqref{eq:RNNHRmetric}. For the far equation, it is easy to see that if one expands the Kerr metric \eqref{eq:Kerrmetric} at large distance, it is has the same form as Reissner-Nordström \eqref{eq:RNmetric}, with the leading order term given by the Schwarzschild black hole.

\subsection{Real Delta}\label{ss:KRDelta}

Before turning to the evaporation rate, we first present a discussion of the types of near-extremal Kerr radiation in the spirit of subsection \ref{ss:class}. We start with the conformal dimension. For certain combinations of the mode numbers $s$, $l$, and $m$, $\Delta$ can be complex on a subset of the frequency range. It turns out that this is almost always the case for the modes with $l=m$, but there are modes where $\Delta$ is complex when $l\neq m$ and there are also modes where it is not complex when $l=m$. As we will see, the $l=m$ modes will be the dominant ones, which means that we always have to consider both real and complex $\Delta$ for the Kerr black hole.\footnote{
    Since the conformal dimension depends on the spheroidal harmonic eigenvalue, which itself depends on and monotonically decreases with the frequency, one only needs to check the value of the spheroidal eigenvalue at $\omega_\text{sr}$ to determine whether a mode has real or complex $\Delta$. For scalar fields, this was done in \cite{Bardeen:1999} for the lowest few modes. We will repeat this for the photon and graviton.}
In other words, here there is no notion of small versus large black holes. In this subsection, we will only compute the rates for the frequencies with real $\Delta$, leaving the remainder for the next subsection.

\begin{figure}[H]
    \centering
    \begin{tikzpicture}
    \draw[gray, thick, dashed] (0,-3) -- (0,+3);
    \filldraw[black] (-0.2,+3.3) node[anchor=west]{$\log m$};
    \draw[gray, thick, dashed] (-5,0) -- (+5,0);
    \filldraw[black] (+5.1,0) node[anchor=west]{$\log a$};
    \draw[black, thick] (-3.2,-1.6) -- (+3.2,+1.6);
    \filldraw[black] (+3.2,+1.8) node[anchor=west]{$m = J$}; 
    \draw[black, thick] (-3.5,0) -- (+3.5,0);
    \filldraw[black] (+3.5,-0.25) node[anchor=west]{$m=1$};
    \draw[black, thick] (-3.3,+1.1) -- (+3.3,-1.1);
    \filldraw[black] (+3.35,-1.2) node[anchor=west]{$\omega_\text{sr} = E_\text{brk}$};
    \draw[<->, blue, semithick] (0:2.55 cm) 
    arc[radius=2.50, start angle=0, end angle=+26.565];
    \filldraw[blue] (2.6,0.7) node[anchor=west]{\bf{Semi-Cl.}};
    \draw[<->, red, semithick] (+26.565:2.5cm) 
    arc[radius=2.55, start angle=+26.565, end angle=+206.565];
    \filldraw[red] (-1.8,+2.2) node[anchor=east]{\bf{Forbidden}};
    \draw[<->, red, semithick] (-180:2.45cm) 
    arc[radius=2.5, start angle=-180, end angle=0];
    \filldraw[red] (+1.7,-2.3) node[anchor=west]{\bf{Forbidden}};
    \end{tikzpicture}
    \captionsetup{width=.93\linewidth}
    \caption{A diagram of $\log m$ vs. $\log a$ parameter space showing the available regions for radiation from a Kerr black hole. In the only available wedge, the emission rate is always given by the semi-classical formula \eqref{eq:Kerrrate}. $\Delta$ can be both real and complex, depending on the mode.}
    \label{fig:logalogm}
\end{figure}
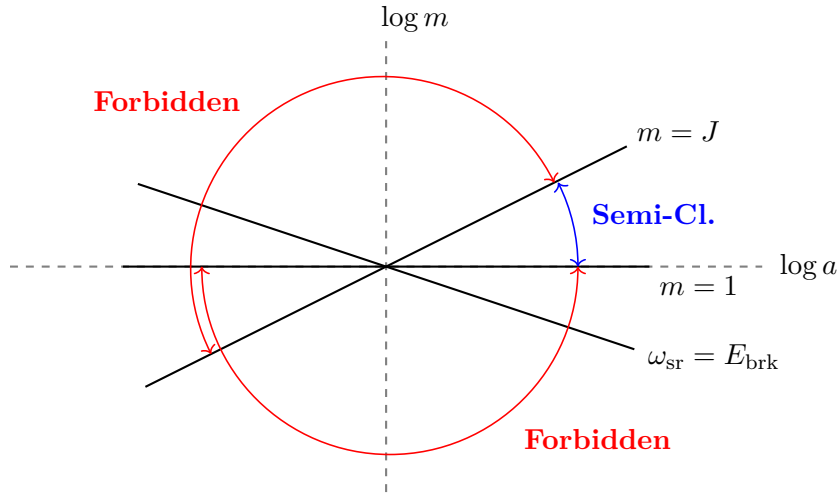

As mentioned before, each mode that carries angular momentum $m$ can be thought of as a charged particle with charge $m$, but there is one important difference between this "charged" particle and the particles in Reissner-Nordström, and that is the fact that we do not get to choose $m$ as we did $e$, instead modes with all possible values of angular momentum need to be considered. In this sense, we have one less free parameter for Kerr, which leads to less diversity in the possible energy regimes and their emission spectra. Keeping this in mind, we can still plot $\log m$ versus $\log a$, which will tell us how $\omega_\text{sr}$ compares to $E_\text{b}$ for different combinations of $a$ and $m$. This is done in figure \ref{fig:logalogm} which has the $\frac{m}{2a} = \omega_{sr} = E_\text{b} = \frac{1}{2a^3}$ line and two addition constraints. First, the black hole must have more angular momentum than the mode it is emitting, so $\sqrt{a} = J \geq m$. And secondly, the smallest angular momentum that can be emitted is $m = 1 $. This second constraint we did not have in Reissner-Nordström.
\noindent What we find is that all of the allowed combinations of $a$ and $m$ must have $\omega_\text{sr} \gg E_\text{b} $, which by the arguments of \ref{ss:class} means that the energy ratio in the quantum evaporation rate \eqref{eq:Echratio} is always greater than one, regardless of the size of $E_\text{}$, which then implies that the quantum rate, even if we computed it, would always reduce to the semi-classical formula, which is why we will only consider the semi-classical rate \eqref{eq:Kerrrate} for Kerr.

There is one more important difference between Reissner-Nordström and Kerr, which is the fact that for near-extremal Kerr there is only a superradiant-dominanted regime. To see this, for the semi-classical rate we need to check how the temperature $T$ compares to the superradiant scale $\omega_\text{sr}$. We are treating $T$ as an independent parameter, but for a near-extremal black hole, it still has to be small, meaning $a T \ll 1$. Comparing $aT\ll1$ to $a \omega_\text{sr} = \frac{m}{2}$ which is at best an order one number, we find that $T \ll \omega_\text{sr}$, which implies that near-extremal Kerr always radiates in the superradiant regime. Therefore, we only need to consider the range of frequencies $0\leq\omega\leq\omega_{sr}$ as shown with equations \eqref{eq:srint1} and \eqref{eq:srint2}.

This also implies that we have to consider both photon and graviton radiation. If there were a neutral-like regime, we would have $\Delta=l+1$, and the photon would dominate, as it has $l=1$ as its lowest mode, while the graviton has $l=2$, which would be suppressed due to a higher $\Delta$. We will see that in the superradiant regime, both fields will have roughly the same $\Delta$ and produce comparable amounts of radiation. Furthermore, we can ignore "neutral" radiation, that is, modes with $m=0$, since their evaporation rates are proportional to a power of temperature, while the rates for the "charged" modes will be constant.

To summarize, for near-extremal Kerr, the dominant modes in the emission specturm are those with $l=m$, most of these having both a real and complex $\Delta$, while all other modes can be ignored, including $m=0$. The spectrum will be given by the semi-classical Hawking formula and will be dominated by the superradiant frequencies. With that in mind, we can turn to computing the evaporation rate.

In the superradiant regime, we can approximate the $F-$dependent Gamma functions in the semi-classical rate \eqref{eq:Kerrrate} expanding in large $|F|$ as in \eqref{eq:Gammaratioapprox}. Furthermore, we can ignore the last factor in \eqref{eq:Kerrrate} at large $|F|$. The rate then reduces to
\begin{equation}\label{eq:Kerrrateapprox}
    \begin{split}
        & \frac{d^2E}{d\omega  dt} \bigg|^{(s)}_{\text{sc}} \!=\!
        2^{8\Delta\!-\!6} 
        \pi^{2\Delta\!-\!3} 
        a^{4\Delta-\!2} 
        T^{2\Delta\!-\!2} 
        \frac{|\Gamma(\Delta \!+\! s \!-\! i\alpha_r)|^2 \,
        |\Gamma(\Delta \!-\! s \!-\! i\alpha_r)|^2}
        {\Gamma(2\Delta)^2 \,
        \Gamma(2\Delta-1)^2}
        \omega^{2\Delta}
        |\omega_\text{eff}|
        \left( \frac{\omega_\text{eff}}{2 \pi T} \right)^{2(\Delta\!-\!1)} \, ,
    \end{split}
\end{equation}
where we used the fact that $\mu_r=-\alpha_r$ to cancel the exponential functions. We perform a change of variables $x = a \omega$, which gives our final result for the Kerr emission rate of a spin$-s$ field for frequencies with real $\Delta$
\begin{equation}\label{eq:Kerrratefin}
    \frac{dE}{dt} \bigg|^{(s)}_{\text{sc}} =
    H_{slm} \,
    \frac{1}{a^2} \, ,
\end{equation}
where the coefficients $H_{slm}$ are now just numbers, not functions, and are given by the following integral
\begin{equation}\label{eq:Hslm}
    \begin{split}
        H_{slm} & = \frac{1}{\pi}
        \int_{0}^{x_*} dx \,
        2^{6\Delta-4} \,
        x^{2\Delta} \,
        \left( \tfrac{m}{2} - x \right)^{2\Delta-1} \,
        \frac{\left| \Gamma(\Delta-s+2ix)  \right|^2 \,
        \left| \Gamma(\Delta+s+2ix)  \right|^2}
        {\Gamma(2\Delta)^2 \,
        \Gamma(2\Delta-1)^2} \,
        ,
    \end{split}
\end{equation}
where the frequency-dependent conformal dimension written as a function of $x$ is given by
\begin{equation}\label{eq:KerrDeltax}
    \Delta_{slm}(x) =
    \tfrac{1}{2} +
    \sqrt{\tfrac{1}{4} + K_{slm}(x) +s(s+1) - 2m^2x^2} \, .
\end{equation}
If $\Delta(x)$ is real on the entire frequency range $0\leq x \leq x_\text{sr} = a \omega_\text{sr} = m/2$, then $x_*=m/2$  and \eqref{eq:Kerrratefin} covers the entire evaporation rate. If $\Delta(x)$ turns complex at some frequency, then the upper bound of integration $x_*$ is the point where this happens, that is where $\beta(x)^2 = 0$ (the expression under the square root in $\Delta(x)$). In that case, \eqref{eq:Kerrratefin} covers only a part of the evaporation rate and the remaining part we leave for the next subsection. To compute the numerical values of $K_{slm}(x)$ we use the Mathematica package Black Hole Perturbation Toolkit \cite{BHPToolkit}.

We can now compute the evaporation rate for $s=0,1,2$, for the lowest few modes. One can check that for the scalar, the dominant $l=m$ modes have complex $\Delta$ starting at $l=m=2$, while for the photon and graviton complex $\Delta$ starts at the lowest $l=m$ mode. Starting with the scalar, we numerically evaluate \eqref{eq:Hslm} for $l=1,2,3$ and all possible $m\neq0$. The results are summarized in the following table
\begin{table}[H]
    \centering
    \begin{tabular}{|m{1.6cm}|m{2.2cm}|m{2.2cm}|m{2.2cm}|}
    \hline
     & \centering $l\!=\!m$ & \centering $l\!=\!m\!-\!1$ & \ \ \ $l\!=\!m\!-\!2$ \\
    \hline
    \ \ \ $l\!=\!1$ & \,$0.0483 \cdot 10^{-3}$ & \centering / & \ \ \ \ \ \ \ / \\
    \hline
    \ \ \ $l\!=\!2$ & \,$0.0729 \cdot 10^{-3}$ & \centering $\sim\! 10^{-9}$ & \ \ \ \ \ \ \ / \\
    \hline
    \ \ \ $l\!=\!3$ & \,$0.0334 \cdot 10^{-3}$ & \centering $\sim\! 10^{-9}$ & \ \ \ \ $\sim\! 10^{-\!14}$ \\
    \hline
    \end{tabular}
    \captionsetup{width=.93\linewidth}
    \caption{Table of energy and angular momentum emission rates for lowest few modes of a scalar field ($s=0$) being radiated by a near-extremal Kerr black hole, obtained by evaluating the integrals in equation \eqref{eq:Hslm}. This only account for the frequencies with real $\Delta$.}
    \label{tab:H0low}
\end{table}
\noindent This shows that at a given $l$, the rates for the modes with different $m$ fall off very quickly with decreasing $m$. The trend continues if we go to higher $l$. This implies that it is safe to ignore modes with $l\neq m$, unless extremely high precision is required. Furthermore, this shows that the contribution of the $l=m$ modes does not drop very quickly,\footnote{
    From this table it looks like the rate for the $l=m$ modes grows and then drops, but this is only true for the $l=m=1$ mode of the scalar.
}
which implies that a number of these must be taken for any reasonable degree of precision, and therefore there is no single dominant mode for near-extremal Kerr radiation.

We can also write down an equation for the angular momentum emission rate, which takes the following form
\begin{equation}\label{eq:KerrrateJfin}
    \frac{dJ}{dt} \bigg|^{(s)}_{\text{sc}} =
    H^{J}_{slm} \,
    \frac{1}{a} \, .
\end{equation}
To compute the coefficients $H^{J}_{slm}$, we replace a power of $\omega$ in \eqref{eq:Kerrrateapprox} with a factor of $m$ and then changing variables, which is straightforward and will not be explicitly shown here. We can now proceed to evaluate the energy and angular momentum emission rates for further scalar $l=m$ modes. These are given in the following table
\begin{table}[H]
    \centering
    \begin{tabular}{|m{1.9cm}|m{1.1cm}|m{1.1cm}|m{1.1cm}|m{1.1cm}|m{1.1cm}|m{1.1cm}|m{1.1cm}|m{1.1cm}|}
    \hline
    $\ \ \, l=m$ &
    \centering1 & \centering2 & \centering3 & \centering4 &
    \centering5 & \centering6 & \centering7 & \ \ \ 8 \\
    \hline
    $H_{0lm}^E \, [10^{-3}]$ &
    0.0483 & 0.0729 & 0.0334 & 0.0115 &
    0.0034 & 0.0010 & 0.0003 & 0.0001 \\
    \hline
    $H_{0lm}^J \, [10^{-3}]$ &
    0.1450 & 0.2007 & 0.0940 & 0.0328 &
    0.0099 & 0.0028 & 0.0007 & 0.0002 \\
    \hline
    \end{tabular}
    \captionsetup{width=.93\linewidth}
    \caption{Table of energy and angular momentum emission rates for the most dominant modes, those with $l=m$, of a scalar field ($s=0$) being radiated by a near-extremal Kerr black hole, obtained by evaluating the integrals in equation \eqref{eq:Hslm}. This only account for the frequencies with real $\Delta$.}
    \label{tab:HRDelta0}
\end{table}
\noindent This table should give a better sense of how quickly the contributions of subsequent modes drop. For the photon and graviton, the same patterns are observed. We compute the energy and angular momentum emission rates for the lowest eight modes, which are given in the following two tables.
\begin{table}[H]
    \centering
    \begin{tabular}{|m{1.9cm}|m{1.1cm}|m{1.1cm}|m{1.1cm}|m{1.1cm}|m{1.10cm}|m{1.10cm}|m{1.10cm}|m{1.10cm}|}
    \hline
    $\ \ \, l=m$ &
    \centering1 & \centering2 & \centering3 & \centering4 &
    \centering5 & \centering6 & \centering7 & \ \ \ 8 \\
    \hline
    $H_{1lm}^E \, [10^{-3}]$ &
    0.6022 & 0.2622 & 0.0832 & 0.0236 &
    0.0063 & 0.0016 & 0.0004 & 0.0001 \\
    \hline
    $H_{1lm}^J \, [10^{-3}]$ &
    1.7135 & 0.7534 & 0.2412 & 0.0687 &
    0.0183 & 0.0047 & 0.0012 & 0.0003 \\
    \hline
    \end{tabular}
    \captionsetup{width=.93\linewidth}
    \caption{Table of energy and angular momentum emission rates for the electromagnetic field ($s=1$). This only accounts for the frequencies with real $\Delta$.}
    \label{tab:HRDelta1}
\end{table}
\begin{table}[H]
    \centering
    \begin{tabular}{|m{1.9cm}|m{1.1cm}|m{1.10cm}|m{1.10cm}|m{1.10cm}|m{1.10cm}|m{1.10cm}|m{1.10cm}|m{1.10cm}|}
    \hline
    $\ \ \, l=m$ &
    \centering2 & \centering3 & \centering4 &
    \centering5 & \centering6 & \centering7 & \centering 8 & \ \ \ 9 \\
    \hline
    $H_{2lm}^E \, [10^{-3}]$ &
    5.0565 & 0.9008 & 0.1690 & 0.0335 &
    0.0068 & 0.0014 & 0.0003 & 0.0001 \\
    \hline
    $H_{2lm}^J \, [10^{-3}]$ &
    17.557 & 2.8857 & 0.5250 & 0.1024 &
    0.0208 & 0.0043 & 0.0009 & 0.0002 \\
    \hline
    \end{tabular}
    \captionsetup{width=.93\linewidth}
    \caption{Table of energy and angular momentum emission rates for the gravitational field ($s=2$). This only accounts for the frequencies with real $\Delta$.}
    \label{tab:HRDelta2}
\end{table}
\noindent Note that there is no $l=m=1$ mode for a spin$-2$ field. As mentioned before, all modes except for the scalar $l=m=1$ mode have $\Delta$ go complex on an upper part of the frequency range. Once we have computed the contribution from that part of the range, we will sum over all modes to obtain the full evaporation rate for a given spin.

\subsection{Complex Delta}\label{ss:KCDelta}

To compute the rate at complex $\Delta $, we need to make the appropriate modifications to the solution of the wave equation and then rederive the greybody factor and evaporation rate. Most of this section will involve the same kinds of arguments used in section \ref{ss:RNCDelta}.

The first step is to make the replacement $\beta \rightarrow i \beta$ in the Kerr matching equation \eqref{eq:KerrCpm}, which results in
\begin{equation}\label{eq:KCpmi}
    \begin{split}
        C_{\pm} =
        \frac{\Gamma(\pm2i\beta) \,
        \Gamma(1-s-2iF+i\mu_r)}
        {\Gamma(\tfrac{1}{2}-s\pm i\beta+i\mu_r) \,
        \Gamma(\tfrac{1}{2}\pm i\beta-2iF)} \,
        \delta r^{+1/2\mp i\beta} \,
        A \, .
    \end{split}
\end{equation}
We need to keep both $C_\pm$, since they involve the same (real) power of the small quantity $\delta r$, and one cannot be ignored relative to the other. Then, the same replacement is made in the amplitude of the incident wave, which gives
\begin{equation}\label{eq:Yinci}
    Y_{\text{inc, }\pm} =
    \frac{\Gamma(1\pm2i\beta)}
    {\Gamma(\tfrac{1}{2}+s\pm i\beta-i\alpha_r)} \,
    (-2i\omega)^{-1/2+s\mp i \beta-i\alpha_r} \,
    C_\pm \, .
\end{equation}
Plugging the matching condition \eqref{eq:KCpmi} into this equation and squaring results in
\begin{equation}\label{eq:Yincisq}
    \begin{split}
        & |Y_{\text{inc, }\pm}|^2 =
        \frac{\delta r}{(2\omega)^{1-2s}}
        \frac{|\Gamma(1+2i\beta)|^2 \,
        |\Gamma(2i\beta)|^2 \,}
        {|\Gamma(\tfrac{1}{2}+s\pm i\beta-i\alpha_r)|^2 \,
        |\Gamma(\tfrac{1}{2}-s\pm i\beta+i\mu_r)|^2} \\
        & \qquad \qquad \ \ \ \ \ \ \ \
        \frac{|\Gamma(1-s-2iF+i\mu_r)|^2}
        {|\Gamma(\tfrac{1}{2}\pm i\beta-2iF)|^2} \,
        e^{\pi(\mp \beta-\alpha_r)} \,
        |A|^2 \, .
    \end{split}
\end{equation}
Since we only need to consider the superradiant regime, we can simplify the fractions that involve $F$ as we did in equation \eqref{eq:Gammaratio2}. This gives
\begin{equation}\label{eq:Gammasalg}
    \begin{split}
        \frac{|\Gamma(1-s-2iF+i\mu_r)|^2}
        {|\Gamma(\tfrac{1}{2}\pm i\beta-2iF)|^2}
        =
        e^{\pi(-\mu_r \pm \beta)} \,
        \prod_{j=0}^{s-1} \,
        \frac{1}
        {j^2 + (2F)^2} \, .
    \end{split}
\end{equation}
The spin product can be extracted using \eqref{eq:Gimag1} and \eqref{eq:Gimag2}. Note that there is no $\mu_r$ next to the $2F$ in the product as in \eqref{eq:spinmanip} since we are in the large $F$ regime. Plugging this into equation \eqref{eq:Yincisq} and using that fact that $-\mu_r = + \alpha_r \geq 0$, we have
\begin{equation}\label{eq:Yincisq2}
    \begin{split}
        & |Y_{\text{inc, }\pm}|^2 =
        \frac{\delta r \, |F|}{2^{-2s}\omega^{1-2s}}
        \frac{|\Gamma(1+2i\beta)|^2 \,
        |\Gamma(2i\beta)|^2}
        {|\Gamma(1/2\pm i\beta-i\alpha_r)|^4} \,
        \prod_{j=0}^{s-1} \,
        \frac{1}
        {j^2 + (2F)^2} \,
        |A|^2 \, .
    \end{split}
\end{equation}
Expanding the Gamma functions in terms of hyperbolic trigonometric functions, we find that this simplifies to
\begin{equation}\label{eq:Yincisq3}
    \begin{split}
        |Y_{\text{inc, }\pm}|^2 =
        \frac{\delta r \, |F|}{2^{-2s}\omega^{1-2s}} \,
        \frac{\cosh^2 \pi (\alpha_r \mp \beta)}
        {\sinh^2 2 \pi \beta} \,
        \prod_{j=0}^{s-1} \,
        \frac{1}
        {j^2 + (2F)^2} \,
        |A|^2 \, .
    \end{split}
\end{equation}
This equation, together with the amplitude of the transmitted wave \eqref{eq:Ytrans}, can be plugged into the greybody equation \eqref{eq:Kerrgrey2}, which results in
\begin{equation}\label{eq:Kgreyi}
    \begin{split}
        \Gamma^{(s)} =
        -
        \frac{\sinh^2 2 \pi \beta}
        {\cosh^2 \pi \, (\alpha_r + \beta) +
        \cosh^2 \pi \, (\alpha_r - \beta)} \, .
    \end{split}
\end{equation}
Even though the spin-dependent products canceled, there is still spin dependence hidden in $\beta(\omega)$.

As in \ref{ss:RNCDelta}, we omitted the cross-term between $Y_{\text{inc} +}$ and $Y_{\text{inc} -}$ when squaring, but for Kerr we do not have a parameter like $eQ$ which we can tune to make one of these terms dominate. It turns out that $\alpha\geq\beta$ is true for almost all modes, and for those that it isn't they are still approximately equal. This means that the first term in the denominator will always dominate the second, and therefore we do get to omit the cross term, as well as the second term in the denominator. The greybody factor then simplifies to
\begin{equation}\label{eq:Kgreyiapprox}
    \begin{split}
        \Gamma^{(s)} =
        - e^{-2 \pi(\alpha-\beta)} \, .
    \end{split}
\end{equation}

After evaluating at $\omega=\omega_\text{sr}$ and ignoring the cross-term, equation \eqref{eq:Kgreyi} agrees with  Starobinskii \cite{Starobinskii:1973a}, who considered an extremal black hole and frequencies around the superradiant bound. Note that this is different from the limit in equation \eqref{eq:Kerrratelim}. Press \& Teukolsky in \cite{Teukolsky:1973,Press:1973,Teukolsky:1974} performed the same calculation for near-extremal Kerr. If we had not made the large $|F|$ approximation in \eqref{eq:Yincisq} we would have recovered their result. They also took the extremal limit and recovered the Starobinskii result. It is not surprising that taking the superradiant large $|F|$ limit or the extremal limit gives the same result, since at extremality, all frequencies that are being radiated are superradiant.

The greybody factor can now be plugged into the Hawking formula, which in the superradiant regime amounts to multiplying by $-1/2\pi$. After a changing of variables $x=a\omega$, we obtain our final result for the Kerr emission rate of a spin$-s$ field for frequencies with complex $\Delta$
\begin{equation}\label{eq:Kerrrateifin}
    \frac{dE}{dt} \bigg|^{(s)}_{\text{sc}} =
    H'_{slm} \,
    \frac{1}{G a^2} \, .
\end{equation}
If we use equation \eqref{eq:Kgreyiapprox} for the greybody factor, we obtain the following for the $H'_{slm}$ coefficients
\begin{equation}\label{eq:Hslmprapprox}
    \begin{split}
        H'_{slm} =
        \frac{1}{2\pi} \,
        \int_{x_*}^{\frac{m}{2}} dx \,
        x \,
        e^{-2\pi(x-\beta)}
        \, .
    \end{split}
\end{equation}
If we want higher precision, we can use equation \eqref{eq:Kgreyi}, which would result in
\begin{equation}\label{eq:Hslmpr}
    \begin{split}
        H'_{slm} =
        \frac{1}{2\pi} \,
        \int_{x_*}^{\frac{m}{2}} dx \,
        x \,
        \frac{\sinh^2 2 \pi \beta}
        {\cosh^2 \pi \, (2x + \beta) +
        \cosh^2 \pi \, (2x - \beta)} \, ,
    \end{split}
\end{equation}
but this would still not give the exact answer, since the cross-terms are missing. The parameter $\beta$ as a function of $x$ is given by
\begin{equation}\label{eq:betaix}
    \beta_{} =
    \sqrt{2m^2x^2 -
    \tfrac{1}{4} -
    K_{slm}(\tfrac{mx}{2}) -
    s(s+1)} \, .
\end{equation}
Note the flipped sign under the square root. The lower bound of integration $x_*$ is given by $\beta(x_*)=0$ as explained in the previous subsection. We can also compute the emission rate of angular momentum as in \ref{ss:KRDelta}.

We compute the coefficients $H'_{sl;m}$ using equation \eqref{eq:Hslmprapprox} for the scalar, photon, and graviton. The $l\neq m$ modes are again highly suppressed compared to the dominant $l=m$ modes whose contribution can be found in the following three tables.
\begin{table}[H]
    \centering
    \begin{tabular}{|m{1.9cm}|m{1.10cm}|m{1.10cm}|m{1.10cm}|m{1.10cm}|m{1.10cm}|m{1.10cm}|m{1.10cm}|m{1.10cm}|}
    \hline
    $\ \ \, l=m$ &
    \centering1 & \centering2 & \centering3 & \centering4 &
    \centering5 & \centering6 & \centering7 & \ \ \ 8 \\
    \hline
    $H_{0lm}^E \, [10^{-3}]$ &
    \centering 0 & 0.0045 & 0.0115 & 0.0108 &
    0.0076 & 0.0047 & 0.0027 & 0.0014 \\
    \hline
    $H_{0lm}^J \, [10^{-3}]$ &
    \centering 0 & 0.0092 & 0.0235 & 0.0222 &
    0.0156 & 0.0095 & 0.0054 & 0.0029 \\
    \hline
    \end{tabular}
    \captionsetup{width=.93\linewidth}
    \caption{Table of energy and angular momentum emission rates for the most dominant modes, those with $l=m$, of a scalar field ($s=0$) being radiated by a near-extremal Kerr black hole, obtained by evaluating the integrals in equation \eqref{eq:Hslmpr}. This only account for the frequencies with complex $\Delta$.}
    \label{tab:HCDelta0}
\end{table}
\begin{table}[H]
    \centering
    \begin{tabular}{|m{1.9cm}|m{1.10cm}|m{1.10cm}|m{1.10cm}|m{1.10cm}|m{1.10cm}|m{1.10cm}|m{1.10cm}|m{1.10cm}|}
    \hline
    $\ \ \, l=m$ &
    \centering1 & \centering2 & \centering3 & \centering4 &
    \centering5 & \centering6 & \centering7 & \ \ \ 8 \\
    \hline
    $H_{1lm}^E \, [10^{-3}]$ &
    0.0141 & 0.0809 & 0.0675 & 0.0420 &
    0.0232 & 0.0121 & 0.0061 & 0.0030 \\
    \hline
    $H_{1lm}^J \, [10^{-3}]$ &
    0.0285 & 0.1664 & 0.1387 & 0.0860 &
    0.0475 & 0.0247 & 0.0124 & 0.0061 \\
    \hline
    \end{tabular}
    \captionsetup{width=.93\linewidth}
    \caption{Table of energy and angular momentum emission rates for the electromagnetic field ($s=1$). This only accounts for the frequencies with complex $\Delta$.}
    \label{tab:HCDelta1}
\end{table}
\begin{table}[H]
    \centering
    \begin{tabular}{|m{1.9cm}|m{1.10cm}|m{1.10cm}|m{1.10cm}|m{1.10cm}|m{1.10cm}|m{1.10cm}|m{1.10cm}|m{1.10cm}|}
    \hline
    $\ \ \, l=m$ &
    \centering 2 & \centering3 & \centering4 &
    \centering5 & \centering6 & \centering7 & \centering8 & \ \ \ 9 \\
    \hline
    $H_{2lm}^E \, [10^{-3}]$ &
    11.5373 & 3.6450 & 1.2039 & 0.4235 &
    0.1566 & 0.0601 & 0.0237 & 0.0096 \\
    \hline
    $H_{2lm}^J \, [10^{-3}]$ &
    24.2331 & 7.5508 & 2.4756 & 0.8668 &
    0.3196 & 0.1224 & 0.0482 & 0.0194 \\
    \hline
    \end{tabular}
    \captionsetup{width=.93\linewidth}
    \caption{Table of energy and angular momentum emission rates for the gravitational field ($s=2$). This only accounts for the frequencies with complex $\Delta$.}
    \label{tab:HCDelta2}
\end{table}
\noindent Note that there is no contribution from the $l=m=1$ mode for the scalar since it has real $\Delta$ on the full frequency range. Another pattern can be observed if we compare these three tables with the ones from the previous subsection. The $H'_{slm}$ start to be significantly bigger than the $H_{slm}$ after a certain mode, which parallels what we found for charged radiation, where the frequencies with complex $\Delta$ (at least for the right half of the frequency interval) dominated the emission spectrum at large charge (see figure \ref{fig:RNlrate1}).

We can now add up the contribution from the $H_{slm}$ and $H'_{slm}$ coefficients for each spin to find the total emission rate. The following is found for the energy and angular momentum emission rates to the third digit. 
\begin{table}[H]
    \centering
    \begin{tabular}{|m{1.4cm}|m{2.2cm}|m{2.2cm}|}
    \hline
    \ & \centering $H^{E} \, [10^{-3}]$ & \ \ $H^{J} \, [10^{-3}]$ \\
    \hline
    \centering $s\!=\!0$ & \centering $0.215$ & \ \ \ \ \ $0.577$ \\
    \hline
    \centering $s\!=\!1$ & \centering $1.23$ & \ \ \ \ \ \ $3.32$ \\
    \hline
    \centering $s\!=\!2$ & \centering $23.2$ & \ \ \ \ \ \ $56.7$ \\
    \hline
    \end{tabular}
    \captionsetup{width=.93\linewidth}
    \caption{Table of the total energy and angular momentum emission rates for the scalar ($s=0$), electromagnetic ($s=1$) and gravitational ($s=2$) fields being radiated by a near-extremal Kerr black hole, obtained by adding up the contributions of all modes and all frequencies (those with real and complex $\Delta$).}
    \label{tab:Hfin}
\end{table}

\section{Discussion}\label{sec:disc}

\subsection*{Summary of Reissner-Nordström}

The semi-classical and quantum-corrected evaporation rates are uniquely specified by the three parameters $Q$, $E$, and $e$. A choice of $Q$ and $e$ places us somewhere in figure \ref{fig:logQloge}  and determines whether we are dealing with a small black hole (bottom two wedges in \ref{fig:logQloge}) and we use equation \eqref{eq:RNscrate} for the semi-classical and \eqref{eq:RNqrate} for the quantum-corrected evaporation rate, as $\Delta$ is real across the entire frequency range, or a large black hole (top wedge in \ref{fig:logQloge}) where we use equation \eqref{eq:RNscrate} for the semi-classical rate for frequencies with real $\Delta$ and \eqref{eq:RNscratei} for frequencies with complex $\Delta$. $Q$ and $e$ also determine the two energy scales $E_\text{b}$ and $\omega_\text{sr}$. Depending on how these two scales compare, we can distinguish two more cases. When $E_\text{b} \ll \omega_\text{sr}$, the radiation will always be semi-classical (top two wedges in figure \ref{fig:logQloge}), and when $\omega_\text{sr} \ll E_\text{b}$ (bottom wedge in figure \ref{fig:logQloge}), the radiation can be either semi-classical or quantum, depending on whether $E_\text{b} \ll E$ or $E \ll E_\text{b}$, respectively. Semi-classical radiation has the quantum-corrected formula reduce to the semi-classical one to leading order, whereas for quantum radiation, the quantum rate gives genuinely different predictions from the semi-classical formula. Furthermore, depending on how $E$ compares to $\omega_\text{sr}$, the radiation will come out as neutral-like or superradiant-dominated. The different ways of ordering these three energy scales and the corresponding regimes they define are summarized in table \ref{tab:Eregimes}. The rates for small black holes are given in \eqref{eq:nlscfin} (semi-classical neutral-like), \eqref{eq:nlqfin} (quantum neutral-like), \eqref{eq:srscfin} (semi-classical superradiant-dominated) and \eqref{eq:srqfin} (quantum superradiant-dominated), and for large black holes in \eqref{eq:lsrscfin}. When the black hole starts at any combination of these three parameters, as it emits radiation, it is driven away from extremality, which means that $E$ grows. This implies that it will move through possibly a number of these regimes during its evolution towards non-extremality. We now turn to a discussion of this evaporation history for each type of black hole.

\subsection*{Evaporation history of near-extremal Reissner-Nordström}

For the remainder of the discussion, we reintroduce $G$ into our formulas. We start with the smallest black holes, those with $\omega_\text{sr} \ll E_\text{b}$, or $ Q \leq \frac{1}{\sqrt[3]{e}}$ (bottom wedge of figure \ref{fig:logQloge}). We will also have the black hole start arbitrarily close to extremality, which means that it starts with $E_\text{} \ll \omega_\text{sr} \ll E_\text{b} $, the quantum superradiant-dominated regime, with its evaporation rate captured by the dominant $l=0$ mode and given by
\begin{equation}\label{eq:smBHfin1}
    \frac{dE}{dt} =
    \frac{32 \, \sqrt{2}}{105 \, \pi^2} \,
    (eQ)^{4} \,
    \bigg( \frac{\sqrt{G} E_\text{b}}{e} \bigg)^{\!\frac{1}{2}}
    \frac{1}{G Q^2} \, .
\end{equation}
We remind the reader that $eQ$ is small in most of this wedge, so we can use the small charge limit of our equations. As the black holes moves away from extremality, the energy eventually surpasses the superradiant scale and the black hole crosses into the quantum neutral-like regime with $\omega_\text{sr} \ll E_\text{} \ll E_\text{b}$ where the dominant contribution to the rate is given by
\begin{equation}\label{eq:smBHfin2}
    \frac{dE}{dt} =
    \frac{16\sqrt{2}}{105 \, \pi^2} \,
    \bigg( \frac{\sqrt{G} E}{Q} \bigg)^{\!2} \,
    \left( \frac{E_\text{}}{E_\text{b}} \right)^{\!\frac{3}{2}}
    \frac{1}{G Q^2} \,.
\end{equation}
The energy continues to rise, and eventually the semi-classical  regime with $\omega_\text{sr} \ll E_\text{b} \ll E_\text{}$ is reached, where the evaporation rate just reduces to the semi-classical answer 
\begin{equation}\label{eq:smBHfin3}
    \frac{dE}{dt} =
    \frac{1}{30 \pi} \,
    \left( \frac{\sqrt{G} E}{Q} \right)^{\!2}
    \frac{1}{G Q^2} \, .
\end{equation}
Eventually, the black hole leaves the region of extremality $\frac{\sqrt{G} E}{Q} \ll 1$. We can see that the evaporation history, at least while it is in the near-extremal region, is significantly different from the one predicted by the semi-classical formula - it starts out emitting superradiant radiation that is enhanced compared to the semi-classical answer, then it emits neutral-like radiation that is suppressed compared to semi-classical, and finally it exits the quantum regime. Given that the quantum superradiant-dominated rate is strongly suppressed compared to the neutral-like rate (their ratio goes like $(\frac{e}{E})^{7/2} \ll 1$), most of the time in the near-extremal regime is spent emitting superradiant radiation, and since this is enhanced compared to semi-classical rate, the black hole spends overall less time in the near-extremal regime than would be expected based on semi-classical theory.

In the above, we assumed that charged emission drives the black hole away from extremality. We can check that this is indeed the case using the energy and charge emission rates. In the black hole parameter space $(M,Q)$, the extremal line is given by $M=Q$. As long as the slope of the evaporation trajectory in this space has a slope smaller than one (the black hole moves towards smaller $Q$ and $M$), the black hole will be moving away from this line, away from extremality. This slope can be found from the energy and charge emission rates as follows\footnote{
    Note the misleading use of notation. $M$ stands for the total black hole energy, $E$ stands for energy above extremality, and $dE/dt$ does not stand for the derivative of $E$, but $M$!}
\begin{equation}\label{eq:ratiocond}
    \frac{dM}{dQ} = \frac{dE}{dt} \bigg/ \frac{dQ}{dt} \leq 1 \, .
\end{equation}
In the quantum superradiant regime, these rates are given by \eqref{eq:srqleq0} and \eqref{eq:srqleq0Q}, which gives the following value for the slope
\begin{equation}\label{eq:ratiosmBH}
    \frac{dM}{dQ} = \frac{4}{7} \, .
\end{equation}
This proves that the black hole is driven away from extremality in this regime. For both the quantum and semi-classical neutral-like regimes, we can see from equations \eqref{eq:nlqleq0Q} and \eqref{eq:nlscleq0Q} that $dQ/dt$ has one less power of $QT$ than $dE/dt$, which implies that the slope $dM/dQ$ is of the order of $QT\ll1$. Therefore, it is nearly horizontal and the black hole moves away from the extremal line.

In addition to charged radiation, the black hole also emits neutral radiation. If we assume this is a universe like ours where there is no (massless) neutral scalar, but there is a photon-like particle in addition to the graviton, we can use the results of \cite{Brown:2024} for the evaporation rates of these particles. In the quantum superradiant regime, due to the fact that $T \ll e$, neutral radiation, which is of the order of a power of $QT$, is completely suppressed compared to charged radiation, which is proportional to $eQ$. In the two neutral-like regimes that follow, both charged and neutral radiation are proportional to a power of $T$, which is determined by the conformal dimension $\Delta$. For the dominant $l=0$ mode, for our charged scalar, $\Delta = \tfrac{1}{2} + \sqrt{\tfrac{1}{4} - (eQ)^2} \approx 1 $ when $eQ \ll 1$. For the photon and graviton, \cite{Brown:2024} found the dominant modes to have $\Delta=l+1=2$ and $\Delta=3$ for the photon and graviton, respectively.\footnote{
    In fact, if the black hole has zero angular momentum, it cannot radiate single photons/gravitons, as they carry away angular momentum. Instead it has to radiation two photons/gravitons at once with equal and opposite angular momentum. Quantum-mechanically this process is highly suppressed, which results in an even higher suppression in temperature for the evaporation rate.}
This means that they can also be safely ignored relative to the emission of charged particles. If there are also heavier charged particles that the black hole can emit, their emission rates will also be suppressed because the conformal dimension grows with mass.

For completeness, we comment on the mid-sized black holes (middle wedge in figure \ref{fig:logQloge}), where the quantum rates for charged radiation just reduce to the semi-classical in all available energy regimes. A mid-sized black hole is one with $E_\text{b} \ll \omega_\text{sr}$ and $Q \leq \frac{1}{2e}$, so $ \frac{1}{\sqrt[3]{e}} \leq Q \leq \frac{1}{2e}$. If we start very close to extremality, the black hole goes from the $E_\text{} \ll E_\text{b} \ll \omega_\text{sr} $ regime to $E_\text{b} \ll E_\text{} \ll \omega_\text{sr}$. Both of these are semi-classical superradiant-dominated regimes. As the energy keeps growing, the semi-classical neutral-like regime $E_\text{b} \ll \omega_\text{sr} \ll E_\text{} $ is reached, and then eventually the black hole leaves the near-extremal region. Note that something new happens for these black holes. As the black hole emits radiation, it also loses charge, which means that its trajectory in figure \ref{fig:logQloge} is a leftward straight line. Therefore, if the black hole starts in the middle wedge, there is a possibility that it crosses into the bottom wedge, while still in the near-extremal regime. For the process we considered, we assumed that it leaves extremality before reaching the bottom wedge, but one could also imagine a more complicated evaporation history, where the initially mid-sized black hole becomes small while still in the near-extremal region, passing along the way through a richer sequence of energy regimes, some of which are possibly quantum.

Finally, for large black holes, those with $Q \geq \frac{1}{2e}$, when $eQ$ is not too big, the story is similar to the mid-sized black holes. In the limit that $eQ$ is big, the Gibbons suppression kicks in for charged emission and the black hole, through neutral emission, has sufficient time to re-enter the near-extremal region, once it has left it. This is just the story of the black holes studied in \cite{Brown:2024}, and was also briefly summarized in the introduction.

\subsection*{Evaporation history of near-extremal Kerr}

For near-extremal Kerr radiation, we argued that for all modes the superradiant scale is always above the breakdown scale, and therefore the quantum-corrected rate is to leading order always given by the semi-classical formula \eqref{eq:Kerrrate}. We also argued that emission is dominated by the superradiant frequencies for all modes. We saw that the dominant modes are those with $l=m$, and for almost all of these we found that the conformal dimension $\Delta$ can be both real and complex on the frequency range $0 \leq \omega \leq \omega_\text{sr}$. The evaporation rates are computed by equations \eqref{eq:Kerrratefin} ($\Delta$ is real) and \eqref{eq:Kerrrateifin} ($\Delta$ is complex) for single modes. The summed evaporation rate (over the first 10 or so modes) can be found in table \ref{tab:Hfin}. For our Universe, where we just have a photon and graviton, the evaporation rate is given by
\begin{equation}\label{eq:Kfin}
    \frac{dE}{dt}\bigg|_\text{}^{(\text{ph+gr})} =
    0.0244 \,
    \frac{1}{G a^2} \, .
\end{equation}
We did not explicitly include the "neutral" $m=0$ mode, since it scales with $T$ and can be ignored. Using \ref{tab:Hfin}, we can also find the ratio of photon versus graviton radiation for near-extremal Kerr. For the emitted energy and angular momentum we find the following ratio
\begin{equation}\label{eq:phvsgrE}
\begin{split}
    \text{energy: }
    5.0\% \ \text{ photon \ vs. }
    \ 95.0\% \ \text{ graviton} \, , \\
    \text{ang.\,mom.: }
    5.5\% \ \text{ photon \ vs. }
    \ 94.5\% \ \text{ graviton} \, .
\end{split}
\end{equation}

We can check that these rates drive the black hole away from extremality, the $M=a$ line in $M$ vs. $a$ parameter space, which is true if
\begin{equation}\label{eq:Kratiocond1}
    \frac{dM}{da} \leq 1 \, .
\end{equation}
We can rewrite the left side using $a = J/M$ in terms of the ratio of the energy and angular momentum flux
\begin{equation}\label{eq:Kratiocond2}
    \frac{dM}{da} = \frac{M \, \frac{dM}{dJ}}{1 - a \, \frac{dM}{dJ}} \, .
\end{equation}
Plugging this into \eqref{eq:Kratiocond1} and using the near-extremal condition $M\approx a$ gives the following bound for $dM/dJ$
\begin{equation}\label{eq:Kratiocond3}
    \frac{dM}{dJ} \leq \frac{1}{2\sqrt{G J}} \, .
\end{equation}
Using the rates from table \ref{tab:Hfin}, we find for the scalar, photon and graviton the following 
\begin{equation}\label{eq:Kratiocond4}
    \frac{dM}{dJ} = 0.373 \, \frac{1}{\sqrt{G J}} \, , \qquad
    \frac{dM}{dJ} = 0.371 \, \frac{1}{\sqrt{G J}} \, , \qquad
    \frac{dM}{dJ} = 0.409 \, \frac{1}{\sqrt{G J}} \, ,
\end{equation}
which confirms that the emission of each of these fields drives the black hole away from extremality confirming the numerical results of \cite{Page:1976b}.

Note that even though the evaporation rate is not corrected, the Schwarzian theory still gives corrections to the thermodynamics, as long as the black hole is below the breakdown scale \cite{Kapec:2023,Rakic:2023}. These evaporation rates can be used to address the following question, which was raised in \cite{Rakic:2023}. The time scale associated with these quantum corrections to the thermodynamics is given by the inverse of the breakdown scale, so $t_\text{thermo} \sim a^3$. The question was whether the black hole spends enough time below the breakdown scale for these corrections to be noticeable. Integrating the rates found in table \ref{tab:Hfin} from $E=0$ to $E=E_\text{b}$, we find the time to be of order $t_\text{evap} \sim 1 / a$, which is very short compared to the thermodynamic time scale.

\subsection*{Instanton in the near-horizon region}

As mentioned in the introduction, an alternative way of deriving the emission rate of charged particles, is using the instanton method \cite{Coleman:1977a, Coleman:1977b, Coleman:1980}. This amounts to computing the rate of decay of the black hole to a lower energy state, through the emission of a particle-antiparticle pair. The rate is given in terms of the on-shell Euclidean action of a geometry called the instanton, which consists of the black hole geometry corrected by the worldline of the pair \cite{Brown:1988}. The instanton associated with the emission of a light charged particle from a large near-extremal Reissner-Nordström black hole was found in the full 4D geometry by \cite{Brown:2024}, which resulted in the same exponentially suppressed Gibbons decay rate found from the wave equation. One benefit of doing the gravitational instanton calculation is that it incorporates the gravitational backreaction to the pair production process, which leads to terms that correct the decay rate.

To lay some groundwork for a future analysis of this effect, we point out an alternative way of doing the calculation. One can work directly in the AdS$_2$ part of the near-horizon region, where the instanton was found in \cite{Brown:1988}. In Appendix B of \cite{Pioline:2005}, it was shown that the decay rate corresponding to this instanton also gives the Gibbons result. We suggest checking whether other geometrical properties of the instanton (its radius, conical angle, etc.) agree with the ones found from the 4D calculation in the near-extremal limit, as these quantities play an important role in the analysis of quantum fluctuations on top of this geometry, including those coming from the Schwarzian mode. The advantage of doing the 2D calculation over the 4D one is that it can easily be adapted to the near-horizon region of Kerr, which also has an AdS$_2$ structure, and where the 4D calculation would be significantly harder. In fact, we saw in subsection \ref{ss:KCDelta} that at complex $\Delta$ we obtain a Gibbons-like exponential suppression with the same form for the emission rate in Kerr \eqref{eq:Kgreyiapprox} as in Reissner-Nordström, which provides further motivation for the study of instantons.

\subsection*{Future directions}

For the "mid-sized" and large black holes in Reissner-Nordström (upper two wedges in figure \ref{fig:logQloge}) and for all Kerr black holes, we argued that the quantum-corrected evaporation rate \eqref{eq:RNqrate} to leading order is always given by the semi-classical formula \eqref{eq:RNscrate}. It would be interesting to find the quantum corrections to this semi-classical leading order term in the relevant expansion parameter, which for the neutral-like radiation was $E_\text{} / E_\text{b}$, and for the superradiant-dominated radiation was $\omega_\text{sr} / E_\text{b}$. Furthermore, the evaporation rate was obtained using the first-order term in a perturbative expansion. This expansion involves the same three energy scales and relies on a certain combination to be small. Therefore, one would also need to check whether the second order term in this perturbative expansion would be comparable to the sub-leading quantum correction at first order perturbation theory. Going to higher order in perturbation theory would also allow us to compute processes involving the emission of two or more quanta, such as the di-photon/graviton emission studied in \cite{Brown:2024}.

In the same spirit, a more interesting question is the following. For large Reissner-Nordström black holes and for certain Kerr modes, we found that $\Delta$ was complex on a subset of frequencies. We only argued that for these modes, the superradiant scale is always above the breakdown scale, which implies that the leading order term is always captured by the semi-classical formula, and therefore we were able to get away without computing the quantum-corrected rate, but only the semi-classical, which only relies on the wave equation, where $\Delta$ is just another parameter, and it being complex does not represent a problem. But we could still ask how would the quantum-corrected rate be computed in this case, i.e. how would Schwarzian theory with a complex $\Delta$ work in the effective string formalism? A complex $\Delta$ is a sign of instabilities in the 2D theory in the AdS$_2$ part of the near-horizon region. These are closely related to the instanton calculation mentioned earlier (see e.g. \cite{Pioline:2005, Kim:2008}). It would be interesting to revisit this instanton calculation with the goal of developing a better understanding of complex $\Delta$ in Schwarzian theory. We plan to explore this question in future work.

\subsection*{Acknowledgements}\label{acknowl}

It is a pleasure to thank my advisor Mukund Rangamani for the valuable discussions we had over the course of this work, and for reading a final draft and providing me with feedback. I was supported by funds from the University of California and U.S. Department of Energy grant DE-SC0009999. I would also like to thank the Kavli Institute for Theoretical Physics (KITP) for their hospitality during the program, “What is string theory? Weaving perspectives together”, which was supported by the grant NSF PHY-2309135 to KITP.

\appendix
\section{Mathematical appendix}\label{app:math}

\subsection*{Ordinary and confluent hypergeometric functions}

The ordinary hypergeometric function is defined as follows
\begin{equation}\label{eq:Fdef}
    F(a,b,c;z) = \sum_{n=0}^{\infty} \frac{(a)_n \, (b)_n}{(c)_n} \frac{z^n}{n!} \, ,
\end{equation}
where $(a)_n$ is the rising Pochhammer symbol. For our purposes $z$ is a negative real number, and for the combination of parameters $a,b,c$ that we encounter, this series is convergent. The asymptotics of this functions is given by
\begin{equation}\label{eq:Flarge}
    \begin{split}
        F &(a,b,c; -x) \rightarrow \,
        \frac{\Gamma(b-a) \, \Gamma(c)}{\Gamma(b) \, \Gamma(c-a)} \, x^{-a}
        +
        \frac{\Gamma(a-b) \, \Gamma(c)}{\Gamma(a) \, \Gamma(c-b)} \, x^{-b} \, ,
        \qquad x \gg 1 \, .
    \end{split}
\end{equation}
The confluent hypergeometric function is defined as follows
\begin{equation}\label{eq:Mdef}
    M(a,b;z) = \sum_{n=0}^{\infty} \frac{(a)_n}{(b)_n} \frac{z^n}{n!} \, .
\end{equation}
For our purposes $z$ is a positive imaginary number. The series is convergent everywhere, as long as $b$ is not zero or a negative integer, which for us is the case. The asymptotics of this functions is given by
\begin{equation}\label{eq:Mlarge}
    \begin{split}
        M(a,b,z) \, \sim \,
        \frac{\Gamma(b)}{\Gamma(a)} \, z^{a-b} \, e^z \, + \,
        \frac{\Gamma(b)}{\Gamma(b-a)} \, (-z)^{-a} \, , \qquad
        |z| \gg 1 \, .
    \end{split}
\end{equation}

\subsection*{Gamma function}

When the Gamma function has a complex argument with an integer real part, it simplifies as follows
\begin{equation}\label{eq:Gimag1}
    |\Gamma(1+n+ib)|^2=
    \frac{\pi b}{\sinh(\pi b)}
    \prod_{k=1}^{n} (k^2 + b^2) \, , \ \ \
    |\Gamma(-n+ib)|^2=
    \frac{\pi}{b \, \sinh(\pi b)}
    \prod_{k=1}^{n} (k^2 + b^2)^{-1} \, ,
\end{equation}
Similarly, when the real part is a half-integer we have
\begin{equation}\label{eq:Gimag2}
    |\Gamma(\tfrac{1}{2}\pm n+ib)|^2=
    \frac{\pi}{\cosh(\pi b)}
    \prod_{k=1}^{n} \left( (k-\tfrac{1}{2})^2 + b^2 \right)^{\pm1} \, .
\end{equation}
The product is absent when $n=0$ in both of the above equations. The asymptotic expansion of the Gamma function when the argument is real is given by the Stirling approximation
\begin{equation}\label{eq:GappStir}
    \Gamma(x) \sim
    \sqrt{2 \pi x } \, x^{(x-1)} \, e^{-(x-1)} \, ,
    \qquad
    x \gg 1 \, .
\end{equation}
The asymptotics of the absolute value when the argument is complex, and the imaginary part is large is given by
\begin{equation}\label{eq:Gappimag}
    |\Gamma(x+iy)|^2 =
    2 \pi \, |y|^{2x-1} \, e^{-\pi |y|} \, ,
    \qquad
    |y|\gg 1 \, .
\end{equation}
This is from the NIST Digital Library of Mathematical Functions \cite{NIST:DLMF}. The asymptotics of the hyperbolic trigonometric functions is given by
\begin{equation}\label{eq:hypapp}
\begin{split}
    \sinh(x) \sim \, \mathrm{sgn}(x) \, e^{|x|} \, ,
    \qquad
    \cosh(x) \sim 2 \, e^{|x|} \, ,
    \qquad
    |x| \gg 1 \, ,
\end{split}
\end{equation}
which appears as a special case of the previous equation.

\section{Determining the normalization constant in the quantum rate}\label{app:Nsq}

To determine the normalization constant $|\mathcal{N}|^2$, we start by isolating the energy-dependent part of evaporation rate \eqref{eq:RNqrate0}, which comes from the density of states and the matrix element
\begin{equation}\label{eq:Edep}
    \sinh \! \left(
    2\pi \sqrt{2E_\text{f}/E_\text{b}}
    \right)
    \left|
    \Gamma\!\left(
    \Delta \!+\!
    i \sqrt{2E_\text{f}/E_\text{b}} \!+\!
    i \sqrt{2E_\text{i}/E_\text{b}}
    \right)
    \right|^2
    \left|
    \Gamma\!\left(
    \Delta \!+\!
    i \sqrt{2E_\text{f}/E_\text{b}} \!-\!
    i \sqrt{2E_\text{i}/E_\text{b}}
    \right)
    \right|^2 \, ,
\end{equation}
where $E_\text{i} = E$ and $E_\text{f} = E - \omega_\text{eff}$. We will be taking two limits in the calculation, the high energy limit $|E-\omega_\text{eff}| \gg E_\text{b}$ and $E  \gg E_\text{b}$, and the low frequency limit $\omega \ll E+e$. We will break up the calculation into three parts, each focusing on one factor in \eqref{eq:Edep}. We start with
\begin{equation}
    \mathcal{A} =
    \sinh \! \left(
    2\pi \sqrt{2E_\text{f}/E_\text{b}}
    \right)
    \left|
    \Gamma\!\left(
    1 \!+\!
    i \sqrt{2E_\text{f}/E_\text{b}} \!+\!
    i \sqrt{2E_\text{i}/E_\text{b}}
    \right)
    \right|^2
    \left|
    \Gamma\!\left(
    1 \!+\!
    i \sqrt{2E_\text{f}/E_\text{b}} \!-\!
    i \sqrt{2E_\text{i}/E_\text{b}}
    \right)
    \right|^2 \, ,
\end{equation}
which is what we would have had if $\Delta = 1$. The Gamma functions can be written in terms of hyperbolic trigonometric functions using \eqref{eq:Gimag1} as follows
\begin{equation}
    \mathcal{A} =
    -\frac{2\pi^2 \omega_{\text{eff}}}
    {E_\text{b}}
    \frac{
    \sinh \! \left(
    2\pi\sqrt{2(E_\text{}\!-\!\omega_{\text{eff}})/E_\text{b}}
    \right)}
    {\sinh \! \Big(
    \pi\sqrt{2(E_\text{}\!-\!\omega_{\text{eff}})/E_\text{b}} \!+\!
    \pi\sqrt{2E_\text{}/E_\text{b}}
    \Big)
    \sinh \! \Big(
    \pi\sqrt{2(E_\text{}\!-\!\omega_{\text{eff}})/E_\text{b}} \!-\!
    \pi\sqrt{2E_\text{}/E_\text{b}}
    \Big)} \, ,
\end{equation}
Using a trigonometric product identity, the denominator can be rewritten as follows
\begin{equation}
    \mathcal{A} =
    \frac{4 \pi^2 \omega_{\text{eff}}}{E_\text{b}}
    \frac{
    \sinh \! \Big(
    2\pi\sqrt{2(E_\text{}-\omega_{\text{eff}})/E_\text{b}}
    \Big)}
    {\cosh \! \Big(
    2\pi \sqrt{2E_\text{}/E_\text{b}} 
    \Big) -
    \cosh \! \Big(
    2\pi \sqrt{2(E_\text{}-\omega_\text{eff})/E_\text{b}}
    \Big)} \, .
\end{equation}
We take the high energy limit and simplify, which gives
\begin{equation}
    \mathcal{A} =
    \frac{4 \pi^2 \omega_{\text{eff}}}{E_\text{b}}
    \frac{1}
    {\exp \! \Big(
    2\pi \sqrt{2E_\text{}/E_\text{b}} -
    2\pi \sqrt{2(E_\text{}-\omega_\text{eff})/E_\text{b}}
    \Big) \ - \
    1} \, .
\end{equation}
In the low frequency limit, we can expand the square root, and rewrite $E$ in terms of $T$ using equation \eqref{eq:drET}, which gives the final form for this factor
\begin{equation}
    \mathcal{A} =
    \frac{4 \pi^2 \omega_{\text{eff}}}{E_\text{b}}
    \frac{1}
    {e^{\omega_{\text{eff}}/T}-1} \, .
\end{equation}
The next factor in \eqref{eq:Edep} is given by
\begin{equation}
    \mathcal{B} =
    \frac
    {\Big| \Gamma \left(
    \Delta +
    i\sqrt{2(E - \omega_{\text{eff}})/E_\text{b}} + 
    i \sqrt{2E/E_\text{b}}
    \right) \Big|^2}
    {\Big| \Gamma \left(
    1 + 
    i\sqrt{2(E-\omega_{\text{eff}})/E_\text{b}} + 
    i\sqrt{2E/E_\text{b}}
    \right) \Big|^2} \, .
\end{equation}
In the low frequency limit, this simplifies to
\begin{equation}
    \mathcal{B} =
    \frac
    {\Big| \Gamma \left(
    \Delta+2i\sqrt{2E/E_\text{b}}
    \right) \Big|^2}
    {\Big| \Gamma \left(
    1+2i\sqrt{2E/E_\text{b}}
    \right) \Big|^2} \, .
\end{equation}
Using the asymptotic Gamma function expansion \eqref{eq:Gappimag}, we can take the high energy limit, which gives
\begin{equation}
    \mathcal{B} =
    \left(
    \frac{2 \sqrt{2E}}{\sqrt{E_\text{b}}}
    \right)^{2(\Delta-1)} \, ,
\end{equation}
where we rewrote $E$ in terms of $T$ using \eqref{eq:drET}, and used the fact that $Q^3 = 1/E_\text{b}$. We note that we also could've taken the limit in the opposite order, which would've resulted in the same expression. The third factor in \eqref{eq:Edep} is given by
\begin{equation}
    \mathcal{C} =
    \frac
    {\left|
    \Gamma \left(
    \Delta+i\sqrt{2(E-\omega_{\text{eff}})/E_\text{b}} - i\sqrt{2E/E_\text{b}}
    \right) \right|^2}
    {\Big|
    \Gamma \left(
    1+i\sqrt{2(E-\omega_{\text{eff}})/E_\text{b}} - i\sqrt{2E/E_\text{b}}
    \right) \Big|^2}
\end{equation}
In the low frequency limit, we can expand the square root, which gives
\begin{equation}
    \mathcal{C} =
    \frac
    {\Big| \Gamma \left(
    \Delta - i\sqrt{2} \, \omega_\text{eff}\,/\sqrt{E_\text{b} E}
    \right) \Big|^2}
    {\Big| \Gamma \left(
    1 - i\sqrt{2} \, \omega_\text{eff}\,/\sqrt{E_\text{b} E}
    \right) \Big|^2} \, .
\end{equation}
Then, we take the high energy limit using \eqref{eq:Gappimag} and find
\begin{equation}
    \mathcal{C} =
    \left(
    \frac{\sqrt{2} \, \omega_\text{eff}}
    {\sqrt{E_\text{b} E}}
    \right)^{2(\Delta-1)} \, .
\end{equation}

Before comparing to the semi-classical rate \eqref{eq:RNscrate}, we also need to take the high energy limit of the $F-$dependent Gamma function term in that equation. On the frequency interval from $0$ to $E+e$, when $E \gg E_\text{b}$, we have
\begin{equation}
    |F| \sim
    \frac{|\omega_\text{eff}|}{T} \sim
    \frac{|\omega_\text{eff}|}{E}
    \sqrt{\frac{E}{E_\text{b}}} \gg 1 \, ,
\end{equation}
where we used \eqref{eq:drET} to write $T$ in terms of $E$. On this interval, $|\omega_\text{eff}|/E$ is a number between $e/E$ and $1$. Therefore, without assuming any relationship between $E$ and $e$, in the high energy limit $|F| \gg 1$. Using this, we can expand the $F-$dependent Gamma functions as follows
\begin{equation}
    \frac{|\Gamma(\Delta-2iF)|^2}
    {|\Gamma(1-2iF+i\mu)|^2} =
    \left(
    \frac{\sqrt{2} \, \omega_\text{eff}}
    {\sqrt{E_\text{b} E}}
    \right)^{2(\Delta-1)}
    e^{-\text{sgn}(\omega_\text{eff}) \mu} \, ,
\end{equation}
where $\text{sgn}$ is the sign function and comes from that fact that $F$ can be positive or negative.

Plugging $\mathcal{A}, \mathcal{B}, \mathcal{C}$ into the quantum rate and comparing to the semi-classical gives the normalization constant
\begin{equation}
    \left| \mathcal{N} \right|^2 =
    \frac{1}{\pi} \,
    2^{2\Delta-2} \,
    \frac{\Gamma(2\Delta)}
    {\Gamma(2\Delta-1)^2} \,
    \left| \Gamma(\Delta + i\mu) \right|^2 \,
    e^{-\text{sgn}(\omega_\text{eff}) \mu} \, .
\end{equation}


\bibliographystyle{JHEP}
\bibliography{Kerr}

\end{document}